\documentclass{aa}

\usepackage{graphicx}
\usepackage{lscape}
\usepackage{txfonts}
\usepackage[pagewise, switch]{lineno}
\nolinenumbers
\begin{document}

   \title{Metallicities and ages for star clusters and their
   surrounding fields in the Large Magellanic Cloud
   \thanks{Full Table~\ref{tab:phot} is available on the Araucaria Project webpage
   (https://araucaria.camk.edu.pl/) and at the CDS via anonymous ftp
   to cdsarc.u-strasbg.fr (130.79.128.5) or via http://cdsarc.u-strasbg.fr/viz-bin/cat/J/A+A/666/A80}}
   \titlerunning{Metallicities and ages for star clusters and their
   surrounding fields in the Large Magellanic Cloud}

   \author{W. Narloch\inst{1}, %\fnmsep\thanks{contact author},
          G. Pietrzy\'nski\inst{1,2}, W. Gieren\inst{1}, A.~E. Piatti\inst{3,4},
          P. Karczmarek\inst{1}, M. G\'orski\inst{1,2}, D. Graczyk\inst{5}, \\
          R. Smolec\inst{2}, G. Hajdu\inst{2}, K. Suchomska\inst{2},
          B. Zgirski\inst{2}, P. Wielg\'orski\inst{2},
          B. Pilecki\inst{2}, M. Taormina\inst{2}, M. Ka\l uszy\'nski\inst{2}, \\
          % et al.
          W. Pych\inst{2}, G. Rojas Garc\'ia\inst{2}
           \and
          M.~O. Lewis\inst{2}
          }
   \authorrunning{W. Narloch et al.}

   \institute{Univesidad de Concepci\'on, Departamento de Astronomia,
              Casilla 160-C, Concepci\'on, Chile\\
              \email{wnarloch@astro-udec.cl}
         \and
             Nicolaus Copernicus Astronomical Center, Polish Academy of Sciences,
              Bartycka 18, 00-716, Warsaw, Poland
             % \thanks{The university of heaven temporarily does not
             %         accept e-mails}
         \and
            Instituto Interdisciplinario de Ciencias B\'asicas (ICB), CONICET-UNCUYO,
            Padre~J. Contreras 1300, M5502JMA, Mendoza, Argentina
         \and
            Consejo Nacional de Investigaciones Cient\'{\i}ficas y T\'ecnicas (CONICET),
            Godoy Cruz 2290, C1425FQB,  Buenos Aires, Argentina
         \and
            Nicolaus Copernicus Astronomical Center, Polish Academy of Sciences,
            Rabia\'nska 8, 87-100 Toru\'n, Poland
             }

   \date{Received 20 February 2022; Accepted 7 July 2022}

% \abstract{}{}{}{}{}
% 5 {} token are mandatory

  \abstract
   {}
   {We study $147$ star clusters in the Large Magellanic Cloud (LMC) in order to
   determine their mean metallicities and ages, as well as the mean metallicities
   of $80$ surrounding fields. We construct an age--metallicity relation (AMR)
   for the clusters in the LMC.}
   {For this purpose, we used Str\"omgren photometry obtained with the SOI camera on
   the 4.1~m SOAR telescope. We derived the metallicities of individual stars utilizing
   a~metallicity calibration of the Str\"omgren $(b-y)$ and $m1$ colors
   from the literature. Cluster ages were determined from the isochrone fitting.}
   {We found the mean metallicity and age for $110$ star clusters. For the
   remaining $37,$ we provide an age estimation only. To the best of our knowledge,
   for $29$ clusters from our sample, we provide both the metallicity and age for the
   first time, whereas for $66$ clusters, we provide a~first determination of the
   metallicity, and for $43$ clusters, the first estimation of the age.
   We also calculated the mean metallicities for stars from $80$ fields around
   the clusters.
   The results were then analyzed for spatial metallicity and age distributions
   of clusters in the LMC, as well as their AMR.
   The old, metal-poor star clusters occur both in and out of the LMC bar region,
   while intermediate-age clusters are located mostly outside of the bar.
   The majority of star clusters younger than $1$~Gyr are located in the bar region.
   We find a good agreement between our AMR and theoretical models of the LMC chemical
   enrichment, as well as with AMRs for clusters from the literature.
   Next, we took advantage of $26$ stellar clusters from our sample which host
   Cepheid variables and used them as an~independent check of the correctness of
   our age determination procedure. We used period-age relations for Cepheids to
   calculate the mean age of a~given cluster and compared it with the age obtained
   from isochrone fitting.
   We find good agreement between these ages, especially for models taking into
   account additional physical processes (e.g., rotation).
   We also compared the AMR of the LMC and Small Magellanic Cloud (SMC) derived in a~uniform
   way and we note that they indicate possible former interaction between these two
   galaxies.
   The Str\"omgren photometry obtained for this study has been made publicly available.}
   {}


   \maketitle
%
%-------------------------------------------------------------------

\section{Introduction}

   The Large Magellanic Cloud (LMC) is a~Magellanic spiral galaxy (SB(s)m type)
   and a~satellite of the Milky Way (MW). It is located even closer to our Galaxy
   than its nearby companion on the sky -- the Small Magellanic Cloud (SMC).
   Its proximity, favorable face-on orientation and miscellaneous stellar population
   make it an attractive target for detailed astrophysical studies.
   The LMC harbors a~large and diverse system of star clusters, whose metallicities
   and ages follow a specific age--metallicity relation (AMR) and, as such, can serve as
   good tracers of the chemical evolution history of their host galaxy.

   Many studies indicate a~much more complicated star formation history
   \citep[SFH; e.g.,][]{PG1998,Carrera2008a,HZ2009,Rubele2012,Meschin2014,Palma2015,
   Perren2017} or AMR for star clusters
   \citep[e.g.,][]{Olszewski1991,Hill2000,Dirsch2000,Rich2001,LR2003,Kerber2007}
   in the LMC than in the SMC.
   Early studies of the cluster AMR in the LMC \citep[e.g., the spectroscopic study
   of][]{Olszewski1991} revealed a~mysterious gap in the star cluster formation
   between $10$ and $3$~Gyr ago \citep[with the sole exception of cluster ESO121-3
   and very recently also KMHK1592, as reported by][]{Piatti2022},
   which follows the initial formation of old, metal-poor globular clusters,
   and precedes more recent active formation of intermediate-age clusters.
   This so-called "age gap,' which is simultaneously a~metallicity gap, was
   later reported also by other authors
   \citep[e.g.,][and references therein]{HZ2009,Sharma2010,Kerber2007}, even though
   it has not been fully confirmed when only the field stars are considered.
   For example, \citet{PiattiGeisler2013} reported that no clear age gap in the
   field star formation is observed in their study of fields in the LMC main body,
   based on Washington photometry.
   Furthermore, based on their spectroscopic analysis of four fields to the north
   of the LMC bar, \citet{Carrera2008a} maintain that the disk and cluster AMR
   are similar. However, unlike clusters, there is no age gap in the field population.
   On the contrary, \citet{HZ2009} claim that such a~gap is evident in the field
   population of the bar, mostly omitted in previous studies.
   Also \citet{Meschin2014}, using optical photometry, identified two main star
-forming epochs with a~period characterized by a~lower activity in between;
   however, in their work that interval of time was shorter and lasted from
   $\sim$$8$ up to $\sim$$4$~Gyr ago.
   All the aforementioned authors agree that star formation in the LMC is continuing
   to this day.

   \citet{Olszewski1991} claimed that the metal abundance of star clusters of age
   $\sim$2~Gyr from inner and outer regions of the LMC are nearly identical ($-0.3$
   and $-0.42$~dex, respectively) and that there is only a~small radial abundance
   gradient visible in the cluster system. The difference in metallicity between
   various LMC regions, was later reported by other authors, for example, by
   \citet{Piatti2003b} based on their study of LMC clusters observed with Washington
   photometry. These authors have claimed that clusters from the inner disk are, on average,
   more metal-rich and simultaneously younger than those from the outer disk.
   Similar conclusions were reached by, for instance, \citet{Livanou2013} (based on
   Str\"omgren photometry), \citet{Pieres2016} (from Sloan bands).
   \citet{PiattiGeisler2013}, and \citet{Meschin2014} observed such an abundance
   gradient in the field population and postulated an outside-in star formation
   and chemical enrichment in the LMC.

   \citet{Olszewski1991} also mentioned, that the burst of cluster formation
   around $2$~Gyr ago in the outer regions could result from the interaction
   between the LMC and the MW, possibly also with the SMC. Similar
   interpretation of this phenomenon was given by, for instance, \citet{HZ2009}.

   A~second period of low cluster formation rate, lasting from about
   $200 - 700$~Myr ago, is reported by \citet[][and references therein]{LR2003}.
   Such a~period is not clearly visible in other works
   \citep[e.g.,][]{Glatt2010,Perren2017}. Nevertheless, some authors refer to multiple
   peaks in recent SFH of the LMC. For example, \citet{HZ2009} found peaks at
   roughly $2$~Gyr, $500$~Myr, $100$~Myr, and $12$~Myr. They also emphasized that
   the peaks at $500$~Myr and $2$~Gyr coincide well with similar peaks seen in
   the SMC. \citet{Glatt2010}, on the other hand, found two periods of enhanced
   cluster formation, but at $125$~Myr and $800$~Myr. The authors of these two
   studies argue that observed peaks in the recent cluster formation rate suggest
   a~common history of the LMC and SMC.

   In \citet[hereafter Paper~I]{Narloch2021}, we analyzed the AMR for $35$ star
   clusters in the SMC. A~similar analysis for the LMC, carried out in a~homogeneous
   manner, would therefore be a~valuable addition to the overall picture of the
   evolution and interaction of these two MW satellites.
   Additionally, most of the results presented in this work come from the central
   regions of the LMC, which has not been well studied previously.

   In this work, we obtain Str\"omgren photometry of stars belonging to the clusters
   and the fields in the LMC, and using the Str\"omgren metallicity
   calibration of \citet{Hilker2000}, we determine their metallicities.
   We use the obtained metallicity values of individual stars to calculate mean
   metallicities of the star clusters and their surrounding fields. We employ
   theoretical isochrones to estimate the age of each cluster.
   The resulting AMR is derived in a~homogeneous way, which allows us to investigate
   the chemical history of the LMC, in order to verify the evolutionary scenarios
   presented in the extensive literature described above. To maintain homogeneity
   in the study, we repeated the reduction and photometry, and then we reanalyzed the clusters
   that have already been reported based on the same dataset
   \citep[e.g.,][]{Piatti2018,Piatti2019,Piatti2020}.
   Moreover, we use recent reddening maps of the Magellanic Clouds
   \citep{Gorski2020,Skowron2021} and the most recent distance determination to
   the LMC, which is precise to $1\%$ \citep{Pietrzynski2019}. The positions and
   radii of the star clusters were taken from the catalog of \citet{Bica1999}.
   A~uniform handling of data allowed us to compare the resulting AMR of the LMC
   to that of the SMC from Paper~I.

   This paper is organized as follows. In Sect.~\ref{sec:obsred}, we describe the
   observations, reduction procedure, and analysis techniques. In Sect.~\ref{sec:results},
   we analyze our results for the spatial metallicity and age distributions of
   star clusters from our sample. We describe our resulting AMR and compare to SFH
   models from the literature. In Sect.~\ref{sec:discussion}, we discuss obtained
   cluster metallicities with the corresponding metallicities obtained with
   different methods in the literature, as well as the AMRs, with those found in
   the literature. We compare ages of the clusters hosting Cepheid variables with
   ages obtained from the period--age (PA) relations. Finally, we compare the AMR
   for the LMC and SMC, as derived in Paper~I.
   Section~\ref{sec:summ} provides the conclusions of this study.

%--------------------------------------------------------------------
\section{Observations and data reduction}
\label{sec:obsred}
%--------------------------------------------------------------------

   Images in three Str\"omgren filters ($v$, $b,$ and $y$) were collected within
   the Araucaria Project \citep{GPB2005} using the $4.1$~m Southern Astrophysical
   Research (SOAR) Telescope placed in Cerro Pach\'on in Chile, equipped with the
   SOAR Optical Imager (SOI) camera (program ID: SO2008B-0917, PI: Pietrzy\'nski).
   Observations were conducted during two runs: $17$, $18$, and $19$ December 2008
   and $16$, $17$, and $18$ January 2009. The SOI camera is a~mosaic of two E2V
   2k$\times$4k CCDs (read by a total of four amplifiers). The field of view is
   $5.26$$\times$$5.26$~arcmin$^2$ and the pixel scale is
   $0.077$~arcsec$\cdot$pixel$^{-1}$.
   During both runs, a $2$$\times$$2$ pixel binning was applied, giving an effective
   pixel scale of $0.154$~arcsec$\cdot$pixel$^{-1}$.
   Single images were taken in the  air mass range of $1.19 - 2.09$, and the average
   seeing was about $0.89$, $0.93$ and $0.99$~arcsec in $y$, $b,$ and $v$ filters,
   respectively. Table~\ref{tab:lmc} summarizes the information about the data set.
   Several fields were observed more than once. The repeated clusters have
   multiple records in Tables~\ref{tab:reddC} and \ref{tab:lmc} .

   During the reduction and analysis of the data, we followed the procedures described
   in Paper~I. After bias subtraction and flatfield correction, we performed profile
   photometry using the standard DAOPHOT/ALLSTAR package \citep{Stetson1987}, where
   the point spread function (PSF) was defined by a~spatially variable Gaussian.
   Very dense fields were divided into smaller overlapping subframes to reduce the
   PSF and background variability, where we were always careful to provide
   a~sufficient number of PSF stars; this number ranged from a~handful to over
   $200$ depending on the frame.
   Photometry was performed iteratively by gradually decreasing the detection threshold.
   In the last step, the images were inspected by eye and stars omitted in the automatic
   procedure were manually added to the final list of stars in a~given frame.
   {This step was performed particularly in dense fields with significant
   background gradient where mostly faint stars ($>17$~mag) or stars located close
   to much brighter companions were added. They accounted for a maximum of $10\%$ of all
   stars on the final list.
   Subsequently, the aperture corrections for each frame were calculated using
   the DAOGROW package \citep{Stetson1990}.

   The instrumental color-magnitude diagrams (CMDs) were then standardized for
   each chip of the camera separately, using the transformation equations from
   Paper~I and coefficients from Table~2 therein.
   The average errors of the photometry from DAOPHOT were $0.02$~mag in $V$, $0.03$~mag
   in $(b-y),$ and $0.05$~mag in $m1$ for stars with brightness $V<20$ mag.
   The astrometric solutions of images in the $y$ band were performed using the
   Gaia EDR3 catalog \citep{Gaia1,Gaia3,Lindegren2021} with subarcsec accuracy.
   The artificial star tests done using the ADDSTAR routine of the DAOPHOT package
   assured that the completeness of our master lists of stars is close to $100\%$
   for stars used for the metallicity determination.

%--------------------------------------------------------------------
\subsection{Selection of cluster members and field stars}
\label{ssec:sel}
%--------------------------------------------------------------------

   In the first step of our selection of cluster members, we rejected galactic
   foreground stars with significant proper motion (PM) values. We applied a~similar
   approach to that presented in \citet{Narloch2017}, where stars are excluded from
   the sample based on their location on the vector point diagram (VPD). To that
   end, we cross-matched our CMDs with the Gaia EDR3 catalog and calculated mean
   values of PMs of all stars from a~given field (M$_{RA}$, M$_{DE}$). These values
   were then subtracted from the individual PMs of stars in order to center the
   VPD. Then, mean and standard deviations of PMs (M$_{RA}$, M$_{DE}$, S$_{RA}$,
   S$_{DE}$) and PM errors (ME$_{RA}$, ME$_{DE}$, SE$_{RA}$, SE$_{DE}$) were
   calculated as well as total PMs ($\mu$) and their errors ($\sigma_{\mu}$).
   Because the number of stars in a field is often small, we decided to not divide
   them into magnitude bins. Next, the stars satisfying the conditions
   $\mu \leq 3 \cdot S$ and $\sigma_{\mu} \leq ME + 3 \cdot SE$ were retained.
   The procedure was iterated twice to ensure reliable removal of stars with high
   PMs.

   We adopted equatorial coordinates and sizes of star clusters taken mostly from
   \citet{Bica1999} and, for the case of the cluster OGLE-CL~LMC~478, from
   \citet{Pietrzynski1999b}.
   Stars lying outside of the cluster radii were classified as field stars.
   We did not perform a~statistical subtraction. Most of the clusters
   are small and placed close to each other in dense fields. Also, the small field
   of view of the camera does not provide good statistics for the field stars.
   All of these obstacles make it difficult to perform statistical subtraction correctly.
   On the other hand, populous star clusters located in sparse fields are marginally
   contaminated by field stars. In the end, the individual metallicites can help
   to disentagle cluster and field stars.

%--------------------------------------------------------------------
\subsection{Determination of reddening toward clusters}
\label{ssec:red}
%--------------------------------------------------------------------

   Reddening for each star cluster was determined using the two recent reddening
   maps of \citet[][hereafter G20]{Gorski2020} and \citet[][hereafter S21]{Skowron2021}.
   Both maps are based on red clump stars in the LMC, where S21 characterize a~much
   larger area of the sky than G20. As most of our clusters are much smaller than
   the field of view of the SOI camera, we calculated the reddening of a~given cluster
   or field in an area centered on the target with the G20 map resolution of three
   and five arcmin, respectively. The resolution of the S21 maps is
   $1.7\,\mathrm{arcmin} \times 1.7\,\mathrm{arcmin}$ in the central parts of
   the LMC and decreases in the outskirts down to about
   $27\,\mathrm{arcmin} \times 27\,\mathrm{arcmin}$.
   For clusters located in the area covered by both reddening studies, we adopted
   the average of both values ($E(B-V)_{GS}$), where $E(V-I)$ from S21 is converted
   into $E(B-V)$ with $E(B-V) = E(V-I)/1.318$.
   Due to the limited area of G20, for the most outlying clusters, the converted
   value from S21 was used directly. For several clusters distant from the LMC center
,   even the reddening data from S21 was not available (clusters marked with a--symbol
   in fourth column of Table~\ref{tab:reddC}), or the calculated reddening seemed to
   be incorrect for a~cluster (the isochrone with this reddening value clearly
   did not fit the CMD of a~given cluster).
   In such cases, we adopted reddening values from other literature sources (see
   Table~\ref{tab:lit}).

   The mean difference between the G20 and S21 reddening values for the star
   clusters and fields studied in this work is about $0.037$~mag, where S21 gives
   systematically smaller values than G20.
   Half of this value, rounded up ($\sigma_{E(B-V)_{GS}} = 0.019$~mag), was
   propagated into the systematic error on the derived metallicities, resulting
   from the reddening. The reddening values for magnitudes and colors were calculated using the following
   equations:
   $A_V = 3.14 \cdot E(B-V)$, $E(b-y) = 0.73 \cdot E(B-V),$ and
   $E(m1) = -0.25 \cdot E(B-V)$ \citep{Cardelli1989,ODonnell1994}.

   Differential reddening, namely, spatially-variable extinction (either
   internal or external to the star clusters), is an effect that can occur in
   studied fields, affecting stellar magnitudes and colors, which influences
   both metallicity and age estimations. It would manifest itself as a broadening
   of the metallicity and age distribution, and for objects with large differential
   reddening, it could cause a~systematic shift of the mean metallicity towards
   higher values, with larger statistical errors. In most of the clusters in our
   sample, we do not expect large reddening variations, although isolated cases
   may occur. The clusters most affected by differential reddening are likely
   located in the LMC bar.
   Although it cannot be precisely estimated solely based on our data, \citet{Milone2018}
   reported that the typical differential reddening in the LMC clusters is
   $\Delta E(B-V) \approx 0.003$~mag, which is considerably smaller than our
   photometric error.

%--------------------------------------------------------------------
\subsection{Metallicity calculation based on Str\"omgren colors}
\label{ssec:metal}
%--------------------------------------------------------------------

   The metallicites for red giants and supergiants in our fields were calculated
   based on the existing calibration of the Str\"omgren colors $(b-y)$ and $m1$
   with metallicity ([Fe/H]). The adopted method gives the metallicities of
   individual stars determined nearly independent of their age
   \citep[e.g.,][]{Dirsch2000}.
   During the analysis, we adopted a~metallicity calibration of the
   Str\"omgren $m1$ versus $(b-y)$ two-color relation derived by \citet{Hilker2000},
   which is valid in the range of colors of $0.5<(b-y)<1.1$, and is given by
   following equation:

   \begin{equation}
   \label{eq:feh}
     \mathrm{[Fe/H]} = \frac{m1_0 + a1 \cdot (b-y)_0 + a2}{a3 \cdot (b-y)_0 + a4}
   ,\end{equation}

  \noindent where

   $$a1 = -1.277 \pm 0.050,\, a2 = 0.331 \pm 0.035,$$
   $$a3 = 0.324 \pm 0.035,\, a4 = -0.032 \pm 0.025.$$

   The metallicity errors of individual stars were calculated by performing a full
   error propagation, as done by \citet{Piatti2019} or as shown in Paper~I.
   For the derivation of the calibration, \citet{Hilker2000} used stars with
   spectroscopic metallicities on the \citet[][]{ZW1984} metallicity scale
   (hereafter the ZW84 scale).

   The procedure of the metallicity determination is described in detail in
   Paper~I. Here we provide a brief summary for reference.
   After dereddening the data (see Section~\ref{ssec:red}) we chose stars from
   the color range $0.5<(b-y)_0<1.1$, having $\sigma_{(b-y)_0}<0.1$ and
   $\sigma_{m1_0}<0.1$ ({calculated from the DAOPHOT error estimates}).
   Next, following \citet{Dirsch2000}, we introduced a~cut at the blue edge of
   the $m1_0$ vs. $(b-y)_0$ relation (marked in the left panels of
   Fig.~\ref{fig:ngc1651} and \ref{fig:ngc1903} with grey dotted line), where
   deviating stars cause a~bias toward metal-poor stars with larger metallicity
   errors.
   For the remaining stars, the mean and unbiased standard deviation were calculated
   and then recalculated after applying $3\sigma$ clipping. The resulting CMDs
   and $m1_0$ vs. $(b-y)_0$ relations were examined by eye to manually reject
   single stars deviating significantly, and the final values of the mean and the
   unbiased standard deviation were obtained.
   The statistical error of the mean metallicity was determined as an unbiased
   standard deviation divided by the square root of the number of stars used for
   the calculation.

   One of the main sources of systematic metallicity error is the reddening.
   A~$0.01$~mag increase in reddening increases the derived metallicity by about
   $0.05$~dex (see Paper~I). A~typical error resulting from the reddening adopted
   in the previous section corresponds to $\sigma_{\mathrm{[Fe/H]}} \approx 0.10$~dex;
   this value is used to calculate the systematic error of the mean metallicity
   of clusters and their surrounding fields.
   The effect of differential reddening in the studied fields was neglected during
   the calculation of the total systematic error, as it cannot be precisely estimated
   with the available data.

   Another source of systematic uncertainty is the precision of the $m1$ and $(b-y)$
   calibration to the standard system. This uncertainty causes a~bias in the
   metallicity of individual stars depending on their color. The effect is larger
   for bluer stars, leading to larger metallicity errors
   \citep[see Fig.~1 in][]{Dirsch2000}.
   To estimate this uncertainty we performed simulations as described in Paper~I.

   Metallicities derived from Str\"omgren colors are also affected by the
   contribution of CN molecules absorption which reduces the flux in $v$ filter.
   CN bands lead then to higher value of $m1$ index and as a~consequence N-enriched
   stars appear to be more metal-rich. The increase of the metallicity would decrease
   the age derived via isochrone fitting. The chemical anomalies were found in
   the ancient, massive globular clusters from our Galaxy
   \citep[e.g.,][]{Richter1999,HR2000}, as well as massive, intermediate-age star
   clusters ($\sim 2$~Gyr and older) of the Magellanic Clouds
   \citep[e.g.,][]{Martocchia2021,Martocchia2019,Hollyhead2018},
   but none were found in younger clusters \citep[e.g.,][]{Martocchia2017,Martocchia2021}.
   \citet{Martocchia2019} showed that chemical anomalies in the form of N spreads
   is a~strong function of age.
  \citet{Martocchia2021} presented a~spectroscopic data for two clusters from our
   sample: NGC1651 and NGC1978.
   In the case of NGC1651 only three stars out of $81$ used for the metallicity calculation
   have a measured CN index, while ten more lie outside the cluster radius that we adopted  and
   are classified as field stars. In the case of NGC1978, 6 stars out of
   $287$ have spectroscopic measurements and another 5 are classified as field stars.
   Rejection of these few N-enriched stars would not change the obtained mean
   metallicities of those clusters. CN bands increase photometric metallicities
   derived from Str\"omgren photometry but without detailed spectroscopic studies,
   we cannot account for this effect.

   The total systematic error of a~given mean metallicity is composed of the
   reddening and calibration errors added in quadrature.
   Tables~\ref{tab:reddC} and \ref{tab:reddF}, which summarize the measurements
   for the clusters and fields analyzed in this work, respectively, contain both
   the statistical and systematic (in parentheses) metallicity errors.

%--------------------------------------------------------------------
\subsection{Age determination}
\label{ssec:age}
%--------------------------------------------------------------------

   We determined the ages of the star clusters in our sample by performing
   an isochrone fitting.
   To that end, we utilized isochrones from the Dartmouth Stellar Evolutionary
   Database\footnote{http://stellar.dartmouth.edu/models/isolf\_new.html}
   \citep[][hereafter the Dartmouth isochrones]{Dotter2008} and the Padova database
   of stellar evolutionary tracks and isochrones available through the CMD 3.3
   interface\footnote{http://stev.oapd.inaf.it/cgi-bin/cmd\_3.3} \citep{Marigo2017}
   calculated with the PARSEC \citep{Bressan2012} and COLIBRI \citep{Pastorelli2019}
   evolutionary tracks (hereafter the Padova isochrones).
   Most of the ages were estimated with the Padova isochrones sets, as they cover
   a~wide range of possible age values, while the Dartmouth isochrones were available
   only for the $1-15$~Gyr range; thus, they were too old for most of our objects.
   Where possible, both isochrones were employed for the determination.

   The isochrones were fitted for a~specific metallicity of a~given cluster at
   a~fixed distance to the LMC $(m-M)_{\mathrm{LMC}} = 18.477$~mag, as reported by
   \citet{Pietrzynski2019}.
   In cases where the isochrones for the reddening calculated from averaging G20
   and S21 maps clearly did not fit the CMD of a~given cluster, we adopted a~value
   from the literature and iterated the procedure.
   The age error of a~given cluster was defined as half the age difference between
   two marginally fitting isochrones selected around the best fitting isochrone.

   Adopting a~fixed distance to the LMC while fitting an isochrone is an approximation,
   as LMC star clusters can be located at different distances along the line of
   sight \citep[see e.g.,][]{Piatti2021}. Placing them all at the same distance
   introduces error in our age calculations. If a~given cluster turns out to be located
   in a~distance different than adopted, then it would have a~different age:
   would be younger or older depending on being farther or closer than adopted
   distance, respectively.
   \citet{SS2009} reported a~significant line of sight depth of the LMC bar
   ($4.0 \pm 1.4$~kpc) and disk ($3.44 \pm 1.16$~kpc). A~change of adopted cluster
   distance of a~half of the LMC bar depth, would result in a~change of log(Age)
   by $\sim$$0.03$. The~typical age error, however, estimated in the previous
   paragraph, is often higher ($\sim$$0.10$), so the error resulting from assuming
   a~fixed distance to the LMC has no significant impact on the final error.

%--------------------------------------------------------------------
\subsection{Str\"omgren photometry}
\label{ssec:phot}
%--------------------------------------------------------------------

   We publish our photometry for more than $600\,000$ stars having measurements
   in all three Str\"omgren $vby$ filters, and the consequently calculated
   $V$, $(b-y)$ and $m1$ values.
   The first five rows of the catalog are presented in Table~\ref{tab:phot}.
   The photometric errors come from the DAOPHOT package as well as the full error
   propagation of the transformation equations with coefficients from Table~2 in
   Paper~I.

%--------------------------------------------------------------------
\section{Results}
\label{sec:results}
%--------------------------------------------------------------------

   Figures~\ref{fig:ngc1651} and \ref{fig:ngc1903} show example two-color diagrams
   and CMDs of star clusters and their surrounding fields from our sample for:
   an intermediate-age stellar cluster, NGC1651, with a~well populated RGB
   (Fig.~\ref{fig:ngc1651}), and a~young star cluster, NGC1903, hosting Cepheid
   variables in its field (Fig.~\ref{fig:ngc1903}).
   Analogous examples where only one or two stars were used for metallicity
   determination are presented in Appendix~\ref{sec:appexample}.
   The left panels of Fig.~\ref{fig:ngc1651} and \ref{fig:ngc1903} present the
   dereddened $m1_0$ vs. $(b-y)_0$ relation, where stars used for the calculation
   of the mean metallicity of a~given cluster (upper panels) or field (lower panels)
   are color coded. The same stars are marked on the CMDs presented in the right
   panels. The best fitting Padova (turquoise) and Dartmouth (grey) isochrones for
   a~given cluster are shown in the upper right panels. The same isochrones are
   also plotted in the lower right panels to show their position relative to the
   field. Figure~\ref{fig:mapfeh} presents the spatial distribution of the mean
   metallicities of clusters and fields from our sample, while Fig.~\ref{fig:mapage}
   shows the on-sky distribution of estimated cluster ages. The histogram of
   metallicities of field stars is presented in Fig.~\ref{fig:fehhist}.
   The metallicity and age determinations done in this work are summarized in
   Tables~\ref{tab:reddC} and \ref{tab:reddF} and shown in Fig.~\ref{fig:am}.
   Multiple measurements for several clusters in Table~\ref{tab:reddC} are averaged
   in Fig.~\ref{fig:am}. Figure~\ref{fig:aminout} presents the AMR in the LMC bar
   and non-bar regions separately.

   Our sample covers $147$ star clusters in the LMC and $80$ fields associated with
   them. We calculated the mean metallicities, together with ages for $110$ clusters
   in total. The remaining $37$ clusters lack metallicity information, as
   there were no suitable stars to estimate it.
   Nevertheless, we were able to estimate cluster ages by
   adopting Padova isochrones for $\mathrm{[Fe/H]} = -0.40$~dex.
   Metallicities were determined for $66$ clusters in our sample for the first
   time, to the best of our knowledge. For $43$ clusters, we provide the first
   estimation of the age and for $29$ clusters, both of these values were obtained
   for the first time.
   Nine star clusters from our sample are ancient globular clusters, older than
   $10$~Gyr, with low metallicities similiar to their Galactic equivalents;   $17$ objects are intermediate-age clusters, with ages in the range of one to
   $10$~Gyr; and the remaining $121$ are clusters younger than one Gyr. There are
   $49$ clusters with at least 5 stars (up to 287) useful for metallicity
   calculation (these are marked in Fig.~\ref{fig:am} with filled and opened
   squares). Another $61$ objects had less than 5, but at least one star for
   metallicity calculation, which gives a~very poor statistic and makes the final
   value less reliable
   (these clusters are marked in Fig.~\ref{fig:am} with opened circles).
   Cluster ages were estimated as described in Sec.~\ref{ssec:age}.

%-------------------------------------------------------------
%                                             Two column Figure
%-------------------------------------------------------------
   \begin{figure*}
   \resizebox{\hsize}{!}
%      {\includegraphics[bb=10 20 100 300,clip]{./ngc339.pdf}}
            {\includegraphics[]{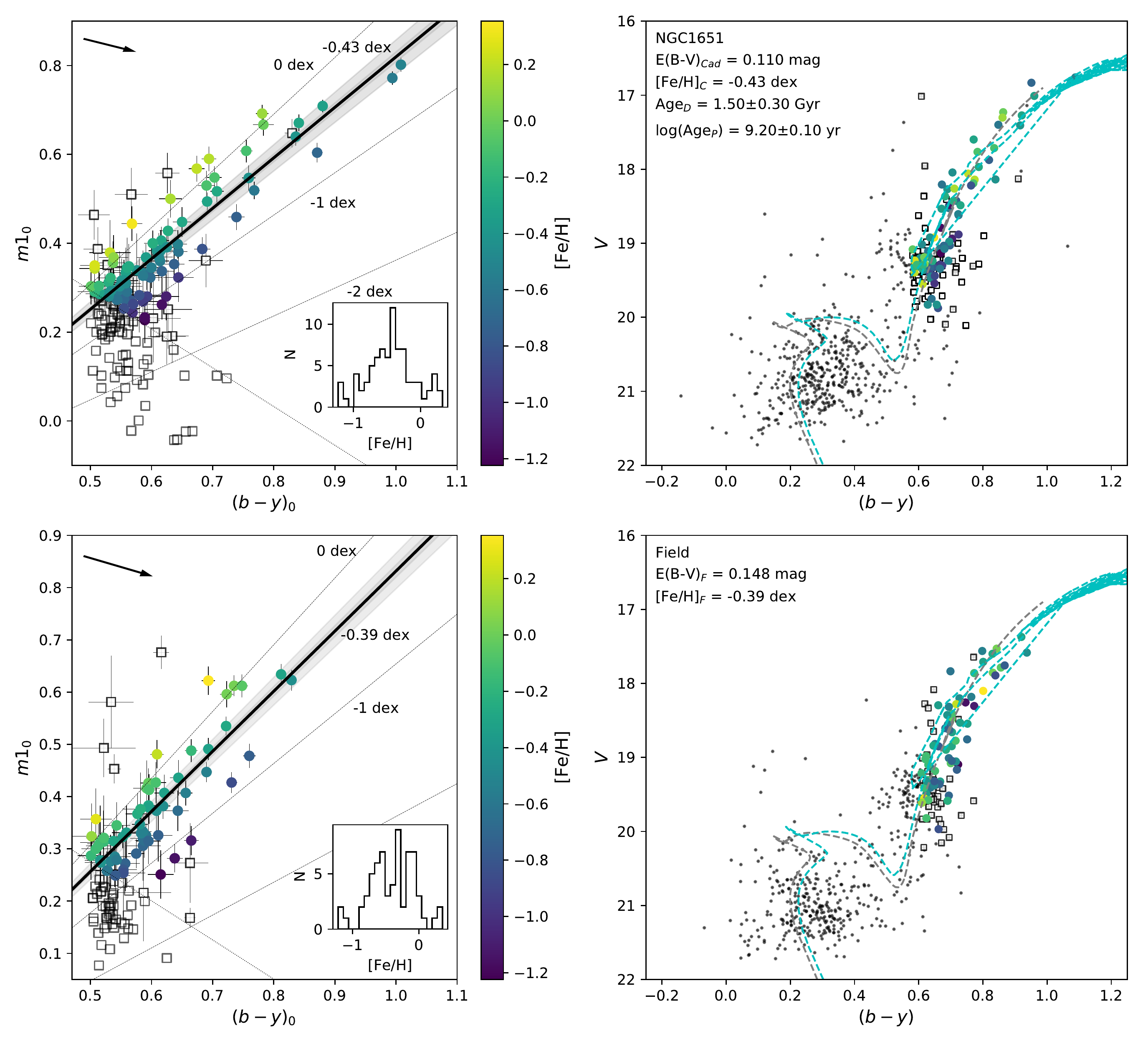}}
      \caption{Reddening-corrected two-color diagrams (left panels) and reddened
               CMDs (right panels) for NGC1651 (upper panels), and the surrounding
               field stars (lower panels).
               Left panels:
               Stars with photometry in the $vby$ filters (black points);
               stars excluded from metallicity determination (open squares);
               stars used to calculate the mean metallicity of a~cluster and
               field (color-coded points, where colors represent the derived
               metallicity value);
               lines of constant metallicity (dashed lines);
               additional selection criteria drawn after visual inspection
               of the plot (dotted line);
               obtained mean metallicities of cluster and field stars (black solid lines);
               the statistical and systematic errors of the mean metallicity of
               the cluster (darker and lighter shaded areas);
               the reddening vectors (black arrows).
               Right panels: Dartmouth and Padova best-fitting isochrones (gray
               and turquoise dashed lines, respectively) superimposed on the
               field CMD (bottom right panel) aimed at illustrating the position
               of the cluster against field stars.
              }
      \label{fig:ngc1651}
   \end{figure*}
%
%-------------------------------------------------------------

%-------------------------------------------------------------
%                                             Two column Figure
%-------------------------------------------------------------
   \begin{figure*}
   \resizebox{\hsize}{!}
            {\includegraphics[]{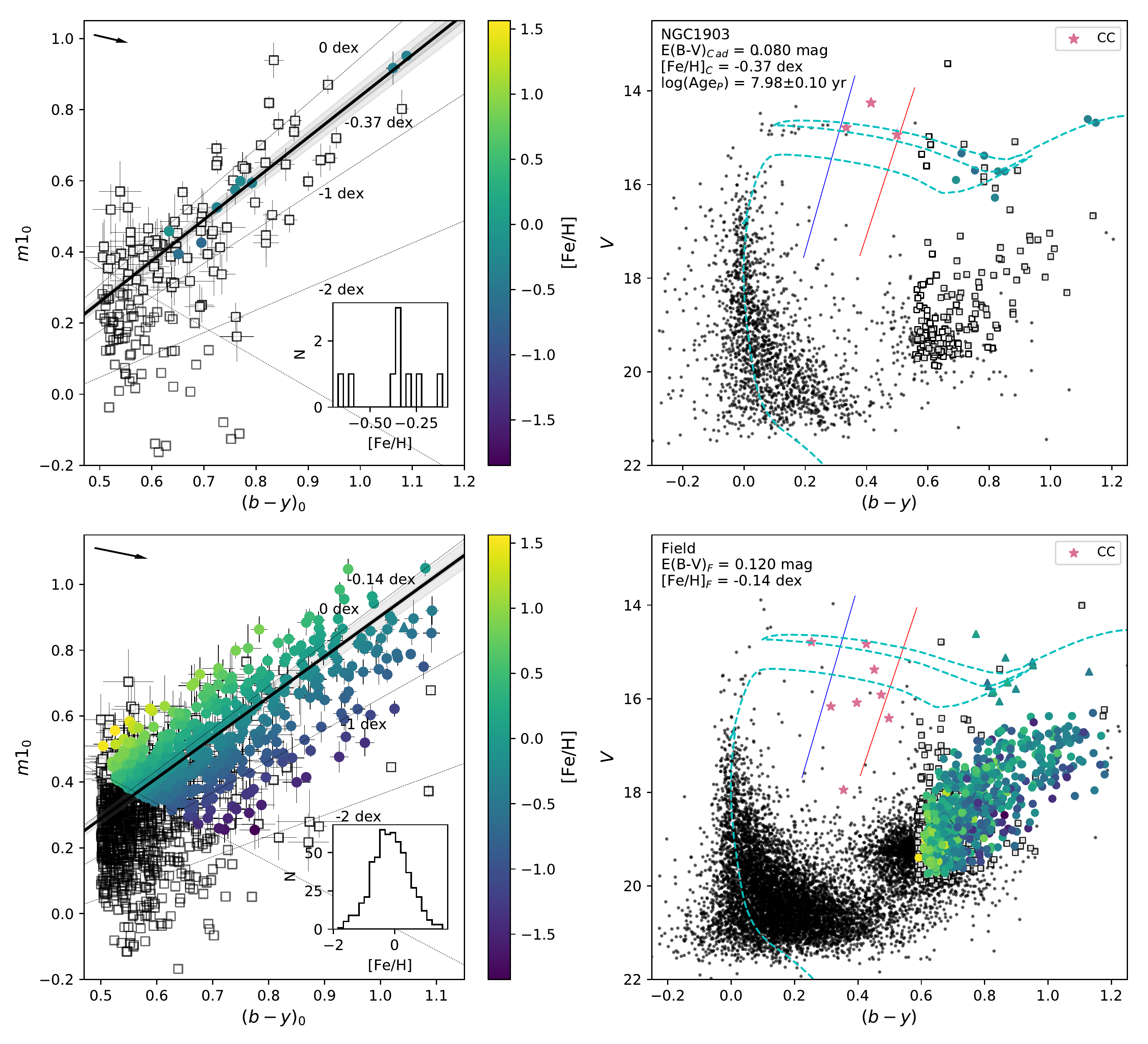}}
      \caption{Reddening-corrected two-color diagrams (left panels) and reddened
               CMDs (right panels) for NGC1903 (upper panels) and the surrounding
               field stars (lower panels).
               Right panels:
               Cepheid variables (pink stars) cross-matched with OGLE catalogs;
               the young field giants (triangles).
               The blue and red lines on the CMDs mark edges of the empirical
               instabillity strip of Cepheids derived by \citet{Narloch2019}.
               The rest of the symbols are the same as in Fig.~\ref{fig:ngc1651}.
              }
      \label{fig:ngc1903}
   \end{figure*}
%
%-------------------------------------------------------------

%-----------------------------------------------------------------
%                                                One column figure
%-----------------------------------------------------------------
   \begin{figure}
   \centering
   \includegraphics[width=\hsize]{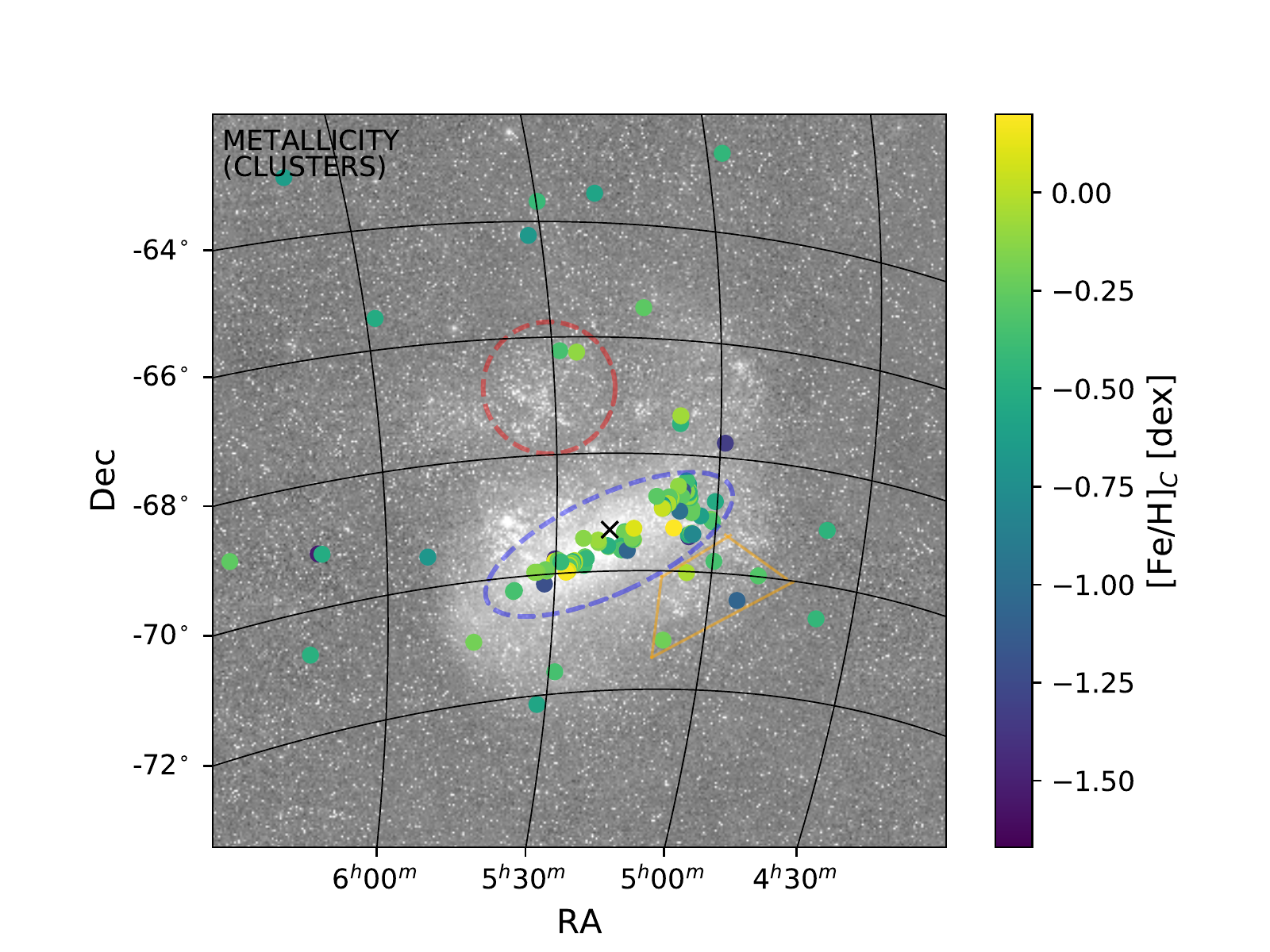}
%   \vskip-10mm
   \includegraphics[width=\hsize]{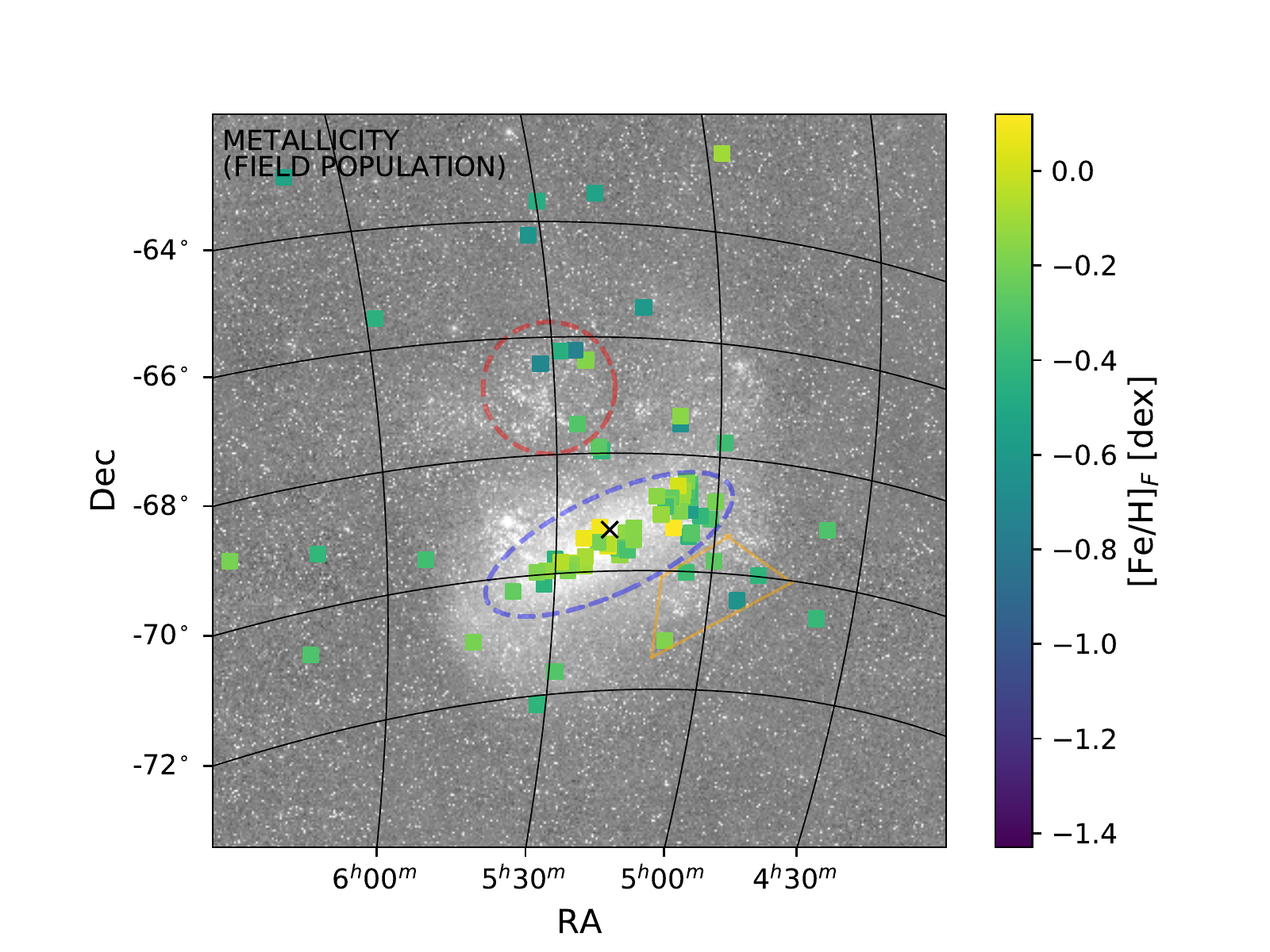}
      \caption{Metallicity map of the star clusters (upper panel) and field stars
               (lower panel) in the LMC. Marked regions: Bar region (blue ellipse),
               Constellation~III (red circle). Orange trapezoid encloses star
               clusters classified as outer bar objects. North is up; east is left.
               Background image originates from the All Sky Automated Survey from
               \citet{Udalski2008a}.
              }
      \label{fig:mapfeh}
   \end{figure}
%-----------------------------------------------------------------

%-----------------------------------------------------------------
%                                                One column figure
%-----------------------------------------------------------------
   \begin{figure}
   \centering
   \includegraphics[width=\hsize]{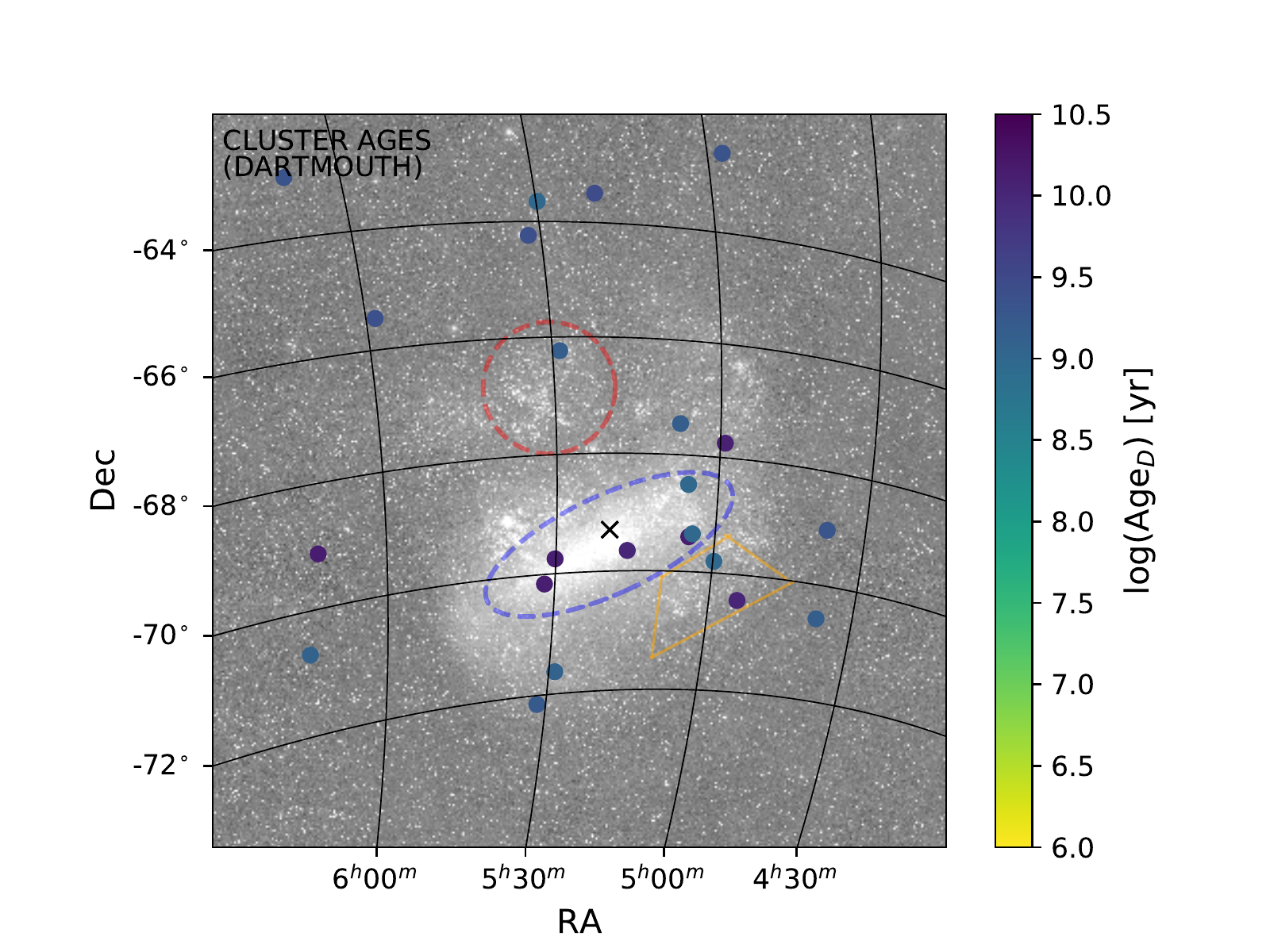}
   \includegraphics[width=\hsize]{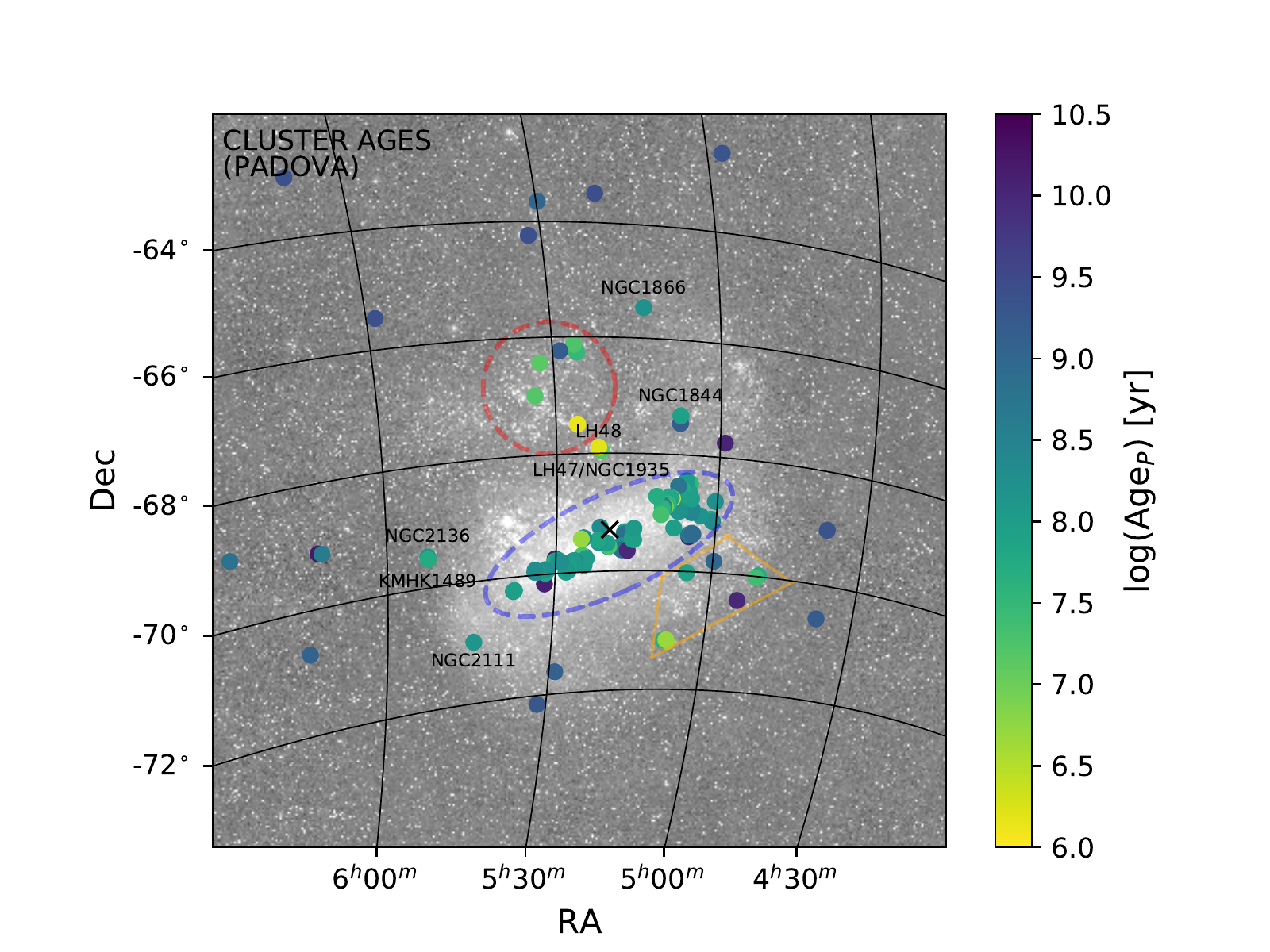}
      \caption{Age map of star clusters in the LMC derived from the Dartmouth
               (upper panel) and Padova (lower panel) theoretical isochrones.
               North is up; east is left. Background image and marked regions as
               in Fig.~\ref{fig:mapfeh}. Orange trapezoid encloses star clusters
               classified as outer bar objects.
              }
      \label{fig:mapage}
   \end{figure}
%-----------------------------------------------------------------

%-----------------------------------------------------------------
%                                                One column figure
%-----------------------------------------------------------------
   \begin{figure}
   \centering
   \includegraphics[width=\hsize]{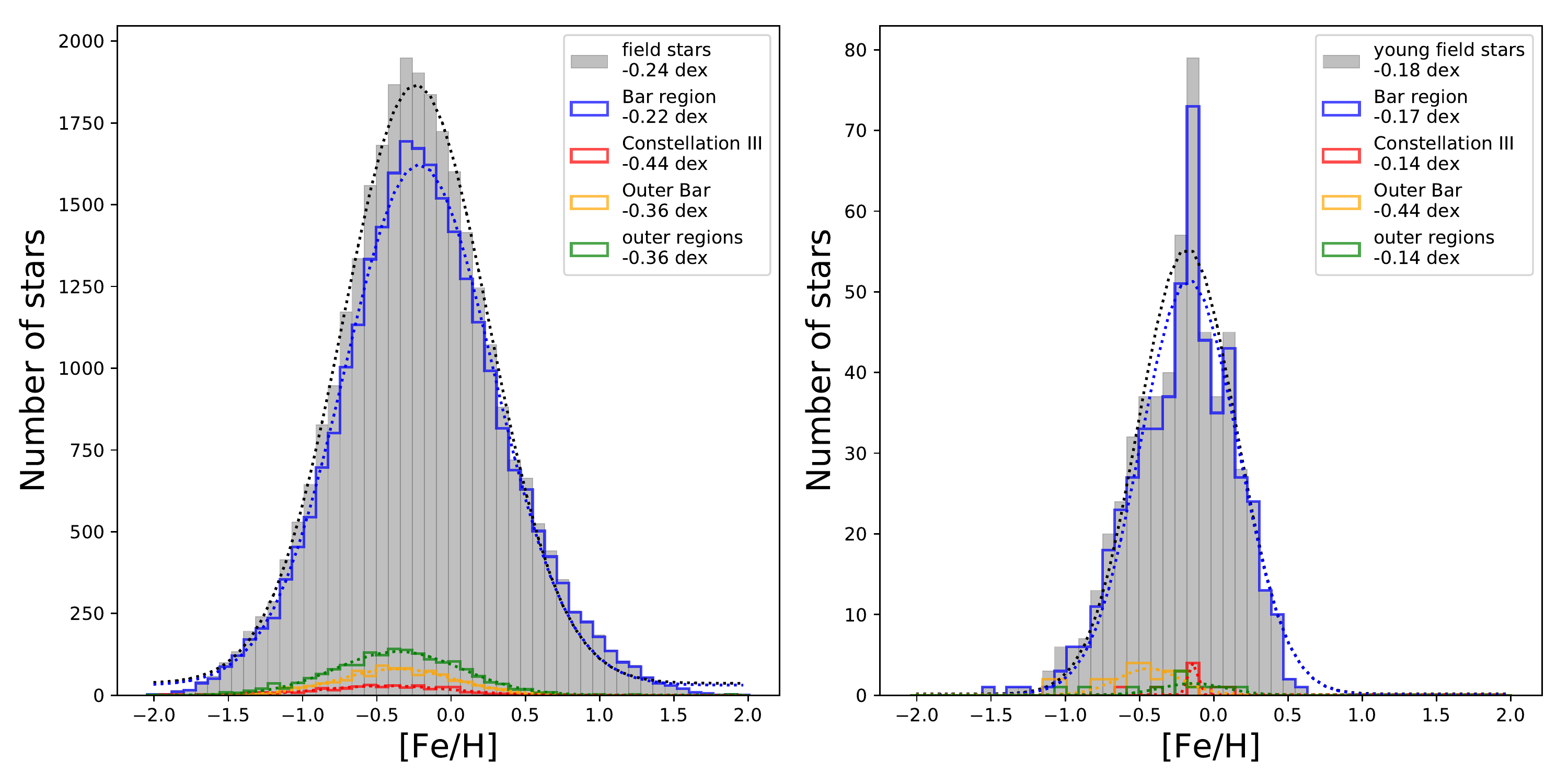}
      \caption{Metallicity distribution of the field stars.
      }
      \label{fig:fehhist}
   \end{figure}
%-----------------------------------------------------------------

%-----------------------------------------------------------------
%                                                One column figure
%-----------------------------------------------------------------
   \begin{figure}
   \centering
   \includegraphics[width=\hsize]{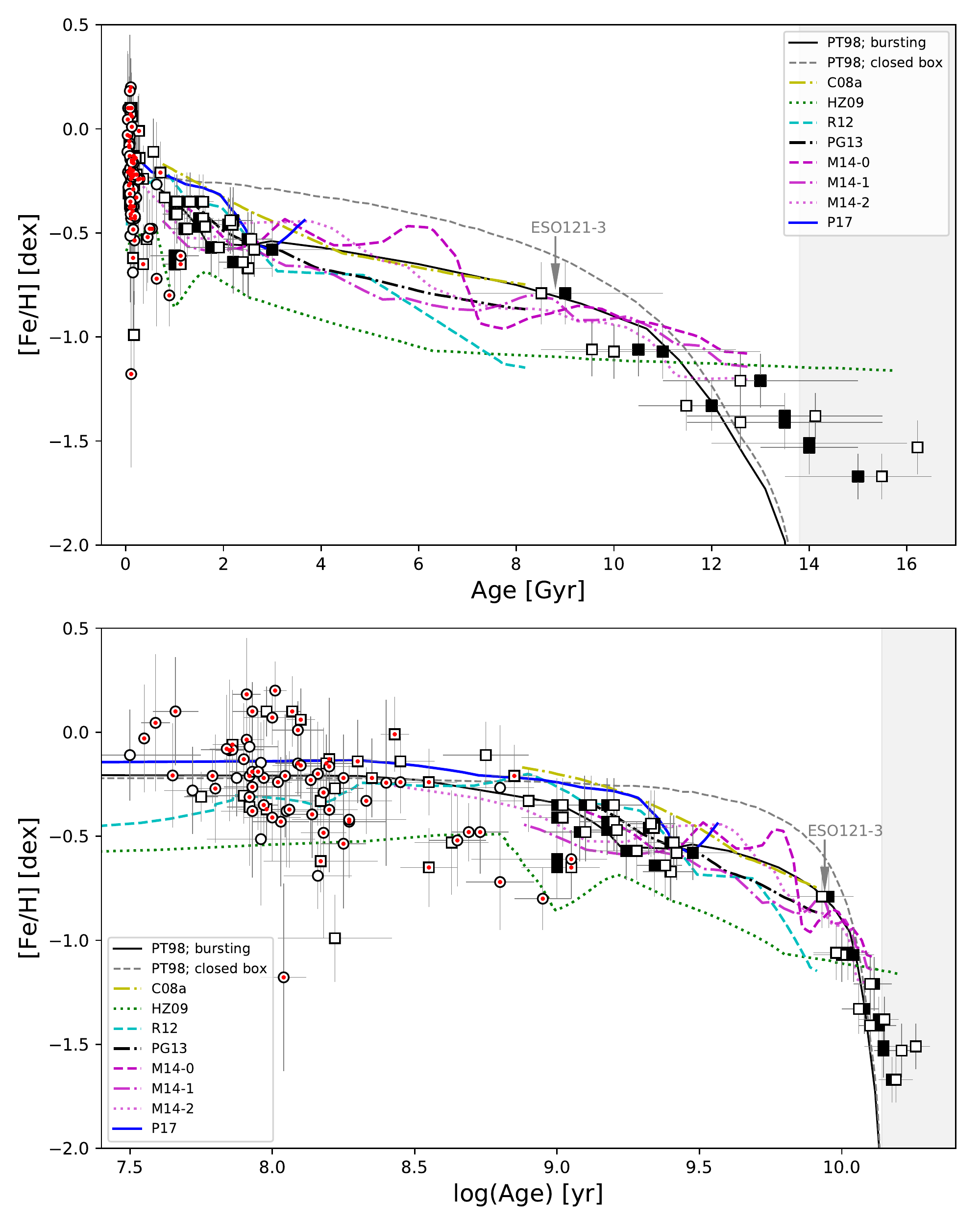}
      \caption{Age-metallicity relation for clusters studied in this work.
               Upper and lower panel show linear and log-age, respectively.
               Clusters with ages derived using Dartmouth isochrones (black squares);
               clusters with reliable number of stars for metallicity determination
               having ages derived from the Padova isochrones (open squares);
               clusters with 1-4 stars for metallicity calculation
               with the Padova ages (open circles). Red dots indicate clusters
               for which metallicity was determined for the first time.
               Overplotted are theoretical models: PT98 bursting model (solid line);
               PT98 closed box model (gray dashed line); C08a (yellow dash-dotted line);
               HZ09 model (green dotted line); R12 (turquoise dashed line);
               PG13 (black dash-dotted line); M14-0, M14-1, M14-2 (magenta dashed,
               dash-dotted, dotted lines, respectively); P17 (blue solid line).
               Grey, shaded area marks ages older than the adopted age
               of the Universe (13.8~Gyr).
              }
      \label{fig:am}
   \end{figure}
%-----------------------------------------------------------------

%-----------------------------------------------------------------
%                                                One column figure
%-----------------------------------------------------------------
   \begin{figure}
   \centering
   \includegraphics[width=\hsize]{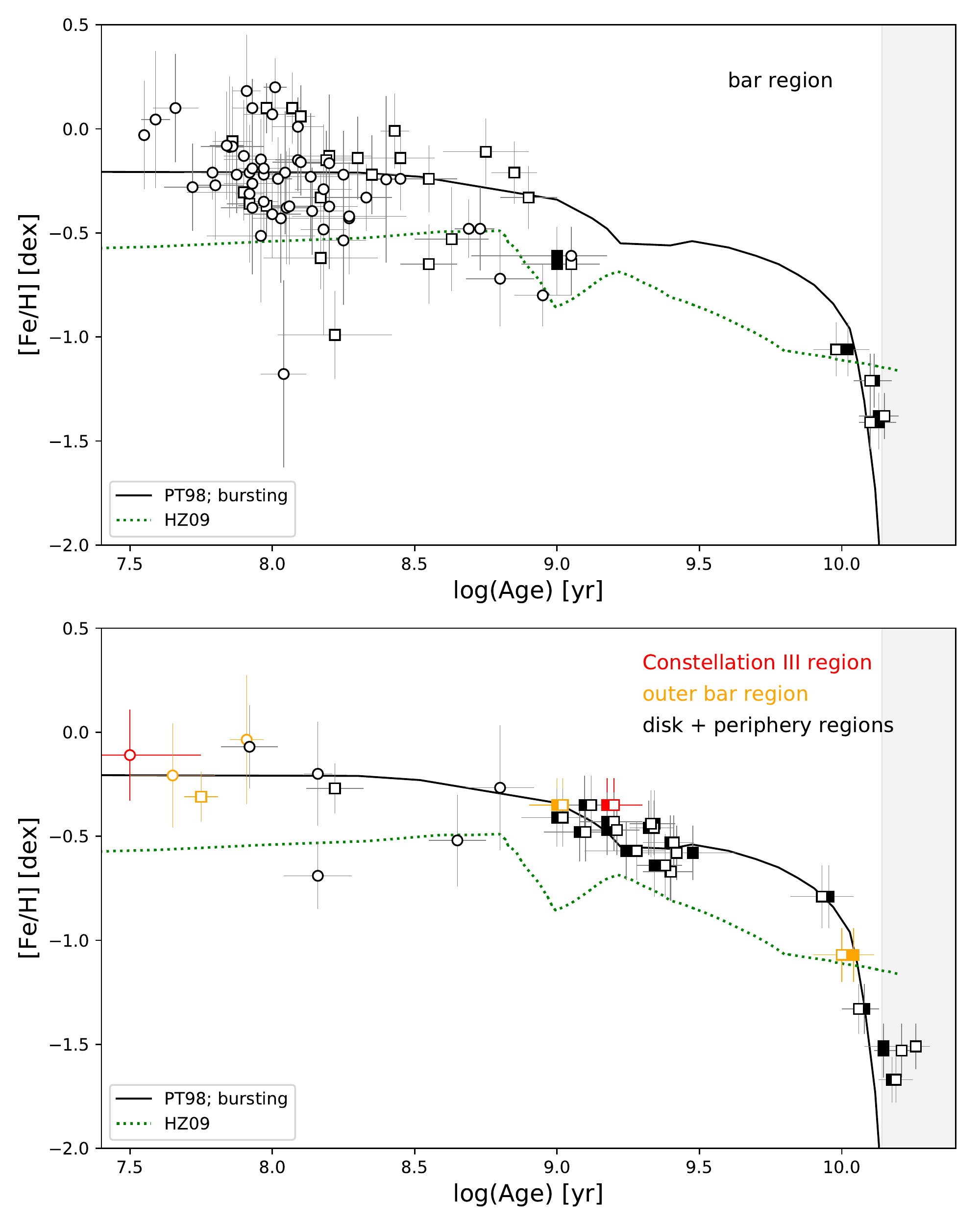}
      \caption{Comparison of the AMR in the inner bar region and outer regions of
      the LMC. Symbols are the same as in Fig.~\ref{fig:am}.
      }
      \label{fig:aminout}
   \end{figure}
%-----------------------------------------------------------------

%-----------------------------------------------------------------
%                                                One column figure
%-----------------------------------------------------------------
   \begin{figure}
   \centering
   \includegraphics[width=\hsize]{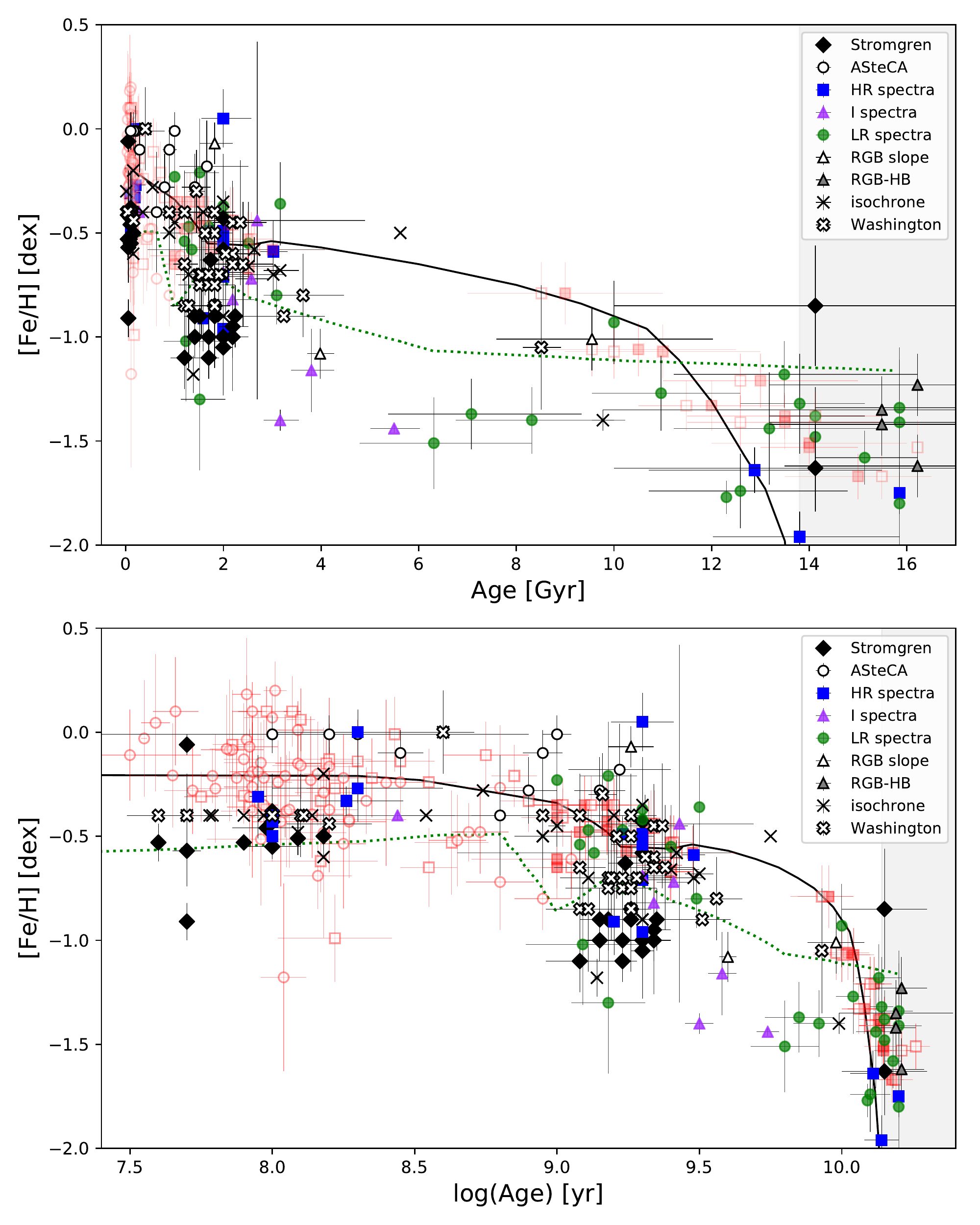}
      \caption{Age-metallicity relation for star clusters studied in this work compared
               with the literature (see Table~\ref{tab:lit}).
               Metallicity values derived from:
               two-color Str\"omgren diagram (black diamonds);
               the ASteCA package by \citet{Perren2017} (open circles);
               high-resolution spectroscopy (blue squares);
               integrated spectroscopy (purple triangles);
               low-resolution spectroscopy (green circles);
               RGB slope method (open triangles);
               RGB--HB method (gray triangles);
               fitting of theoretical isochrones to optical data (crosses);
               Washington photometry (open crosses).
               Red squares and open circles indicate measurements from this work
               presented in Fig.~\ref{fig:am} for comparison.
               Overplotted theoretical models: PT98 bursting model (black solid line);
               HZ09 model (green dotted line).
              }
      \label{fig:amlit}
   \end{figure}
%-----------------------------------------------------------------

%-----------------------------------------------------------------
%                                                One column figure
%-----------------------------------------------------------------
   \begin{figure}
   \centering
   \includegraphics[width=\hsize]{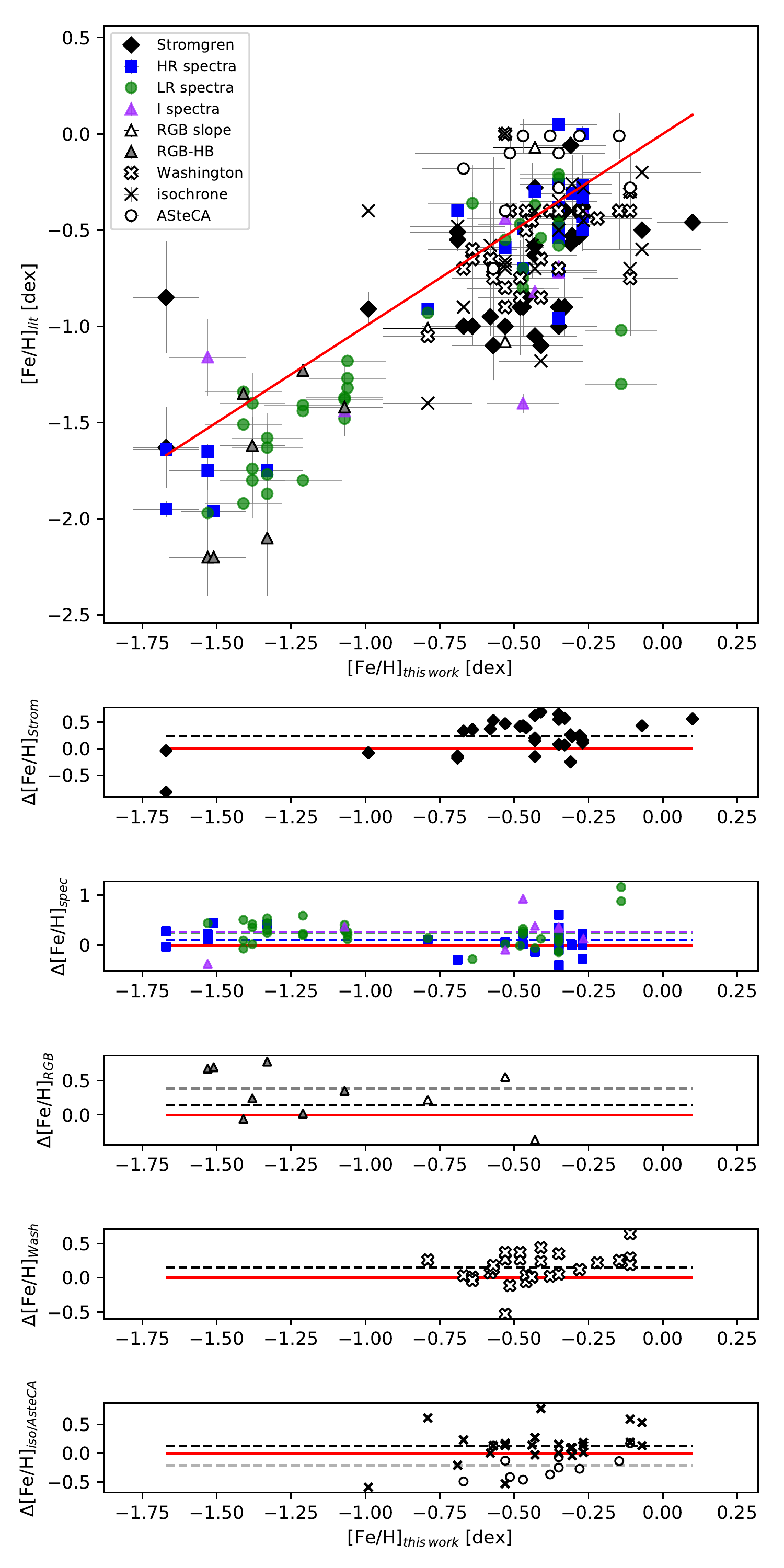}
      \caption{Comparison of the metallicities for star clusters obtained in
      this work with the literature values from Fig.~\ref{fig:amlit}. Red solid
      line represents the 1:1 relation. Dashed lines mark average metallicity
      difference between this work and a~given method from the literature.
      }
      \label{fig:fehlit}
   \end{figure}
%-----------------------------------------------------------------

%-----------------------------------------------------------------
%                                                One column figure
%-----------------------------------------------------------------
   \begin{figure}
   \centering
   \includegraphics[width=\hsize]{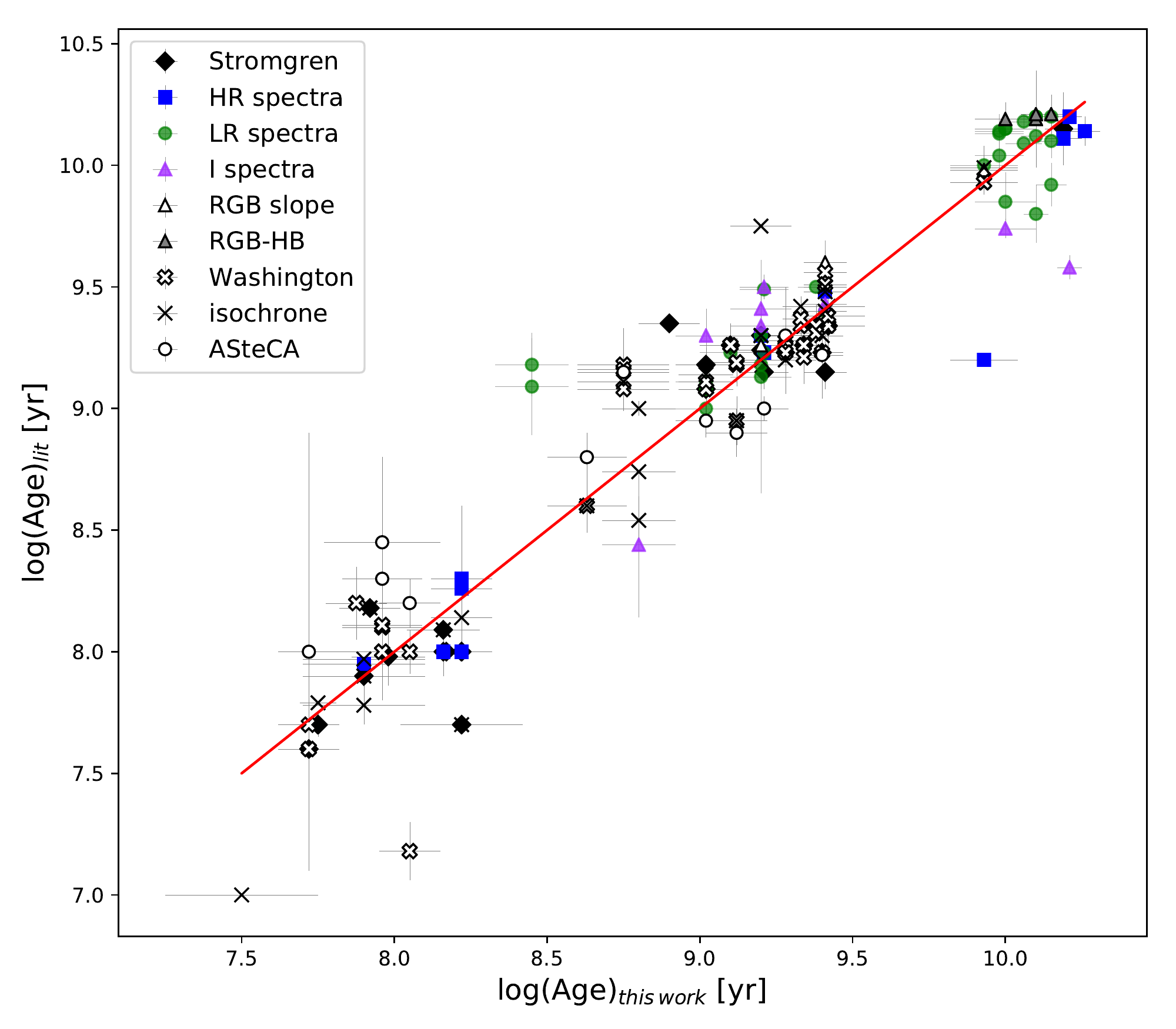}
      \caption{Comparison of the logarithm of ages for star clusters obtained in
      this work based on Padova isochrones with the literature values from
      Fig.~\ref{fig:amlit}. The red solid line represents the 1:1 relation.
      }
      \label{fig:agelit}
   \end{figure}
%-----------------------------------------------------------------

%--------------------------------------------------------------------
\subsection{Spatial metallicity distribution of cluster and field stars in the LMC}
\label{ssec:metalmap}
%--------------------------------------------------------------------

   Figure~\ref{fig:mapfeh} shows the spatial map of the cluster (upper panel) and
   field (lower panel) metallicities given in Tables~\ref{tab:reddC} and \ref{tab:reddF}.
   The black cross marks the center of the LMC adopted from \citet{Pietrzynski2019}.
   The structure of the LMC is more complicated than that of the SMC, which is
   reflected in the metallicity distribution in this galaxy. \citet{HZ2009} defined
   several distinct LMC regions (see their Fig.~6), the description of which we
   follow.

   Most of the star clusters from our sample are located in the bar, the elongated
   structure near the center of the galaxy (marked in Fig.~\ref{fig:mapfeh} and
   \ref{fig:mapage} with blue dotted ellipse).
   The spread of their mean metallicities is large and ranges between $-1.41$~dex
   up to as high as $0.20$~dex, with an average of $-0.32$~dex ($\sigma = 0.33$~dex).
   The majority of the clusters are metal-rich. One possible explanation for the
   presence of low-metallicity clusters in this region is that they might actually
   be in front of or behind it. Such a~scenario is supported by the fact that the
   fields around these clusters are more metal-rich, and coincide better with the
   peak of the metallicity distribution of the field stars in the bar ($-0.22$~dex,
   $\sigma = 0.50$~dex, see Fig.~\ref{fig:fehhist}).
   Alternatively, they might be remnants of an ancient bar history, and their
   fields became chemically enriched while the clusters remained untouched.
   Such speculation is supported by the fact that the AMR of bar and non-bar
   regions are qualitatively similar, as we discuss in the following sections.

   A~star-forming region to the north of the bar is called Constellation~III
   (marked in Figs.~\ref{fig:mapfeh} and \ref{fig:mapage} with red dotted circle).
   We were able to determine metallicities in this region for two clusters:
   NGC1978 is an intermediate-age globular cluster and NGC1948 is a~much younger
   and more metal-rich association of stars. The rest of the clusters were too
   young and did not have enough stars for metallicity estimation.
   The peak of the metallicity distribution of the field stars located in the
   Constellation~III region is significantly lower than in the bar ($-0.44$~dex,
   $\sigma = 0.43$~dex). This may suggest that this structure is formed from the
   unenriched material of the LMC disk, where star formation started recently and
   has not yet enriched the environment.

   The star cluster NGC1754 in the outer bar has much lower metallicity than other
   clusters in this region (clusters enclosed in an orange trapezoid in the
   Figs.~\ref{fig:mapfeh} and \ref{fig:mapage}), while also being quite different
   from the peak of the metallicity distribution of field stars ($-0.36$~dex,
   $\sigma = 0.39$~dex).
   This suggests that NGC1754 is not a~part of the outer bar or that it is a~remnant of
   ancient star-forming activity that had occured there. Other clusters in this region
(except one) have mean metallicities similar to the peak value of the field stars,
   which seem to confirm their affiliation with this structure. On the other hand, KMHK521, seems to be associated with a~nearby H$\alpha$ region, and its
   subsolar metallicity (calculated however based on one star only) seems to be
   confirming this. Interestingly, the average field metallicity of the outer bar
   coincides better with the field metallicity of LMC outer regions as opposed to
   the bar region.

   The clusters lying in the disk arms and periphery of the LMC are characterized
   by a~wide range of metallicities (from $-1.67$ to $-0.07$~dex). The average
   value found for these clusters is $-0.65$~dex ($\sigma = 0.44$~dex) and the
   peak of the field star distribution is $-0.36$~dex ($\sigma = 0.42$~dex).
   There are few stars in the fields of the periphery clusters, which causes
   problems in characterizing the surroundings of the clusters and might be the
   reason of the discrepancy between these two values. The field stars number
   grows in the fields located closer to the denser, more metal-rich, inner regions
   and these stars contribute more to the field star distribution.
   The overall agreement of our spatial metallicity maps from Fig.~\ref{fig:mapfeh}
   with those presented by, for instance, \citet{Choudhury2015,Choudhury2016} is highly
   satisfactory.

   The right panel of Fig.~\ref{fig:fehhist} shows the metallicity distributions
   of the young giants and supergiants from the fields (stars such as the ones marked
   with triangles in the lower right panel in Fig.~\ref{fig:ngc1903}), analogous
   to those for the older field red giants in the left panel. The peak values of
   the metallicity distributions for the former are usually higher than
   corresponding values for the latter. However, there are few such stars in outer
   regions, so this comparison might be misleading.

%--------------------------------------------------------------------
\subsection{Spatial age distribution of clusters in the LMC}
\label{ssec:agemap}
%--------------------------------------------------------------------

   Figure~\ref{fig:mapage} shows the spatial map of the cluster ages estimated
   from Dartmouth (upper panel) and Padova (lower panel) isochrones, given in
   Table~\ref{tab:reddC}. This map is tightly correlated with the metallicity map.
   The low-metallicity star clusters are simultaneously older ones and they mostly
   occur in regions outside of the the bar, with several exceptions mentioned in
   a~previous section.
   Numerous metal-rich star clusters in the LMC bar region have log(Ages) between
   $6.4$ and $10.15$ ($2.5$~Myr to $14.1$~Gyr), with an average of $8.14$ and
   $\sigma = 0.58$ ($\sim$$138$~Myr).
   Such scatter in the ages supports the scenario put forward in the previous
   section that the oldest, low-metallicity star clusters are in fact not part
   of the bar. Also, considering the prediction that the bar was formed about
   $5.5$~Gyr ago \citep{BekkiChiba2005}, the LMC bar cannot be the birthplace
   of those clusters.

   Another region occupied mostly by young star clusters is Constellation~III,
   with objects having logarithm of ages of $6.15 - 7.50$ ($\sim$$1.4 - 31.6$~Myr),
   with the sole exception of cluster NGC1978 which has a~log(Age) of
   $9.20$ ($\sim$$1.6$~Gyr). The presence of this intermediate-age globular cluster
   in a~young star forming region is quite puzzling, as its metallicity does not
   deviate much from the average metallicity of the field.

   The outer bar cluster NGC1754 has an age of about $10$~Gyr, which deviates
   significantly from other clusters in this region, and supports the claim made
   in the previous section that it might not belong to this structure, but instead
   lies in front of or behind the bar. NGC1795, on the other hand, is a~$1$~Gyr
   old intermediate-age cluster and might be a~product of earlier star formation
   (as its metallicity coincide with metallicity of the field). The remaining
   outer bar clusters from our sample are young and could be associated with
   a~nearby H$\alpha$ region.

   The logarithm of ages of star clusters from the LMC disk and peripheries varies
   between $6.22$ and $10.26$ ($\sim$$1.7$~Myr -- $18.2$~Gyr). The youngest
   ages clearly stand out in the bottom panel of Fig.~\ref{fig:mapage}, and belong
   to: a) young association of stars LH47/NGC1935 and LH48 located between
   the LMC bar and Constellation~III; b) clusters located to the east
   of the LMC bar and the $30$~Doradus regions (NGC2136 and KMHK1489); c)
   NGC2111 located in the southeast arm; d) NGC1844 and NGC1866 from
   northwest arm.
   The rest of the outermost stellar clusters are older than
   $\mathrm{log(Age)} = 8.65$ ($\sim$$447$~Myr), reaching up to the age of the
   Universe. The spatial age distribution obtained in this work is in good agreement
   with previous results, for instance, the recent star formation activity in the LMC
   presented by \citet[see their Fig.~10]{HZ2009}, as well as the spatial age
   distribution depicted by \citet[see their Fig.~9]{Glatt2010}.

%--------------------------------------------------------------------
\subsection{Age--metallicity relation for star clusters in the LMC}
\label{ssec:amrel}
%--------------------------------------------------------------------

   The resulting age--metallicity relation for $110$ star clusters studied in
   this work that have both metallicity and age determinations, is illustrated in
   Fig.~\ref{fig:am}. Most of the clusters have ages between $32$~Myr and $3$~Gyr,
   but the oldest ones reach as far back as the age of the Universe (adopted here as $13.8$~Gyr).
   The measured metallicities range between $-1.67$ and $0.20$~dex.
   The young star clusters are characterized by quite a broad spread in metallicity.
   As already mentioned in previous sections, and clearly visible in
   Fig.~\ref{fig:aminout}, most of the young, metal-rich stellar clusters tend
   to lie in the LMC bar region, with a~few exceptions. Both Constellation~III
   and the outer bar regions contain young and intermediate-age clusters.
   The disk and outermost regions of the LMC contain old, metal-poor objects, as
   well as most of the intermediate-age clusters and some young clusters too.

   In Figs.~\ref{fig:am} and \ref{fig:aminout} (bottom panels), we can see a~rapid
   chemical enrichment, which applies mostly to the non-bar areas of the LMC,
   followed by a quiescent period, dubbed "the age gap"\ in the literature,
   when the star formation rate in the galaxy was very low and chemical
   enrichment very modest. After this period, we can see a~burst which started
   about $3$~Gyr ago, causing enrichment mostly in the outer regions (see bottom
   panel of Fig.~\ref{fig:aminout}). This burst was followed by another in the
   bar region, which started about $1.5$~Gyr ago and triggered significant star
   cluster formation from the still quite metal-poor material (about $-0.60$~dex),
   causing enrichment of the environment up to about $0.0$~dex (see top panel of
   Fig.~\ref{fig:aminout}). The period between $\mathrm{log(Age)} = 8.5 - 8.9$
   ($\sim$$316 - 794$~Myr) in the LMC bar still seems to be characterized by a low
   cluster formation rate which increases for younger ages. In the case of the LMC
   disk and periphery, we do not observe clusters in the age range of
   $\mathrm{log(Age)} = 8.3 - 8.6$ ($\sim$$200 - 400$~Myr).
   The age difference between the two clusters in the Constellation~III (NGC1978
   and NGC1948) is $\sim$$1.5$~Gyr, while in the LMC outer bar, we see the presence
   of every age group.

   There are several star clusters younger than about $1$~Gyr with
   $\mathrm{[Fe/H]} < -0.5$~dex, manifesting themselves in Figs.~\ref{fig:am}
   and \ref{fig:aminout}. Two such clusters in the non-bar regions (NGC2136 and
   KMHK1679) are located east to the bar (see Fig.~\ref{fig:mapage}), while bar
   clusters tend to clump in the west end of the bar. The metallicity of the
   most deviating cluster (OGLE-CL~LMC~111) is based on only one star and could
   be an error of selection, however, for the rest of the clusters, this is unlikely.
   The reason of their specific location is unknown, however, areas of lower
   metallicity in this part of the bar could be also identified in the metallicity
   maps of \citet{Choudhury2015,Choudhury2016}.

 %--------------------------------------------------------------------
 \subsection{Age-metallicity relation for star clusters in the LMC in comparison
 to the SFH from the literature}
 \label{ssec:amrel}
 %--------------------------------------------------------------------

   A~literature review shows that there exist many models describing the global
   enrichment history of the LMC that nonetheless cover different age ranges.
   In Fig.~\ref{fig:am}, ten models from seven works are marked:
   \citet[hereafter PT98, bursting and closed box models]{PG1998},
   \citet[C08a, average of four disk frames]{Carrera2008a},
   \citet[HZ09, calcium triplet spectroscopy of individual red giants in four LMC
   fields]{HZ2009},
   \citet[R12, four tiles averaged from the VISTA near-infrared YJKs survey of
   the Magellanic system, VMC]{Rubele2012},
   \citet[PG13, Washington photometry of 21 fields]{PiattiGeisler2013},
   \citet[M14-0, M14-1, M14-1, $VI$ photometry of three fields]{Meschin2014},
   and \citet[P17, Washington photometry of star clusters]{Perren2017}.

   All models predict the initial increase of the metallicity level in the LMC, but
   only the PT98 models trace it over the first $3$~Gyr, when it is rapid and
   steep. Our results also show this rapid increase of the metallicity for old
   clusters (up to about $-1$~dex), although the oldest ones tend to have ages
   older than the age of the Universe \citep[this problem is known also in the
   literature, e.g.,][]{Dirsch2000,Beasley2002}.
   Other models (e.g., M14, HZ09 or R12) do not reach the oldest ages, which means
   that only the PT98 models reproduce the rapid growth of chemical enrichment
   observed in our results well.

   Models in the range of ages between $9$~Gyr up to $3$~Gyr ago cannot be
   unambiguously confirmed by our data, because we have only one cluster in this
   range, which might be equally well fit  to the PT98 bursting model, C08a, M14-0, or
   PG13 models. \citet{Meschin2014} indicated that a~peak at $\sim$$7.5$~Gyr in
   the M14-0 model is very uncertain, which is related to low star formation activity
   in that period of time. Nevertheless, it can be confidently stated that the
   chemical enrichment in the galaxy at that time was slow and mild. The authors
   describe this interval of time as a~quiescence. The period itself is referred
   to as the age gap and our results confirm this gap. Moreover, what seems already
   evident at this point is that the PT98 closed box model tends to fail at
   describing the chemical history of the LMC.

   Many models more or less clearly predict a~burst of chemical enrichment at
   intermediate ages. The PT98 bursting model shows such a~burst at the age of
   $\sim$$1.6$~Gyr, when the metallicity abruptly increases from about $-0.55$~dex
   to the current level of about $-0.20$~dex. The M14 models show a~general trend
   of increasing metallicity over time, but we may distinguish mild bursts about
   $2.2$, $1.6,$ and $1.3$~Gyr ago for M14-0, M14-1, and M14-2, respectively. The
   first two models predict a~growth of chemical abundances up to about $-0.40$~dex
   around 1~Gyr ago and M14-2 up to about $-0.25$~dex currently.
   Model R12, after its initial growth is followed by $\sim$$1.7$~Gyr of stagnation, and shows
   an ultimate big burst $\sim$$3.1$~Gyr ago, which preceded two smaller ones
   $\sim$$1.9$ and $1.3$~Gyr ago.
   Model P17 shows a~major burst about $2.9$~Gyr ago after which the metallicity
   increased from around $-0.58$~dex up to $-0.14$~dex.
   On the other hand, there are three models (C08a, PG13, and PT98 closed box),
   which predict a~rather monotonic growth of metallicity over a~time, up to about
   $-0.30$ to $-0.20$~dex without any bursts in the history. Moreover,
   \citet{Carrera2008a} emphasize that, on average, the bar is slightly more
   metal-rich than the inner disk, which we also confirmed and commented on in
   previous sections.

   The HZ09 model presents a~very different scenario from those described above.
   First, the metallicities predicted by this model are much lower than any
   other model in our list. After an initial growth of chemical enrichment from
   a~starting value of about $-1.1$~dex $\sim$$1.6$~Gyr ago (in contrast to any of the
   other models), a~decline of about $-0.15$~dex occurs instead of an increase.
   The final burst appeares $\sim$$1$~Gyr ago and raises the chemical abundances
   up to current level of about $-0.55$~dex.

   Our results agree well with the burst in chemical enrichment at the intermediate
   ages predicted by bursting PT98, M14,  R12, and PG13 models. The start of
   the burst in the outer regions of the LMC occured $\sim$$2.8$~Gyr ago, raising
   the metallicity by about $0.20$~dex. Such a~burst in the LMC bar took place with
   a~delay of some $1.7$~Gyr. The spread of metallicity in our data means that
   none of the listed models are favorable at the present day, although the overall
   agreement of our total AMR for star clusters with the PT98 bursting model is
   highly satisfactory.
   Nevertheless, this model tends to fail for the bar region where, at least at
   the beginning of the burst, clusters follow the HZ09 model better.
   The latter seems to reflect our resulting AMR of the bar region quite well,
   at least for some of the star clusters from the sample. This model, however,
   does not agree with the AMR for outer regions at all.
   This comparison confirms our belief that the chemical history of the LMC is
   complicated and cannot be described by a~single SFH model.

   \citet{HZ2009} speculate that after an initial epoch of star formation, during
   which old populous star clusters were formed, and the era of stagnation that
   follows it, some dramatic event took place that led to a resumption of star formation processes.
   These authors propose an explanation based on a~merger of a~gas-rich dwarf galaxy or a~tidal encounter
   with the SMC (where a~similar resumption of star formation has been observed).
   Our measured AMR for star clusters strongly corroborates this idea.
   Additionally, \citet{HZ2009} indicate qualitatively similar SFH of the LMC
   bar and non-bar regions, suggesting that the stars in the former have most likely always been part of the LMC. Again, we confirm such a~conclusion, as we
   too observe a~similar behavior of the AMR in these regions. Moreover, the
   growth of the metallicity in the LMC bar starts at an almost identical level
   as in the non-bar regions. The star formation here could have proceeded inwards
   as suggested by \citet{PiattiGeisler2013}. We cannot, however, unequivocally
   confirm (or deny) enhanced star formation activity at $12$~Myr, $100$~Myr,
   $and 500$~Myr as reported by \citet{HZ2009}, as the spread in the metallicities
   of young star clusters is high. We can, however, confirm a~period of enhanced
   star formation that happened $2$~Gyr ago, although we argue that it started
   even earlier.

%--------------------------------------------------------------------
\subsection{Comments on individual star clusters}
\label{ssec:commindv}
%--------------------------------------------------------------------

  The five clusters in Table~\ref{tab:reddC} are marked with the appropriate
  comment. Two of them (NGC1858 and NGC1948) are young clusters for which only age
  estimates could be made. The given age values are mean values as both clusters
  appear to have multiple stellar populations of different ages or the age range
  is large. NGC2136 and NGC1850 are defined as binary or even triple (in the case
  of NGC1850) systems as reported in the SIMBAD database -- and a~possible second population
  of stars of different metallicity but similar age is visible in both of them.
  In the case of NGC2136, it may be explained as field stars, but NGC1850 requires
  deeper reflecion. It has been proven that this cluster has variations of reddening
  across the field of view \citep[e.g.,][]{Correnti2017}, also its companion
  clusters (BRHT5b and NGC1850A) are projected very close to the cluster center.
  Moreover, for instance, \citet{Milone2018} and \citet{Bastian2017} showed the presence of
  an extended main sequence explained as a~separation of fast- and slow-rotating stars.
  Taken together, these factors can explain the additional population.
  Finally, the ESO121-3 cluster may host more populations with different metallicites,
  although in this case it could be, for example, the effect of N-enriched stars.

%--------------------------------------------------------------------
\section{Discussion}
\label{sec:discussion}
%--------------------------------------------------------------------

  Figure~\ref{fig:amlit} presents our AMR with overplotted literature measurements
  of the metallicity and age for some of the star clusters considered in this work,
  obtained with various methods: Str\"omgren photometry (black diamonds); spectroscopy
  (high and low-resolution shown as blue squares and green dots, respectively; integrated
   as  purple triangles); RGB slope (open triangles) and RGB-HB (grey triangles)
  methods, both expressed on the ZW84 scale; Washington photometry (open crosses),
  theoretical isochrone fitting method (crosses), and, finally, metallicities determined
  with ASteCA package from \citet{Perren2017} (open circles).
  Metallicities calculated in this work are compared to the literature in
  Fig.~\ref{fig:fehlit}. We note that the metallicity scales used in each method may
  be different. Ages derived from isochrone fitting are compared with corresponding
  literature values in Fig.~\ref{fig:agelit}. As in Fig.~\ref{fig:am}, multiple
  entries of metallicity and age for several clusters from Table~\ref{tab:reddC}
  are averaged. Table~\ref{tab:lit} provides an overview of selected literature
  parameters of the analyzed clusters.

%--------------------------------------------------------------------
\subsection{Age-metallicity relation for clusters in the LMC in comparison to
the literature}
\label{ssec:AMRlit}
%--------------------------------------------------------------------

  The top panel of Fig.~\ref{fig:fehlit} shows a~direct comparison of the mean
  metallicity values of clusters obtained in this work with metallicities from
  the literature. The lower panels show the residuals of the metallicity values
  obtained in this work for a~given star cluster and the corresponding literature
  values from different methods. Each panel presents a~single method or a~set of
  methods. The dashed lines mark the means of differences. A~similar comparison
  for logarithms of ages estimated on the basis of Padova isochrones with literature
  values is presented in Fig.~\ref{fig:agelit}.

  We noticed a~significant difference in the mean metallicity of our results with
  values obtained with Str\"omgren photometry from
  \citet{SHR1994,Hill1995,Dirsch2000,Piatti2019,PB2019}, and \citet{Piatti2020}
  % (about $0.23\pm0.33$~dex, where the error is unbiased standard deviation).
  ($\sim$$0.23$~dex).
  Moreover, the mean difference for the last three works is $\sim$$0.34$~dex,
  % $0.34\pm0.27$~dex
  while for the rest it is close to zero (about $-0.07$~dex). This may seem
  % $-0.07\pm0.33$~dex
  surprising, especially given that in the last three works the same data were
  used. The reason for the existing disaccord has already been indicated in
  Paper~I, and is a~consequence of the use of the metallicity calibration
  of the Str\"omgren colors from \citet{Calamida2007}, as well as slightly different
  data processing in the three last papers. This calibration does not cover the
  higher metallicity regime (over $-0.70$~dex) and as a~result, it significantly
  underestimates the obtained metallicities.

  The difference between our results and metallicities derived from high-resolution
  spectroscopy is satisfactory and amounting to $\sim$$0.10$~dex.
  % $0.10\pm0.24$~dex.
  A worse agreement exists for metallicities based on low-resolution spectroscopy, whereby
  our results are more metal-rich by $\sim$$0.25$~dex.
  % $0.25\pm0.28$~dex
  In the case of low-metallicity clusters (lower than $-0.70$~dex), this discrepancy
  is $\sim$$0.29$~dex and for more metal-rich clusters $\sim$$0.19$~dex.
  Our metallicities show a~similar offset from results coming from integrated
  spectroscopy, about $\sim$$0.26$~dex.
  % ($0.26\pm0.38$~dex).

  There are only a~few metallicity measurements obtained with the RGB slope method
  for which the mean difference is relatively small ($\sim$$0.14$~dex), although
  the spread of the results is significant.
  A~large difference is also noted with respect to the RGB-HB method ($\sim$$0.38$~dex).
  % ($0.38\pm0.34$~dex).
  Metallicities for three of the studied clusters were estimated with this method
  by \citet{Brocato1996} and they deviate notably from our results
  ($\Delta \mathrm{[Fe/H]} = 0.71$~dex), while the values from \citet{Olsen1998}
  are much closer to our determinations ($\Delta \mathrm{[Fe/H]} = 0.14$~dex); this indicates that perhaps metallicities are underestimated in \citet{Brocato1996}.

  There is comparatively good agreement between our results and Washington
  photometry. The mean difference of metallicities is $\sim$$0.15$~dex.
  % $0.15\pm0.20$~dex).
  The same is true for the isochrone fitting method ($\sim$$0.13$~dex).
  % $0.13\pm0.26$~dex.
  Finally, the ASteCA package from \citet{Perren2017} is the only method
  (on average) that gives more metal-rich results than what is obtained in this work (about
  $-0.21$~dex). As noted by the authors themselves, this method estimates on
  average larger metallicities than those in the literature by about $0.18$~dex.

  In conclusion, on average metallicities derived in this work are higher than
  most literature values calculated by various methods, but the overall agreement
  is satisfactory. Additionally, our choice to use the \citet{Hilker2000}
  metallicity calibration is validated by the wide range of metallicities
  found in the LMC.
  The calibration still has some problems in the low metallicity regime, but it
  gives satisfying results for higher metallicities.
  Other possible causes of non-compliance with the literature may be the use of
  a~higher reddening value in many cases; differential reddening; or the presence
  of N-enriched stars that are found in old and intermediate-age clusters, as their
  presence may lead to an overestimation of the calculated metallicities.

  A~similar comparison of ages shows very satisfactory agreement of our estimations
  with literature values. There are only a~few points deviating significantly
  from our results in Fig.~\ref{fig:agelit}, but the same points also deviate
  from other measurements from the literature.

  A comparison with AMRs from the literature suggests very similar conclusions
  as those described in previous sections. Many authors ascribe ages that are
  older than the presently accepted age of the Universe to the clusters
  \citep[e.g.,][]{Olsen1998,Dirsch2000,LR2003}, which we also obtain from the
  isochrone fitting. \citet{Olszewski1991}, for this example, arbitrarily assigned
  ages of $12$~Gyr for oldest clusters.

  Multiple authors report the existence of an age gap between $\sim$$3$ to
  $\sim$$12$~Gyr when almost no star clusters were formed
  \citep[e.g.,][]{Olszewski1991,Hill2000,Sharma2010}, with a~sole exception of
  the cluster ESO121-3 \citep{Olszewski1991,Bica1998}; and more recently, there
  is also KMHK1592 \citep{Piatti2022}.
  The gap is also evident in our AMR.  However, we do not confirm the specific
  position of ESO121-3 (marked with grey arrow in Fig.~\ref{fig:am}) in the relation
  (noting that for KMHK1592, we do not have the data). The cluster fits the PT98 bursting
  model prediction very well and it seems to complete the initial active epoch of star formation
  ($\sim$$9$~Gyr ago). This period is also characterized by slow increase of the
  metallicity value seen not only in the cluster, but also in the field AMR
  \citep[e.g.,][]{Carrera2008a,PiattiGeisler2013}. This phase came to an end
  $\sim$$3$~Gyr ago, finished by a~burst of chemical enrichment. We confirm this
  result, as we also see an increase of star cluster formation around this age.

  \citet[and references therein]{LR2003} notice a~second minimum in the cluster
  formation rate between $\mathrm{log(Age)}=8.3-8.8$ ($200-700$~Myr).
  Such a~gap, however, is not present in the AMR from, for instance,  \citet[see their Fig.~20]{Palma2015} or \citet[see their Fig.~13]{Perren2017},
  based on star clusters studied using the Washington photometry and coming from
  many fields, that is, mostly regions beyond the LMC bar.
  We, too, tend to see a~lower cluster formation rate in the mentioned age range
  in our AMR. In the LMC bar, the burst in intermediate ages started later than
  in the non-bar regions, but efficient formation of clusters started there earlier
   than in the outer regions (where we do not observe clusters between
  $\mathrm{log(Age)} = 8.3 - 8.6$). Just as it seems reasonable to
  claim that the end of the first age gap indicates some dramatic event in the
  history of the LMC (such as an interaction with another galaxy), the later
  increase in cluster formation rate appears to have a~different cause, as it
  apparently propagates in the inside-outside direction. The latter, however, might be
  caused by an observing bias, resulting from the poor coverage of the various
  LMC regions.

  \citet{Piatti2003b} described a~dual behavior among clusters in the inner and outer
  disk of the LMC. They note that outer disk clusters formed up to $\sim$$1$~Gyr
  ago, reaching [Fe/H] values of about $-0.35$~dex. We agree with this statement,
  as we also see in our AMR that the effective creation of intermediate-age star
  clusters ends $\sim$$1$~Gyr ago at the metallicity level of about $-0.35$~dex,
  growing to about $-0.20$~dex until the present time. They also note that in the
  inner LMC disk, clusters have formed up until the present time and they have
  higher metallicities, where some of them even reach solar abundances. Our results
  confirm that observation.

  In the AMR of \citet{Palma2015}, there is broad dispersion in metallicity for
  young clusters (about $0.50$~dex). We notice a similar spread in our data.
  Moreover, these authors report a~tendency for younger clusters to be more metal-rich than
  intermediate ones (clusters older than $\sim$$1.2$~Gyr have
  $\mathrm{[Fe/H] \leq -0.40}$~dex). Our intermediate-age clusters have metallicities
  between about $-0.35$ and $-0.65$~dex, but we do observe some lower values for
  younger clusters too. Nonetheless, qualitatively, the two relationships are similar.

  \citet{Perren2017} reported a~drop of the metallicity value in their AMR from about
  $-0.45$~dex $\sim$$3.8$~Gyr ago to about $-0.60$~dex $\sim$$3$~Gyr ago.
  We do not observe this in our AMR, as in this age range the metallicites of star
  clusters in our sample increase with time. Furthermore, the authors describe
  a~steep increase in metallicity between $3-2$~Gyr ago up to
  $\mathrm{[Fe/H] \sim -0.30}$~dex, after which the metallicity reaches the present
  day value of about $-0.15$~dex. The increase in our AMR is shallower and longer,
  occurring between $\sim$$3-1$~Gyr ago, and the level of metallicity reaches about
  $-0.35$~dex. Nevertheless, we note that many of our young clusters have mean
  metallicities on the level of $-0.15$~dex or even higher.

  An interesting and completely independent comparison of our results is undertaken with
  \citet{Graczyk2018}, where the authors derived their AMR from analysis of
  spectroscopic and photometric observations of $20$ eclipsing binary systems
  from the field. Figure~\ref{fig:amrDarek} is analogous to their Fig.~7, with
  overplotted values for our clusters. The authors claim that, on average, the
  metallicity of stars older than about $0.6$~Gyr is noticeably smaller than the
  metallicity of the younger population, which we also observe in the case of
  stellar clusters. However,  the authors do go on to argue that from $2$~Gyr to about
  $0.6$~Gyr, the metallicity shows a~large scatter, indicating a~flat relation at
  a~constant level. The eclipsing binaries that were studied lie in the LMC bar,
  as well as the non-bar regions; these are two areas that we find ought to be
  considered separately. Furthermore, these conclusions were reached based on
  a~very low number of studied systems. As we show in the zoom panel in
  Fig.~\ref{fig:amrDarek}, the average metallicity of star clusters from non-bar
  regions (calculated in $0.3$~Gyr age bins) tend to increase in time, so we do
  not confirm the flat relation for those clusters. However, a~large scatter of
  metallicity and poor statistics on bar clusters in the considered age range do
  not allow us to either confirm or refute this interpretation.

%-----------------------------------------------------------------
%                                                One column figure
%-----------------------------------------------------------------
   \begin{figure}
   \centering
   \includegraphics[width=\hsize]{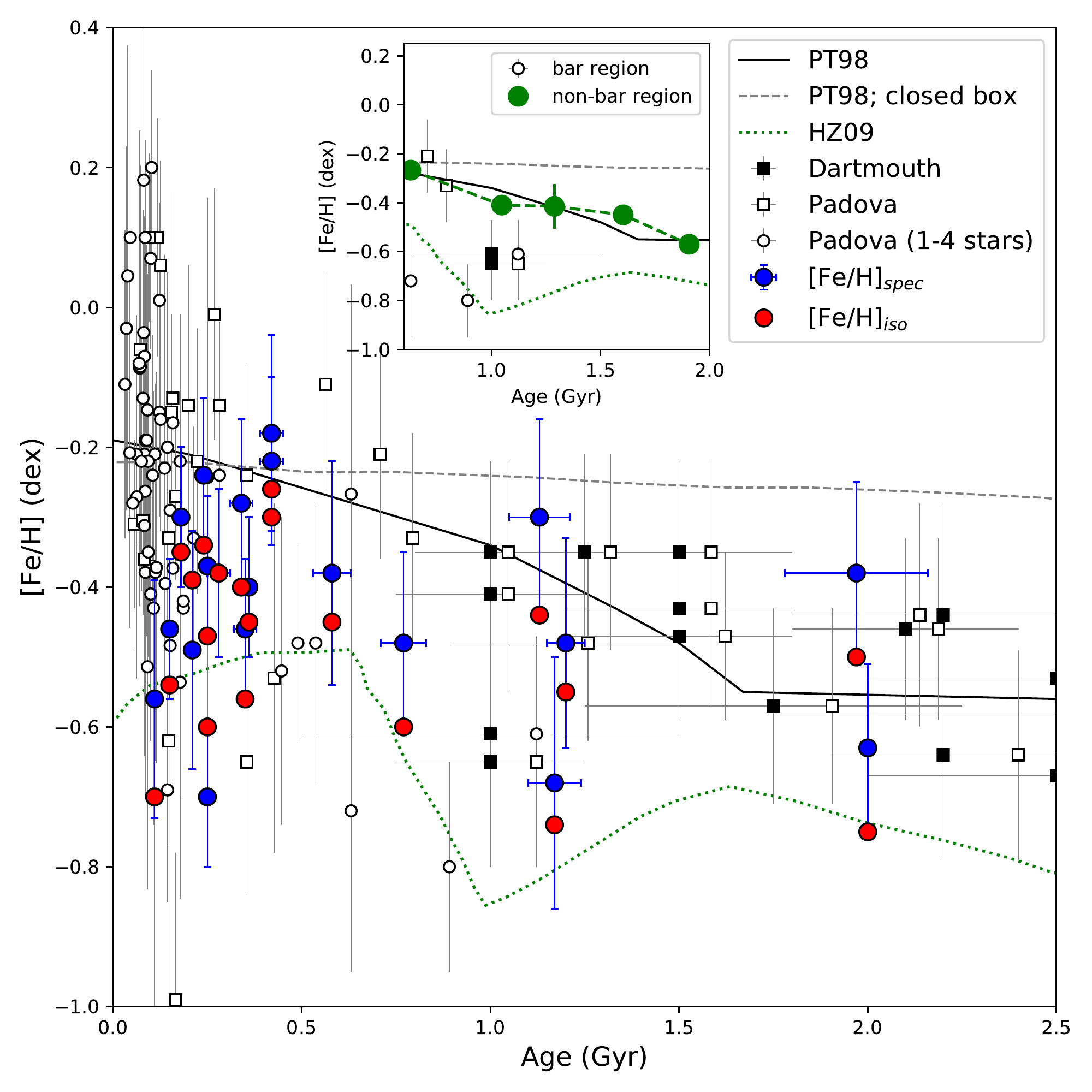}
      \caption{Comparison of the AMR derived in this work with the one provided
      by \citet{Graczyk2018}. Metallicity values obtained from spectroscopy
      (blue) and isochrones (red) from \citet{Graczyk2018}, and this work (black
      and open squares; open circles). Green points in the zoom panel mark average
      metallicites of star clusters from non-bar regions calculated in age bins
      of $0.3$~Gyr in the age range between $0.6$ to $2$~Gyr.
      }
      \label{fig:amrDarek}
   \end{figure}
%-----------------------------------------------------------------

%--------------------------------------------------------------------
\subsection{Ages of star clusters based on Cepheids}
\label{ssec:agecep}
%--------------------------------------------------------------------

 Overall, $26$ open clusters from our sample contain Cepheid variables in their fields.
  We decided to take advantage of these stars, and compute their ages using
  theoretical period-age (PA) relations from the literature, to obtain an
  independent estimation of ages of the clusters hosting Cepheids.
  We first checked the PMs and parallaxes of the identified Cepheids with the
  Gaia EDR3 catalog to check if their PMs are consistent with PMs of other stars
  in the area of the considered cluster. Otherwise, they could be excluded from
  potential cluster members.
  Periods of fundamental and first-overtone Cepheids were adopted from the OGLE
  Collection of Variable Stars \citep{Soszynski2015a,Soszynski2017}.
  Figure~\ref{fig:ageScep} shows a~comparison of the mean cluster ages estimated
  by us from isochrone fitting, as well as PA relations for the LMC metallicity,
also  summarized in Table~\ref{tab:ageCep}.

  The average cluster ages derived from PA relations presented by \citet{Bono2005}
  are systematically smaller than our results obtained from isochrone fitting,
  where the shift is $\Delta \mathrm{log(Age)} \sim 0.29$ ($\sigma = 0.11$).
  \citet{Anderson2016}, unlike former authors, included in their models the effect
  of rotation. When we apply their formula to the same sample of Cepheids, the
  resulting average cluster ages are systematically older for the relation of
  \citet{Anderson2016}. This is due to rotational mixing at the edge of the
  convective core during the main sequence, which extends this evolutionary phase.
  The average cluster ages calculated based on their PA relations with average
  rotation ($\omega = 0.5$, i.e., average initial rotation) and averaged over the
  2nd and 3rd crossing and instabillity strip width, are much closer to our
  estimations, although a~bit older ($\Delta \mathrm{log(Age)} \sim -0.08$,
  $\sigma = 0.09$).
  The recent work of \citet{DeSomma2021} presents new PA relations derived for
  canonical (case~A) and non-canonical (case~B) models of the mass-luminosity
  relation. Canonical models neglect the existence of any physical process able
  to increase the size of the convecting core of a~star (for example, core
  convective overshooting) contrary to the non-canonical models. Their case~A
  gives values very similar to \citet{Bono2005}, while case~B results in older
  ages. The mean difference between cluster ages obtained in this work and case~A
  of \citet{DeSomma2021} is $\Delta \mathrm{log(Age)} \sim 0.24$ ($\sigma = 0.09$)  and for case~B, it is $\sim 0.10$ ($\sigma = 0.08$).

  The above comparison illustrates the differences between various models, where
  the age spread between them is quite large. That means that we still do not have
  a~precise method for calculating the ages of stars and star clusters, and all
  methods are burdened with errors. Our age values are closest to ages obtained
  with Cepheid PA relations that \citet{Anderson2016} derived for models including
  rotation and the non-canonical models from \citet{DeSomma2021}. These models
  are more physically justified than the two others as mixing beyond the edge of
  convective core during its main sequence evolution, which leads to longer main
  sequence phase and older Cepheids -- this is expected. Without additional mixing the
  so-called Cepheid mass discrepancy problem is manifested in masses of classical
  Cepheids, as predicted by evolution theory, that are overestimated as compared
  to pulsation masses \citep[see][]{Keller2008}. Additional mixing, whether due
  to overshooting, or due to rotation, alleviates or removes the discrepancy
  \citep[see e.g.,][]{PradaMoroni2012,Anderson2014}. Also, the width of the main
  sequence in the turn-off region most probably is explained by a~large population
  of fast rotating stars in LMC clusters \citep[e.g.,][]{Bastian2016}.
  % which further supports the greater physicality of non-canonical models.
  Additional mixing is also necessary to reproduce properties of helium burning
  double-lined eclipsing binary systems \citep[see e.g.,][]{ClaretTorres2016} or
  to reproduce the width of the main sequence \citep[see e.g.,][]{MaederMermilliod1981}.

  %-----------------------------------------------------------------
  %                                                One column figure
  %-----------------------------------------------------------------
     \begin{figure}
     \centering
     \includegraphics[width=\hsize]{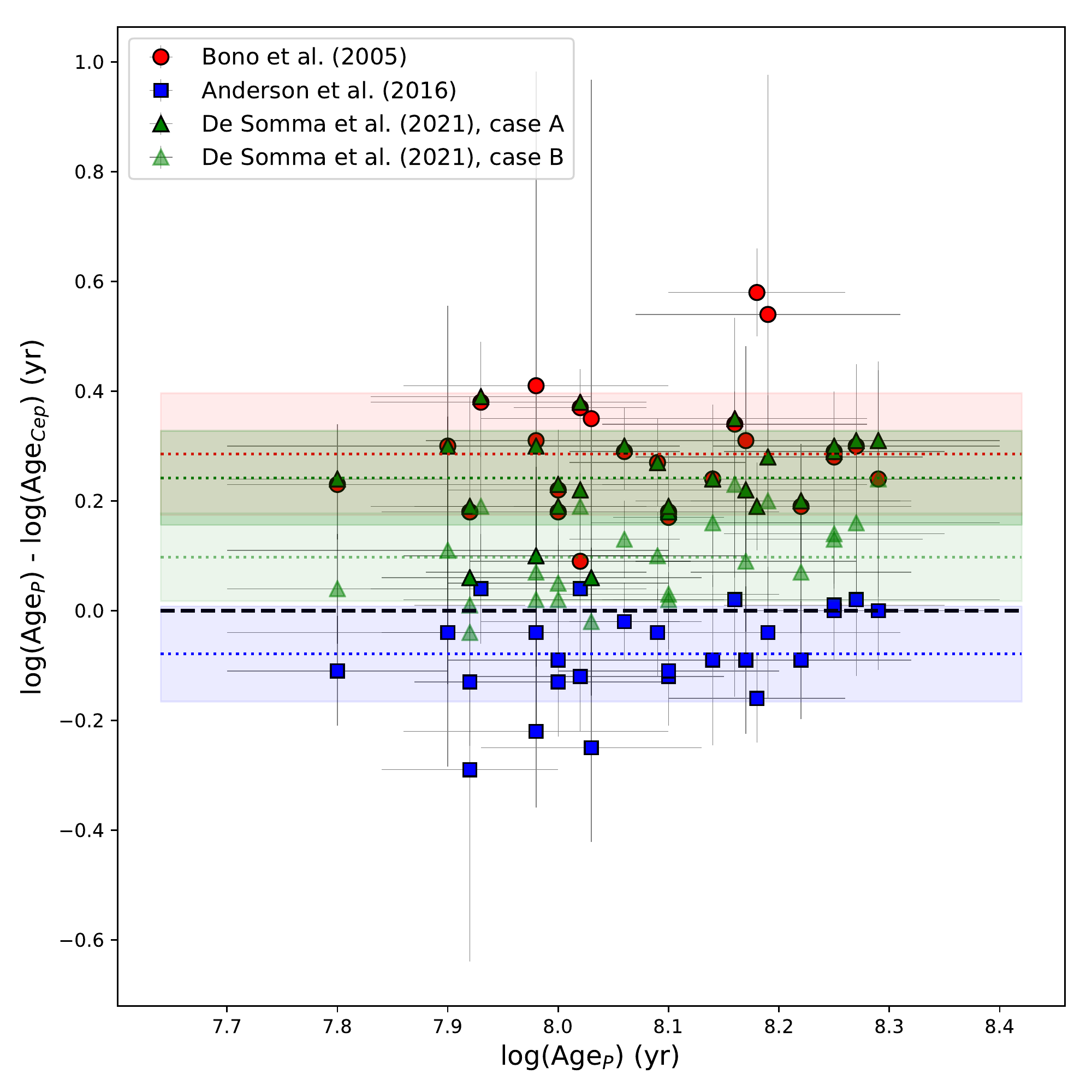}
        \caption{Comparison of the ages of star clusters obtained with two independent
        methods:
        isochrone fitting to the Str\"omgren CMDs (with Padova isochrones) and PA
        relations for Cepheids from the literature. Dotted lines and shaded areas
        mark average differences of ages and their standard deviations, respectively.
        }
        \label{fig:ageScep}
     \end{figure}
  %-----------------------------------------------------------------

%--------------------------------------------------------------------
\subsection{Comparison of age--metallicity relation for clusters in the LMC and SMC}
\label{ssec:AMRlmcsmc}
%--------------------------------------------------------------------

  The uniform analysis of the AMR in the LMC presented in this work and in the
  SMC from Paper~I provides the opportunity to compare them, as shown in
  Fig.~\ref{fig:AMRlmcsmc}. Firstly, we can conclude that these two galaxies show
  a~different chemical enrichment history. Rapid chemical enrichment has occured
  in the LMC since the very beginning of its history and lasted for about $3$~Gyr.
  During that time, many old globular clusters were formed, that survived till
  this day. On the other hand, in the SMC, star clusters older than $10$~Gyr are
  not observed. For the next $\sim$$6$~Gyr there is a~gap in the cluster formation
  history of the LMC while in the AMR of the SMC two bursts are visible at about
  $7.5$ and $3.5$~Gyr ago, causing star cluster formation and chemical enrichment
  in the outer regions of this galaxy. The latter burst is followed by another
  in the LMC's outer regions, which might indicate the ancient event referred to
  in \citet{HZ2009} as the major triple-interaction between these two galaxies
  and the MW. The last enrichment raised the current metallicity value in the SMC
  up to about $-0.70$~dex, and the average metallicity of numerous young star
  clusters in the LMC located mostly in the bar region rose to about $-0.25$~dex.

%-----------------------------------------------------------------
%                                                One column figure
%-----------------------------------------------------------------
   \begin{figure}
   \centering
   \includegraphics[width=\hsize]{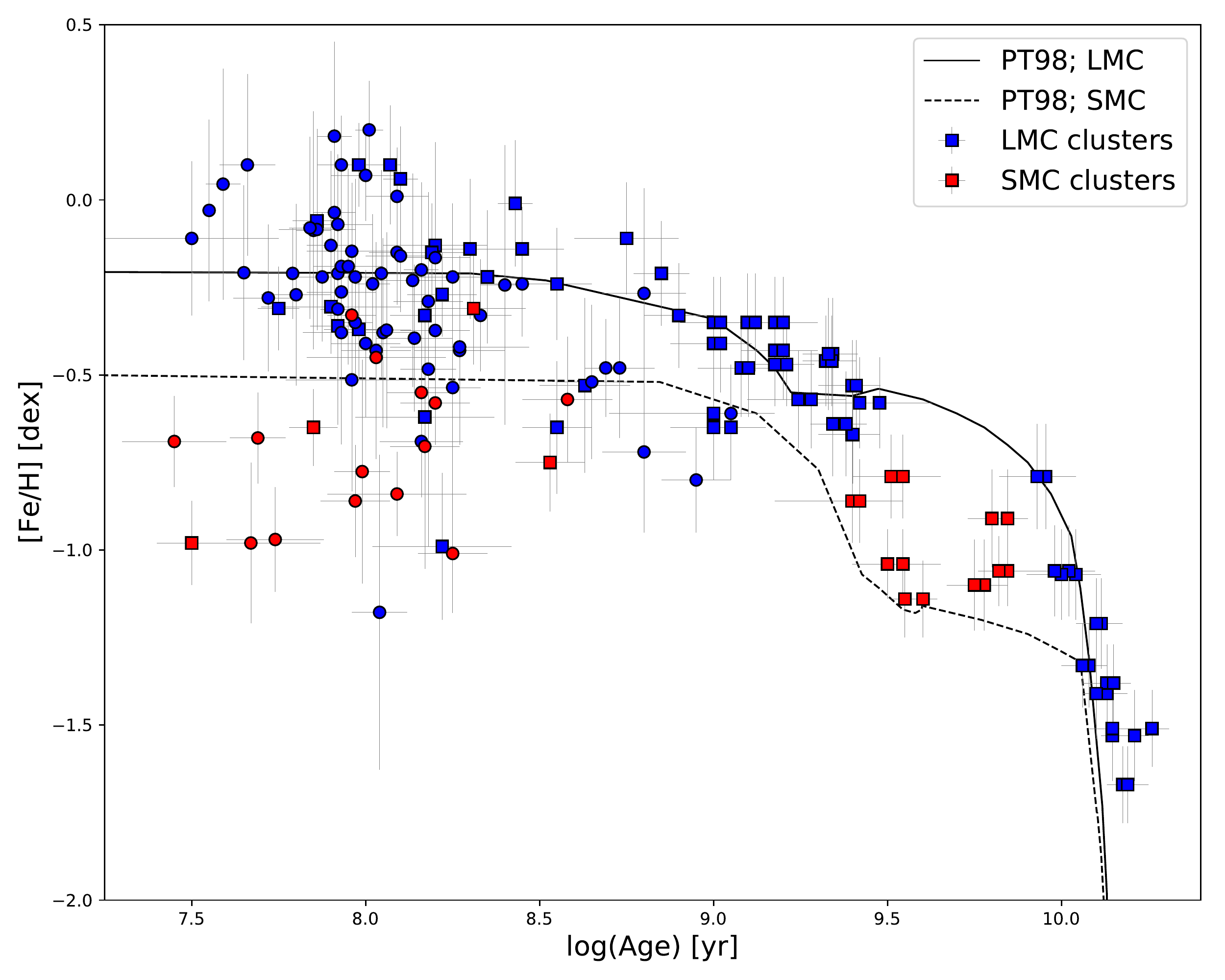}
      \caption{Comparison of the AMR of the LMC (blue) and the SMC (red).
      Overplotted are the PT98 bursting models (solid and dashed lines for the
      LMC and the SMC, respectively). The meaning of squares and circles as in
      Fig.~\ref{fig:am}.
      }
      \label{fig:AMRlmcsmc}
   \end{figure}
%-----------------------------------------------------------------

%--------------------------------------------------------------------
\section{Summary and conclusions}
\label{sec:summ}
%--------------------------------------------------------------------

   In this work, we present Str\"omgren photometry of $80$ fields in the LMC,
   where we identified $147$ star clusters for which we derived either mean
   metallicities and ages or only ages. We also calculated mean metallicities
   for the fields around these clusters. To obtain metallicities of individual
   stars, we took advantage of the metallicity calibration of the Str\"omgren
   colors presented by \citet{Hilker2000}, derived for a~wide range of
   metallicities (from $-2.2$ up to $0.0$~dex). We estimated ages of clusters
   from our sample using theoretical isochrones from the Dartmouth and Padova
   groups. During the~analysis we utilized the~recent reddening maps of G20 and
   S21, as well as the distance to the LMC obtained by \citet{Pietrzynski2019}.

   As a~result, we obtained both metallicities and ages for $110$ star clusters
   from various regions of the LMC. For the remaining $37$ clusters, we provided
   the ages only. To the best of our knowledge, for $66$ clusters, this is the
   first-ever estimation of the metallicity, for $43$ of them age was provided
   for the first time, and, finally, in case of $29$ clusters (having metallicites
   from the range between $-0.62$ and $0.20$~dex and log-ages between $7.59$ and
   $9.05$), both values are given here for the first time. These results allowed
   us to trace the metallicity and age distribution across the LMC, and to construct
   the AMR of the LMC star clusters, from which we were able to deduce the following
   chemical enrichment history in the LMC:
   \begin{itemize}
     \item An initial, ancient burst created old, populous star clusters.
     This period was relatively short and the metallicity increased from
     $\mathrm{[Fe/H]} \sim -1.67$~dex up to about $-0.80$~dex.
     \item This is followed by a~long period of stagnation with hardly any star
     cluster formation, and the chemical enrichment was very minor.
     This age gap lasted from $\sim$$9$~Gyr to $\sim$$3$~Gyr ago.
     \item After that epoch, about $\sim$$3$~Gyr ago, a~burst of formation of
     intermediate-age clusters in the non-bar regions lasted for $\sim$$2$~Gyr,
     enriching the environment from about $-0.65$~dex to $-0.35$~dex. This was
     followed by another minimum in star cluster formation lasting for
     $\sim$$200$~Myr. This, however, is highly uncertain and might be a~consequence
     of the poor coverage of the outer regions of the LMC. The first large burst
     was probably a~result of galaxy-galaxy interaction, while the second one,
     if true, might possibly have a~different cause. The chemical abundance in
     the non-bar region rose to the present-day value of about about $-0.20$~dex.
     \item an analogous burst appeared $\sim$$1$~Gyr ago in the bar region, enriching
     the environment from about $-0.65$~dex up to $-0.50$~dex (with a~few clusters
     having about $-0.20$~dex). After the burst the cluster formation rate was
     relatively low, until $\sim 300$~Myr ago, since when many young and metal-rich
     clusters were formed, some of them even having solar-like abundances. These
     young clusters are also characterized by a~large spread in metallicity (of
     about $0.50$~dex). The bar star clusters (except two) with metallicities
     of about $-0.5$~dex and lower are located at the western end of the bar.
   \end{itemize}

   We compared our AMR with literature SFHs and other AMRs. The best fitting SFH
   model in the non-bar regions is the PT98 bursting model, which reproduces the
   initial rapid enrichment very well, followed by a certain stagnation period and
an   intermediate-age burst, which raised the metallicity to the current level of
   $\mathrm{[Fe/H]} \sim -0.20~$dex. Our results fit this scenario well. The AMR
   of the LMC bar seems to fit the HZ09 model up until $\sim$$300$~Myr ago, but
   fails for young, metal-rich clusters. The AMRs from the literature qualitatively
   agree well with our results and are also characterized by quite a~large
   metallicity dispersion for a~given age, which is especially evident for young
   clusters. It is also worth noting that on average we provided higher metallicity
   values than reported by various authors.

   As an~independent test of the correctness of our isochrone-derived ages, we
   compared them with mean ages calculated for the clusters hosting Cepheid
   variables. For the latter, we used PA relations from literature along with
   pulsation periods from OGLE catalogs. We note systematic shifts between our
   values and a~given PA relation, which show that the ages derived from PA
   relations are model-dependent. This implies that
   a~precise method of calculating ages of stars, and therefore star clusters as well, is lacking.

   Finally, we compared the AMR for the LMC from this work with the AMR for the SMC
   from \citet{Narloch2021}. We claim that these two relations show distinct
   chemical enrichment histories, which, however, became entangled in intermediate
   ages, suggesting a~former interaction of these two galaxies. The photometric
   catalog of Str\"omgren photometry used in this work has been made publicly available.

\begin{acknowledgements}
   We thank the anonymous referee for valuable comments which improved this paper.

   The research leading to these results has received funding from the European
   Research Council (ERC) under the European Union’s Horizon 2020 research and
   innovation program (grant agreement No 695099). We also acknowledge support
   from the National Science Center, Poland grants MAESTRO UMO-2017/26/A/ST9/00446,
   BEETHOVEN UMO-2018/31/G/ST9/03050 and DIR/WK/2018/09 grants of the Polish Ministry
   of Science and Higher Education.

   We also acknowledge financial support from UniverScale grant financed by the
   European Union's Horizon 2020 research and innovation programme under the grant
   agreement number 951549.

   We gratefully acknowledge financial support for this work from the BASAL Centro
   de Astrofisica y Tecnologias Afines (CATA) AFB-170002 and the Millenium Institute
   of Astrophysics (MAS) of the Iniciativa Cientifica Milenio del Ministerio de
   Economia, Fomento y Turismo de Chile, project IC120009.

   B.P. gratefully acknowledge support from the Polish National Science Centre
   grant SONATA BIS 2020/38/E/ST9/00486.

   R.S. acknowledges support from SONATA BIS grant, 2018/30/E/ST9/00598, from the
   National Science Centre, Poland.

   P.W. gratefully acknowledges financial support from the Polish National Science
   Centre grant PRELUDIUM2018/31/N/ST9/02742.

   This work has made use of data from the European Space Agency (ESA) mission
   {\it Gaia} (\url{https://www.cosmos.esa.int/gaia}), processed by the {\it Gaia}
   Data Processing and Analysis Consortium (DPAC,
   \url{https://www.cosmos.esa.int/web/gaia/dpac/consortium}).
   Funding for the DPAC has been provided by national institutions, in particular
   the institutions participating in the {\it Gaia} Multilateral Agreement.

\end{acknowledgements}

%-------------------------------------------------------------
%                      A rotated Two column Table in landscape
%-------------------------------------------------------------
%\begin{sidewaystable*}
\begin{table*}
\caption{Str\"omgren photometry of fields in the LMC.}
\label{tab:phot}
\centering
\begin{tabular}{ccrrcrccrcc}
\hline\hline
RA & DEC & X & Y & Field
& $V$ & $\sigma_{V_{DAO}}$ & $\sigma_{V}$
& $(b-y)$ & $\sigma_{(b-y)_{DAO}}$ & $\sigma_{(b-y)}$ \\
(deg) & (deg) & (pixel) & (pixel) & &
(mag) & (mag) & (mag) & (mag) & (mag) & (mag) \\
\hline
 69.469470 & -70.587620 &  6.071 &  887.872 & 1 & 21.118 & 0.056 & 0.056 & 0.236 & 0.085 & 0.086 \\
 69.469389 & -70.566830 &  7.703 & 1375.837 & 1 & 20.969 & 0.052 & 0.053 & 0.416 & 0.080 & 0.081 \\
 69.469200 & -70.572246 &  8.944 & 1248.710 & 1 & 20.386 & 0.031 & 0.032 & 0.303 & 0.047 & 0.049 \\
 69.469042 & -70.580556 &  9.779 & 1053.677 & 1 & 19.167 & 0.012 & 0.015 & 0.605 & 0.019 & 0.025 \\
 69.468859 & -70.570779 & 11.681 & 1283.150 & 1 & 21.404 & 0.074 & 0.074 & 0.266 & 0.107 & 0.107 \\
 \multicolumn{11}{l}{(...)}\\ % To combine 13 columns into a single one \\
\hline
\end{tabular}
\begin{tabular}{rccrr}
\hline\hline
$m1$ & $\sigma_{m1_{DAO}}$ & $\sigma_{m1}$ & CHI & SHARP  \\
(mag) & (mag) & (mag) & & \\
\hline
 0.001 & 0.195 & 0.200 & 0.779 & -0.696 \\
-0.134 & 0.153 & 0.158 & 0.816 & -0.418 \\
 0.013 & 0.082 & 0.085 & 0.792 & -0.262 \\
 0.385 & 0.059 & 0.067 & 0.778 &  0.187 \\
 0.031 & 0.185 & 0.190 & 0.790 & -1.177 \\
 \multicolumn{5}{l}{(...)}\\ % To combine 13 columns into a single one \\
\hline
\end{tabular}
\tablefoot{A~complete table is presented in its entirety in electronic
form on the Araucaria Project webpage and the CDS. A~portion is shown
here for guidance regarding its form and content.}
\end{table*}
%\end{sidewaystable*}
%-------------------------------------------------------------

%-------------------------------------------------------------
%                                             Two column Table
%-------------------------------------------------------------
%
\begin{table*}
\caption{\label{tab:ageCep} Ages of star clusters hosting Cepheid variables.}
\centering
\begin{tabular}{llcccclc}     % 7 columns
\hline\hline
Cluster & log(Age$_P$) & log(Age$_{B05}$) & log(Age$_{A16}$) & log(Age$_{DS2021}$)
& log(Age$_{DS2021}$) & N$_{F}$ & N$_{1O}$ \\
 &  &  &  & canonical & noncanonical & \\
 & (yr) & (yr) & (yr) & (yr) & (yr) & \\
\hline
KMHK421          & 8.25$\pm$0.08 & 7.97$\pm$0.01 & 8.25$\pm$0.007 & 7.96$\pm$0.009 & 8.12$\pm$0.008 & 2 & - \\
BSDL581          & 8.02$\pm$0.06 & 7.65 & 7.98 & 7.64 & 7.83 & 1 & - \\
OGLE-CL~LMC~113  & 7.93$\pm$0.10 & 7.55 & 7.89 & 7.54 & 7.74 & 1 & - \\
NGC1850          & 7.90$\pm$0.20$^{(*)}$ & 7.60$\pm$0.16 & 7.94$\pm$0.14 & 7.60$\pm$0.16 & 7.79$\pm$0.14 & 3$^{(a)}$ & - \\
H88~165          & 7.80$\pm$0.10 & 7.57 & 7.91 & 7.56 & 7.76 & 1 & - \\
NGC1854          & 7.92$\pm$0.08 & 7.74$\pm$0.14 & 8.21$\pm$0.34 & 7.86$\pm$0.32 & 7.96$\pm$0.19 & 1 & 1 \\
% NGC1858          & 7.85$\pm$0.10 & 8.00$\pm$0.10 & 8.42$\pm$0.11 & 8.10$\pm$0.05 & 8.17$\pm$0.05 & 1 & 1 \\
NGC1866          & 8.22$\pm$0.10 & 8.03$\pm$0.03 & 8.31$\pm$0.04 & 8.02$\pm$0.03 & 8.15$\pm$0.05 & 14 & 2 \\
NGC1894          & 8.00$\pm$0.10 & 7.78 & 8.09 & 7.77 & 7.95 & 1 & - \\
NGC1903          & 7.98$\pm$0.10 & 7.67$\pm$0.09 & 8.02$\pm$0.09 & 7.68$\pm$0.10 & 7.91$\pm$0.14 & 2 & 1 \\
OGLE-CL~LMC~321  & 8.27$\pm$0.13 & 7.97$\pm$0.05 & 8.25$\pm$0.05 & 7.96$\pm$0.05 & 8.11$\pm$0.05 & 2 & - \\
H88~283          & 8.18$\pm$0.08$^{(*)}$ & 7.60 & 8.34 & 7.99 & - & - & 1 \\
% H88~283          & 8.17$\pm$0.07 & 7.60 & 8.34 & 7.99 & - & - & 1 \\
% H88~283          & 8.19$\pm$0.07 & 7.60 & 8.34 & 7.99 & - & - & 1 \\
OGLE-CL~LMC~407  & 8.09$\pm$0.08 & 7.82 & 8.13 & 7.82 & 7.99 & 1 & - \\
OGLE-CL~LMC~431  & 8.03$\pm$0.10 & 7.68$\pm$0.61 & 8.28$\pm$0.14 & 7.97$\pm$0.02 & 8.05$\pm$0.09 & 1 & 1 \\
% OGLE-CL~LMC~438  & 8.09$\pm$0.10 & 7.84 & 8.14 & 7.83 & 8.00 & 1 & - \\
NGC1950          & 8.00$\pm$0.10 & 7.82 & 8.13 & 7.81 & 7.98 & 1 & - \\
NGC1969          & 8.10$\pm$0.05 & 7.93 & 8.22 & 7.92 & 8.08 & 1 & - \\
BSDL1759         & 8.06$\pm$0.05 & 7.77$\pm$0.05 & 8.08$\pm$0.05 & 7.76$\pm$0.05 & 7.93$\pm$0.05 & 2 & - \\
NGC1971          & 8.02$\pm$0.10 & 7.93 & 8.14 & 7.80 & - & - & 1 \\
BSDL1821         & 7.92$\pm$0.05 & 7.74 & 8.05 & 7.73 & 7.91 & 1 & - \\
NGC1986          & 7.98$\pm$0.12 & 7.57$\pm$0.56 & 8.20$\pm$0.07 & 7.88$\pm$0.10 & 7.96$\pm$0.21 & 1 & 1 \\
OGLE-CL~LMC~512  & 8.14$\pm$0.11$^{(*)}$ & 7.90$\pm$0.08 & 8.23$\pm$0.11 & 7.90$\pm$0.08 & 7.98$\pm$0.04 & 1 & 1 \\
% OGLE-CL~LMC~512  & 8.18$\pm$0.10$^{(*)}$ & 7.90$\pm$0.08 & 8.23$\pm$0.11 & 7.90$\pm$0.08 & 7.98$\pm$0.04 & 1 & 1 \\
% OGLE-CL~LMC~512  & 8.10$\pm$0.05$^{(*)}$ & 7.90$\pm$0.08 & 8.23$\pm$0.11 & 7.90$\pm$0.08 & 7.98$\pm$0.04 & 1 & 1 \\
NGC2016          & 8.19$\pm$0.12 & 7.65$\pm$0.42 & 8.23$\pm$0.004 & 7.91$\pm$0.04 & 7.99$\pm$0.15 & 1 & 1 \\
BSDL2205         & 8.25$\pm$0.10 & 7.96 & 8.24 & 7.95 & 8.11 & 1 & - \\
OGLE-CL~LMC~585  & 8.10$\pm$0.10 & 7.92 & 8.21 & 7.91 & 8.07 & 1 & - \\
OGLE-CL~LMC~591  & 8.29$\pm$0.10 & 8.05$\pm$0.09 & 8.29$\pm$0.04 & 7.98$\pm$0.08 & 8.05$\pm$0.19 & 1$^{(b)}$ & - \\
NGC2065          & 8.17$\pm$0.10 & 7.86$\pm$0.14 & 8.26$\pm$0.09 & 7.95$\pm$0.08 & 8.08$\pm$0.08 & 7$^{(a)}$ & 3 \\
NGC2136          & 8.16$\pm$0.12 & 7.82$\pm$0.14 & 8.14$\pm$0.13 & 7.81$\pm$0.14 & 7.93$\pm$0.12 & 2$^{(c)}$ & 1 \\
\hline
\end{tabular}
\tablefoot{
  Cluster: name of the cluster;
  log(Age$_P$): logarithm of age derived from the Padova isochrones;
  log(Age$_{B05}$): logarithm of age derived from PA relation from \citet{Bono2005};
  log(Age$_{A16}$): from \citet{Anderson2016};
  log(Age$_{DS2021}$): from \citet{DeSomma2021};
  N$_{F}$, N$_{1O}$: number of Cepheids used for calculation (fundamental and
  first-overtone modes, respectively).
  \begin{itemize}\scriptsize
  \itemsep0em
  \item [$^{(*)}$] Average age from multiple measurements (with maximum error) from
  Table~\ref{tab:reddC}.
  \item [$^{(a)}$] One of the Cepheids is classified as F/1O in the OGLE catalogs.
  Its average age is calculated from the average of the fundamental and first-overtone
  PA relations.
  \item [$^{(b)}$] Cepheid classified as F/1O in OGLE catalogs. Its age was calculated
  from the average of the fundamental and first-overtone PA relations.
  \item [$^{(c)}$] There are two additional Cepheids within the cluster radius,
  but they were rejected because \citet{Mucciarelli2012} mark them as field objects
  and their ages are noticeably younger than the three variables used for mean age
  calculation.
  \end{itemize}}
\end{table*}
%
%-------------------------------------------------------------


\begin{appendix} %First appendix

\section{Astrophysical properties of star clusters and fields}

Tables~\ref{tab:reddC} and \ref{tab:reddF} summarize the results for star clusters
and fields studied in this work.

%-------------------------------------------------------------
%                                             Two column Table
%-------------------------------------------------------------
%
\longtab[1]{
\begin{longtable}{lcclcccc}
\caption{\label{tab:reddC} Star clusters in the LMC.} \\
\hline\hline
Cluster & D$_{maj}$;D$_{min}$ & P$_A$ & $E(B-V)_C$ & $\mathrm{[Fe/H]}_C$ & N$_C$
& log(Age$_{P}$) & Age$_{D}$\\
 & (arcmin) & (deg) & (mag) & (dex) & & (yr) & (Gyr) \\
\hline
\endfirsthead
\caption{continued.}\\
\hline\hline
Cluster & D$_{maj}$;D$_{min}$ & P$_A$ & $E(B-V)_C$ & $\mathrm{[Fe/H]}_C$ & N$_C$
& log(Age$_{P}$) & log(Age$_{D}$)\\
 & (arcmin) & (deg) & (mag) & (dex) & & (yr) & (yr) \\
\hline
\endhead
\hline
%\endfoot
%%
NGC1651         & 2.70;2.70 & 0 & 0.110$^{(a)}$ & -0.43$\pm$0.04 (0.13) & 81 & 9.20$\pm$0.10 &
1.50$\pm$0.30 \\
KMHK21          & 1.50;1.50 & 0 & 0.070$^{(a)}$ & -0.46$\pm$0.04 (0.12) & 25 & 9.34$\pm$0.07 &
2.10$\pm$0.30 \\
NGC1841         & 4.00;4.00 & 0 & 0.160$^{(a)}$ & -1.51$\pm$0.02 (0.11) & 52 & 10.26$\pm$0.05 &
14.0$\pm$2.00 \\
NGC1711         & 3.50;3.30 & 40 & 0.125 & -0.31$\pm$0.03 (0.11) & 6 & 7.75$\pm$0.06 & - \\
KMHK156$^{(**)}$ & 0.90;0.80 & 120 & 0.109 & - & - & 7.38$\pm$0.03 & - \\
NGC1754         & 1.60;1.60 & 0 & 0.112 & -1.07$\pm$0.04 (0.13) & 50 & 10.00$\pm$0.10 &
11.0$\pm$2.00 \\
ESO85-21       & 1.30;1.30 & 0 & 0.040$^{(S)}$ & -0.44$\pm$0.09 (0.13) & 7 & 9.33$\pm$0.06 &
2.20$\pm$0.40 \\
NGC1786         & 2.00;2.00 & 0 & 0.089 & -1.33$\pm$0.03 (0.12) & 35 & 10.06$\pm$0.06 &
12.0$\pm$1.50 \\
NGC1795         & 1.40;1.30 & 170 & 0.110 & -0.35$\pm$0.04 (0.12) & 28 & 9.02$\pm$0.10 &
1.00$\pm$0.20 \\
KMHK421$^{(*)}$ & 1.00;0.90 & 130 & 0.082 & -0.54$\pm$0.29 (0.12) & 1 & 8.25$\pm$0.08 & - \\
H88~87$^{(*)}$  & 0.85;0.75 & 90 & 0.086 & -0.33$\pm$0.11 (0.12) & 2 & 8.33$\pm$0.09 & - \\
NGC1804$^{(*)}$ & 0.95;0.85 & 170 & 0.106 & -0.21$\pm$0.06 (0.11) & 3 & 7.79$\pm$0.10 & - \\
SL191$^{(*,**)}$   & 1.10;1.00 & 160 & 0.089 & -0.62$\pm$0.10 (0.11) & 6 & 8.17$\pm$0.20 & - \\
H88~104$^{(*)}$ & 0.55;0.50 & 170 & 0.075 & -0.80$\pm$0.07 (0.13) & 4 & 8.95$\pm$0.10 & - \\
H88~107$^{(*)}$ & 0.50;0.40 & 20 & 0.075 & -0.65$\pm$0.09 (0.12) & 5 & 9.05$\pm$0.10 &
1.00$\pm$0.25 \\
BRHT3b$^{(*)}$  & 0.75;0.65 & 30 & 0.073 & -0.72$\pm$0.19 (0.13) & 2 & 8.80$\pm$0.12 & - \\
NGC1830         & 1.30;1.20 & 60 & 0.075 & -0.14$\pm$0.03 (0.12) & 8 & 8.45$\pm$0.12 & - \\
SL211$^{(*,**)}$   & 1.00;0.85 & 50 & 0.071 & -0.24$\pm$0.37 (0.14) & 1 & 8.40$\pm$0.07 & - \\
BSDL555$^{(*,**)}$ & 0.65;0.55 & 100 & 0.088 & -0.24$\pm$0.08 (0.12) & 2 & 8.45$\pm$0.12 & - \\
KMHK521$^{(*)}$ & 0.60;0.55 & 10 & 0.119 & -0.04$\pm$0.28 (0.12) & 1 & 7.91$\pm$0.06 & - \\
BSDL565         & 0.85;0.50 & 90 & 0.079 & - & - & 7.96$\pm$0.05 & - \\
NGC1835         & 2.30;2.00 & 80 & 0.088 & -1.38$\pm$0.03 (0.11) & 38 & 10.15$\pm$0.05 &
13.5$\pm$2.00 \\
H88~119$^{(**)}$ & 0.50;0.45 & 140 & 0.081 & - & - & 8.31$\pm$0.10 & - \\
H88~120$^{(*,**)}$ & 0.70;0.65 & 140 & 0.085 & -0.52$\pm$0.11 (0.14) & 3 & 8.73$\pm$0.05 & - \\
                &  &  &  & -0.44$\pm$0.14 (0.14) & 2 & 8.73$\pm$0.10 & - \\
BSDL577         & 0.60;0.45 & 30 & 0.077 & - & - & 7.97$\pm$0.05 & - \\
BSDL581         & 0.60;0.50 & 140 & 0.077 & - & - & 8.02$\pm$0.06 & - \\
% BSDL580$^{(*)}$ & 0.55;0.50 & 140 & 0.108 & -0.39$\pm$0.25 (0.11) & 1 & 7.81$\pm$0.08 & - \\
BSDL582$^{(**)}$ & 0.95;0.80 & 160 & 0.111 & - & - & 7.50$\pm$0.12 & - \\
HS107$^{(*)}$   & 1.10;0.90 & 120 & 0.076 & -0.37$\pm$0.28 (0.12) & 1 & 8.20$\pm$0.10 & - \\
SOI343          & 0.95;0.90 & 10 & 0.110 & - & - & 7.32$\pm$0.05 & - \\
BSDL591$^{(*,**)}$ & 1.20;0.60 & 140 & 0.082 & -0.32$\pm$0.24 (0.12) & 2 & 7.92$\pm$0.07 & - \\
                &  &  &  & -0.38$\pm$0.19 (0.12) & 2 & 7.92$\pm$0.07 & - \\
                &  &  &  & -0.44$\pm$0.24 (0.11) & 1 & 7.94$\pm$0.11 & - \\
NGC1836         & 1.50;1.40 & 50 & 0.085 & -0.59$\pm$0.06 (0.13) & 13 & 8.63$\pm$0.13 & - \\
                &  &  &  & -0.42$\pm$0.12 (0.13) & 5 & 8.63$\pm$0.13 & - \\
                &  &  &  & -0.57$\pm$0.13 (0.13) & 9 & 8.64$\pm$0.13 & - \\
HS109$^{(*)}$   & 1.00;0.90 & 170 & 0.075 & -0.09$\pm$0.31 (0.12) & 1 & 7.85$\pm$0.05 & - \\
BRHT4b          & 1.00;0.90 & 140 & 0.083 & -0.40$\pm$0.25 (0.11) & 1 & 7.86$\pm$0.10 & - \\
                &  &  &  & -0.51$\pm$0.08 (0.12) & 2 & 8.00$\pm$0.11 & - \\
                &  &  &  & -0.63$\pm$0.08 (0.12) & 3 & 8.03$\pm$0.11 & - \\
HS111$^{(*,**)}$   & 0.60;0.50 & 0 & 0.122 & -0.61$\pm$0.05 (0.13) & 3 & 9.05$\pm$0.10 &
1.00$\pm$0.50 \\
BSDL599$^{(*,**)}$ & 1.80;1.50 & 80 & 0.091 & -0.05$\pm$0.23 (0.11) & 1 & 7.86$\pm$0.08 & - \\
                &  &  &  & -0.12$\pm$0.23 (0.11) & 1 & 7.86$\pm$0.11 & - \\
                % &  &  &  & -0.47$\pm$0.24 (0.11) & 1 & 7.86$\pm$0.04 & - \\
BSDL603         & 1.10;0.95 & 170 & 0.091 & - & - & 7.92$\pm$0.15 & - \\
                % &  &  &  & - & - & 7.92$\pm$0.15 & - \\ % field 2
                % &  &  &  & - & - & 7.92$\pm$0.15 & - \\ % field 3
NGC1839         & 1.60;1.60 & 0 & 0.076 & -0.10$\pm$0.10 (0.12) & 2 & 7.95$\pm$0.12 & - \\
                &  &  &  & -0.17$\pm$0.11 (0.12) & 2 & 7.96$\pm$0.13 & - \\
                &  &  &  & -0.17$\pm$0.08 (0.12) & 2 & 7.96$\pm$0.13 & - \\
NGC1838         & 1.30;1.20 & 20 & 0.104 & -0.38$\pm$0.24 (0.13) & 1 & 8.05$\pm$0.10 & - \\
BSDL623         & 0.90;0.80 & 120 & 0.091 & - & - & 8.00$\pm$0.10 & - \\
OGLE-CL~LMC~111$^{(*)}$ & 1.20;1.10 & 60 & 0.098 & -1.18$\pm$0.42 (0.15) & 1 & 8.04$\pm$0.08 & - \\
BSDL646$^{(**)}$ & 1.50;0.80 & 160 & 0.102 & - & - & 7.79$\pm$0.05 & - \\
OGLE-CL~LMC~113$^{(*)}$ & 1.10;1.00 & 70 & 0.080 & -0.26$\pm$0.28 (0.14) & 1 & 7.93$\pm$0.10 & - \\
NGC1847         & 1.80;1.60 & 0 & 0.113 & -0.99$\pm$0.17 (0.11) & 6 & 8.22$\pm$0.20 & - \\
NGC1848$^{(**)}$ & 2.20;2.00 & 140 & 0.103 & - & - & 6.70$\pm$0.10 & - \\
BSDL664$^{(**)}$ & 1.10;1.00 & 110 & 0.119 & - & - & 8.20$\pm$0.20 & - \\
NGC1844         & 1.60;1.60 & 0 & 0.087 & -0.07$\pm$0.14 (0.14) & 4 & 7.92$\pm$0.10 & - \\
NGC1846         & 3.80;3.80 & 0 & 0.070 & -0.47$\pm$0.03 (0.11) & 188 & 9.21$\pm$0.08 &
1.50$\pm$0.30 \\
KMHK565         & 1.00;0.85 & 140 & 0.089 & - & - & 6.70$\pm$0.10 & - \\
SL244           & 1.00;1.00 & 0 & 0.106 & -0.11$\pm$0.05 (0.15) & 18 & 8.75$\pm$0.15 & - \\
H88~152$^{(*,**)}$ & 1.00;0.85 & 140 & 0.121 & 0.20$\pm$0.03 (0.14) & 2 & 8.01$\pm$0.04 & - \\
SL256$^{(*)}$   & 1.00;0.95 & 50 & 0.099 & -0.21$\pm$0.22 (0.12) & 1 & 7.65$\pm$0.15 & - \\
NGC1850A        & 0.50;0.45 & 0 & 0.110 & - & - & 6.40$\pm$0.20 & - \\
NGC1850$^{(1)}$ & 3.00;3.00 & 0 & 0.111 & -0.31$\pm$0.05 (0.11) & 16 & 7.90$\pm$0.15 & - \\
                &  &  &  & -0.30$\pm$0.05 (0.12) & 17 & 7.90$\pm$0.20 & - \\
BRHT5b$^{(*)}$  & 1.10;1.00 & 40 & 0.090$^{(a)}$ & -0.09$\pm$0.06 (0.12) & 7 & 7.86$\pm$0.07 & - \\
                &  &  &  & -0.03$\pm$0.06 (0.12) & 8 & 7.86$\pm$0.07 & - \\
H88~165$^{(*)}$ & 1.10;1.10 & 0 & 0.127 & -0.27$\pm$0.23 (0.13) & 1 & 7.80$\pm$0.10 & - \\
NGC1854         & 2.30;2.30 & 0 & 0.103 & -0.36$\pm$0.02 (0.12) & 6 & 7.92$\pm$0.08 & - \\
BSDL745         & 1.10;0.80 & 100 & 0.104 & - & - & 7.35$\pm$0.10 & - \\
BSDL748$^{(*)}$ & 0.80;0.70 & 20 & 0.094 & -0.03$\pm$0.23 (0.12) & 1 & 7.55$\pm$0.10 & - \\
NGC1858$^{(2)}$ & 4.40;2.60 & 170 & 0.085 & - & - & 6.96$\pm$0.31 & - \\
H88~177$^{(*)}$ & 1.20;1.00 & 120 & 0.081 & -0.29$\pm$0.08 (0.15) & 3 & 8.18$\pm$0.10 & - \\
BRHT48b         & 0.80;0.65 & 170 & 0.079 & - & - & 8.40$\pm$0.10 & - \\
H88~180$^{(*,**)}$ & 0.90;0.75 & 100 & 0.099 & 0.05$\pm$0.28 (0.16) & 1 & 7.59$\pm$0.05 & - \\
BRHT48a$^{(*)}$ & 0.60;0.65 & 130 & 0.079 & -0.65$\pm$0.12 (0.16) & 6 & 8.55$\pm$0.10 & - \\
OGLE-CL~LMC~185 & 0.85;0.85 & 0 & 0.115 & - & - & 7.35$\pm$0.10 & - \\
NGC1863         & 1.40;1.20 & 50 & 0.092 & -0.28$\pm$0.17 (0.12) & 4 & 7.72$\pm$0.10 & - \\
NGC1866         & 5.50;5.50 & 0 & 0.046 & -0.27$\pm$0.03 (0.11) & 41 & 8.22$\pm$0.10 & - \\
BRHT8b$^{(*)}$  & 1.00;1.00 & 0 & 0.068 & -0.13$\pm$0.25 (0.12) & 1 & 7.90$\pm$0.07 & - \\
OGLE-CL~LMC~273$^{(*,**)}$ & 0.75;0.75$^{(P)}$ & 0 & 0.087 & 0.10$\pm$0.11 (0.13) & 5 & 8.07$\pm$0.05 & - \\
NGC1894         & 1.40;1.20 & 60 & 0.065 & -0.41$\pm$0.17 (0.12) & 4 & 8.00$\pm$0.10 & - \\
H88~236$^{(*,**)}$ & 0.80;0.65 & 10 & 0.073 & -0.19$\pm$0.24 (0.13) & 1 & 7.95$\pm$0.10 & - \\
% BSDL1096        & 0.55;0.50 & 120 & 0.065 & - & - & 8.13$\pm$0.03 & - \\
NGC1898         & 1.60;1.60 & 0 & 0.068 & -1.06$\pm$0.05 (0.12) & 47 & 9.98$\pm$0.08 &
10.5$\pm$2.00 \\
NGC1903$^{(*)}$ & 1.90;1.90 & 0 & 0.080$^{(a)}$ & -0.37$\pm$0.06 (0.13) & 9 & 7.98$\pm$0.10 & - \\
BRHT9b$^{(*)}$  & 1.40;1.20 & 80 & 0.121 & -0.21$\pm$0.07 (0.14) & 40 & 8.85$\pm$0.08 & - \\
H88~255$^{(*)}$ & 0.95;0.85 & 120 & 0.123 & -0.21$\pm$0.18 (0.12) & 2 & 7.92$\pm$0.10 & - \\
OGLE-CL~LMC~318 & 1.30;1.30 & 0 & 0.069 & -0.33$\pm$0.06 (0.13) & 29 & 8.90$\pm$0.10 & - \\
OGLE-CL~LMC~321$^{(*)}$ & 0.65;0.60 & 140 & 0.066 & -0.43$\pm$0.25 (0.12) & 1 & 8.27$\pm$0.13 & - \\
% BSDL1178        & 0.65;0.55 & 140 & 0.075 & - & - & 8.05$\pm$0.10 & - \\
% OGLE-CL~LMC~333 & 0.70;0.55 & 140 & 0.095 & - & - & 7.65$\pm$0.12 & - \\
ESO85-72        & 1.70;1.70 & 0 & 0.040 & -0.58$\pm$0.06 (0.11) & 8 & 9.42$\pm$0.12 &
3.00$\pm$1.25 \\
BSDL1291$^{(**)}$ & 0.80;0.65 & 120 & 0.075 & - & - & 7.30$\pm$0.20 & - \\
OGLE-CL~LMC~369$^{(*)}$ & 1.00;0.90 & 70 & 0.073 & -0.22$\pm$0.11 (0.15) & 6 & 8.35$\pm$0.10 & - \\
H88~283$^{(*,**)}$ & 0.80;0.70 & 120 & 0.072 & -0.44$\pm$0.46 (0.16) & 1 & 8.17$\pm$0.07 & - \\
                &  &  &  & -0.52$\pm$0.44 (0.15) & 1 & 8.19$\pm$0.07 & - \\
NGC1926         & 1.40;1.20 & 120 & 0.068 & - & - & 8.00$\pm$0.10 & - \\
NGC1935$^{(**)}$ & 1.20;1.20 & 0 & 0.090 & - & - & 7.20$\pm$0.30 & - \\
OGLE-CL~LMC~404 & 1.00;0.90 & 110 & 0.075 & - & - & 8.35$\pm$0.10 & - \\
% H88~293         & 0.80;0.70 & 110 & 0.077 & - & - & 7.80$\pm$0.10 & - \\
LH47            & 7.30;5.50 & 160 & 0.105 & - & - & 6.97$\pm$0.10 & - \\
OGLE-CL~LMC~407$^{(*)}$ & 1.20;1.10 & 80 & 0.057 & -0.15$\pm$0.02 (0.12) & 4 & 8.09$\pm$0.08 & - \\
BSDL1411$^{(*,**)}$ & 1.00;0.90 & 100 & 0.059 & -0.08$\pm$0.23 (0.14) & 2 & 7.84$\pm$0.06 & - \\
NGC1937         & 3.20;2.00 & 70 & 0.131 & - & - & 6.22$\pm$0.10 & - \\
OGLE-CL~LMC~431$^{(*)}$ & 0.70;0.65 & 130 & 0.067 & -0.43$\pm$0.28 (0.13) & 2 & 8.03$\pm$0.10 & - \\
OGLE-CL~LMC~438$^{(*,**)}$ & 1.40;1.40 & 0 & 0.068 & -0.24$\pm$0.06 (0.15) & 5 & 8.55$\pm$0.10 & - \\
BSDL1576$^{(*,**)}$ & 0.95;0.95 & 0 & 0.086 & -0.37$\pm$0.25 (0.13) & 1 & 8.06$\pm$0.07 & - \\
BSDL1592$^{(*)}$ & 0.80;0.65 & 110 & 0.095 & -0.22$\pm$0.24 (0.15) & 3 & 7.97$\pm$0.07 & - \\
BSDL1588$^{(**)}$  & 0.95;0.70 & 140 & 0.086 & - & - & 8.30$\pm$0.05 & - \\
OGLE-CL~LMC~446$^{(*)}$ & 1.30;1.20 & 100 & 0.083 & -0.13$\pm$0.07 (0.15) & 5 & 8.20$\pm$0.15 & - \\
BSDL1597$^{(*,**)}$ & 1.10;0.85 & 100 & 0.115 & -0.19$\pm$0.24 (0.12) & 1 & 7.93$\pm$0.05 & - \\
NGC1950$^{(*)}$ & 1.70;1.70 & 0 & 0.085 & 0.07$\pm$0.18 (0.12) & 3 & 8.00$\pm$0.10 & - \\
BSDL1601        & 1.00;0.90 & 70 & 0.088 & - & - & 7.30$\pm$0.20 & - \\
SL457$^{(**)}$  & 1.20;1.10 & 70 & 0.090 & - & - & 6.70$\pm$0.10 & - \\
NGC1948$^{(3)}$ & 7.00;5.70 & 30 & 0.103 & -0.11$\pm$0.18 (0.12) & 3 & 7.50$\pm$0.25 & - \\
NGC1955         & 4.00;3.60 & 20 & 0.082 & - & - & 6.25$\pm$0.10 & - \\
BSDL1674        & 0.85;0.65 & 30 & 0.082 & - & - & 6.15$\pm$0.15 & - \\
LH53            & 7.00;5.70 & 30 & 0.099 & - & - & 7.25$\pm$0.15 & - \\
KMK88~56$^{(*)}$ & 0.80;0.70 & 150 & 0.085 & -0.01$\pm$0.12 (0.14) & 7 & 8.43$\pm$0.05 & - \\
NGC1969$^{(*)}$ & 1.20;1.20 & 0 & 0.075 & 0.06$\pm$0.08 (0.13) & 10 & 8.10$\pm$0.05 & - \\
OGLE-CL~LMC~478$^{(*)}$ & 0.50;0.50$^{(P)}$ & 0 & 0.080 & 0.10$\pm$0.08 (0.12) & 4 & 7.93$\pm$0.05 & - \\
BSDL1759$^{(**)}$ & 0.45;0.45 & 0 & 0.081 & - & - & 8.06$\pm$0.05 & - \\
NGC1971$^{(*)}$ & 1.10;0.95 & 0 & 0.073 & -0.24$\pm$0.16 (0.13) & 2 & 8.02$\pm$0.10 & - \\
                % &  &  &  & -0.54$\pm$0.29 (0.12) & 1 & 8.07$\pm$0.10 & - \\
NGC1972         & 0.90;0.80 & 100 & 0.071 & -0.17$\pm$0.04 (0.12) & 3 & 7.87$\pm$0.10 & - \\
                &  &  &  & -0.27$\pm$0.08 (0.12) & 3 & 7.88$\pm$0.10 & - \\
KMK88~57$^{(*,**)}$ & 0.60;0.55 & 60 & 0.073 & -0.48$\pm$0.03 (0.14) & 3 & 8.69$\pm$0.05 & - \\
BSDL1783$^{(*,**)}$ & 0.65;0.60 & 110 & 0.065 & -0.17$\pm$0.29 (0.16) & 1 & 8.20$\pm$0.10 & - \\
BSDL1785$^{(*,**)}$ & 0.85;0.70 & 110 & 0.068 & -0.13$\pm$0.01 (0.16) & 3 & 8.14$\pm$0.06 & - \\
                &  &  &  & -0.33$\pm$0.22 (0.13) & 2 & 8.13$\pm$0.10 & - \\
BSDL1807$^{(*,**)}$ & 1.00;0.70 & 30 & 0.061 & 0.01$\pm$0.02 (0.14) & 2 & 8.09$\pm$0.09 & - \\
BSDL1821$^{(*)}$ & 0.70;0.70 & 120 & 0.062 & -0.31$\pm$0.29 (0.15) & 1 & 7.92$\pm$0.05 & - \\
NGC1986         & 2.80;2.40 & 140 & 0.088 & 0.10$\pm$0.03 (0.12) & 13 & 7.98$\pm$0.12 & - \\
BSDL1858$^{(**)}$ & 1.00;0.80 & 110 & 0.088 & - & - & 8.06$\pm$0.05 & - \\
OGLE-CL~LMC~500$^{(*)}$ & 0.95;0.95 & 0 & 0.084 & 0.18$\pm$0.24 (0.13) & 1 & 7.91$\pm$0.05 & - \\
% BSDL1906        & 0.55;0.55 & 0 & 0.076 & - & - & 8.10$\pm$0.07 & - \\
OGLE-CL~LMC~512$^{(*)}$ & 0.85;0.85 & 0 & 0.076 & -0.46$\pm$0.09 (0.12) & 2 & 8.18$\pm$0.10 & - \\
                &  &  &  & -0.33$\pm$0.06 (0.13) & 2 & 8.10$\pm$0.05 & - \\
NGC1978         & 4.00;2.70 & 160 & 0.056 & -0.35$\pm$0.02 (0.12) & 287 & 9.20$\pm$0.10 &
1.50$\pm$0.30 \\
KMHK960         & 1.20;1.10 & 160 & 0.103 & -0.35$\pm$0.05 (0.13) & 16 & 9.12$\pm$0.10 &
1.25$\pm$0.25 \\
BSDL1928$^{(*,**)}$ & 1.30;1.00 & 130 & 0.082 & -0.42$\pm$0.07 (0.12) & 3 & 8.27$\pm$0.20 & - \\
OGLE-CL~LMC~525$^{(*,**)}$ & 1.40;1.10 & 70 & 0.095 & -0.27$\pm$0.15 (0.12) & 3 & 8.06$\pm$0.15 & - \\
                &  &  &  & -0.15$\pm$0.24 (0.14) & 3 & 8.03$\pm$0.08 & - \\
ESO85-91        & 1.90;1.80 & 150 & 0.047 & -0.41$\pm$0.06 (0.13) & 20 & 9.02$\pm$0.09 &
1.00$\pm$0.25 \\
NGC2005         & 1.60;1.60 & 0 & 0.082 & -1.41$\pm$0.04 (0.12) & 30 & 10.10$\pm$0.04 &
13.5$\pm$2.00 \\
OGLE-CL~LMC~540$^{(*,**)}$ & 1.50;1.30 & 20 & 0.103 & 0.10$\pm$0.23 (0.11) & 1 & 7.66$\pm$0.08 & - \\
NGC2016$^{(*,**)}$ & 1.80;1.80 & 0 & 0.108 & -0.15$\pm$0.07 (0.12) & 5 & 8.19$\pm$0.12 & - \\
KMHK1046        & 1.40;1.40 & 0 & 0.088 & -0.57$\pm$0.04 (0.13) & 22 & 9.28$\pm$0.10 &
1.75$\pm$0.50 \\
BSDL2205$^{(*,**)}$ & 1.10;1.00 & 10 & 0.102 & -0.22$\pm$0.14 (0.15) & 4 & 8.25$\pm$0.10 & - \\
BSDL2212$^{(**)}$ & 1.10;0.90 & 80 & 0.105 & - & - & 7.84$\pm$0.15 & - \\
NGC2019         & 1.50;1.50 & 0 & 0.079 & -1.21$\pm$0.04 (0.12) & 57 & 10.10$\pm$0.05 &
13.0$\pm$2.00 \\
KMHK1013        & 1.40;1.40 & 0 & 0.056 & -0.67$\pm$0.05 (0.13) & 14 & 9.40$\pm$0.07 &
2.50$\pm$0.50 \\
LH72            & 8.00;4.00 & 160 & 0.052 & - & - & 7.15$\pm$0.10 & - \\
OGLE-CL~LMC~585$^{(*,**)}$ & 1.20;1.20 & 0 & 0.114 & -0.16$\pm$0.09 (0.12) & 3 & 8.10$\pm$0.10 & - \\
LH77            & 35.0;12.0 & 70 & 0.042 & - & - & 7.20$\pm$0.10 & - \\
OGLE-CL~LMC~591$^{(**)}$ & 1.00;0.80 & 100 & 0.113 & - & - & 8.29$\pm$0.10 & - \\
NGC2028$^{(*,**)}$ & 1.10;1.00 & 60 & 0.101 & -0.14$\pm$0.13 (0.15) & 7 & 8.30$\pm$0.10 & - \\
BSDL2624$^{(*,**)}$ & 1.30;1.00 & 20 & 0.126 & -0.35$\pm$0.06 (0.12) & 2 & 7.97$\pm$0.05 & - \\
NGC2065         & 2.30;2.30 & 0 & 0.135 & -0.33$\pm$0.03 (0.12) & 9 & 8.17$\pm$0.10 & - \\
NGC2111$^{(*)}$ & 0.45;0.45 & 0 & 0.152 & -0.20$\pm$0.20 (0.15) & 3 & 8.16$\pm$0.05 & - \\
KMHK1489        & 0.95;0.85 & 150 & 0.112 & - & - & 7.75$\pm$0.10 & - \\
NGC2136$^{(4)}$ & 2.80;2.50 & 140 & 0.121 & -0.69$\pm$0.10 (0.12) & 4 & 8.16$\pm$0.12 & - \\
NGC2155         & 2.40;2.40 & 0 & 0.040$^{(a)}$ & -0.53$\pm$0.04 (0.12) & 53 & 9.41$\pm$0.07 &
2.50$\pm$0.50 \\
ESO121-3$^{(5)}$ & 2.10;2.10 & 0 & 0.030$^{(a)}$ & -0.79$\pm$0.07 (0.13) & 18 & 9.93$\pm$0.11 &
9.00$\pm$2.00 \\
ESO86-61        & 1.70;1.70 & 0 & 0.050 & -0.64$\pm$0.07 (0.13) & 18 & 9.38$\pm$0.06 &
2.20$\pm$0.30 \\
KMHK1679$^{(*,**)}$ & 0.60;0.55 & 30 & 0.069 & -0.52$\pm$0.17 (0.14) & 2 & 8.65$\pm$0.10 & - \\
NGC2210         & 3.30;3.30 & 0 & 0.067 & -1.53$\pm$0.03 (0.12) & 62 & 10.22$\pm$0.04 &
14.0$\pm$1.00 \\
ESO57-75        & 1.70;1.70 & 0 & 0.119 & -0.48$\pm$0.06 (0.13) & 13 & 9.10$\pm$0.10 &
1.20$\pm$0.30 \\
NGC2249         & 2.30;2.30 & 0 & 0.074 & -0.27$\pm$0.28 (0.12) & 1 & 8.80$\pm$0.12 & - \\
NGC2257         & 4.00;4.00 & 0 & 0.040$^{(a)}$ & -1.67$\pm$0.02 (0.11) & 72 & 10.19$\pm$0.06 &
15.0$\pm$1.50 \\
\hline
\end{longtable}
\tablefoot{
  Cluster: name of the cluster;
  D$_{maj}$;D$_{min}$:  major and minor axes of the ellipse encapsulating the cluster
  from the catalog of \citet{Bica1999};
  P$_A$: position angle;
  $E(B-V)_C$:  reddening adopted for a~given star cluster;
  $\mathrm{[Fe/H]}_C$: mean cluster metallicity calculated in this work (systematic
  errors are given in the parentheses);
  N$_C$: number of stars used for metallicity calculation;
  log(Age$_P$): logarithm of ages derived from the Padova isochrones;
  Age$_D$: ages derived from the Dartmouth isochrones.
  \begin{itemize}\scriptsize
  \itemsep0em
  \item [$^{(a)}$] Adopted based on CMD.
  \item [$^{(S)}$] Reddening from \citet{Skowron2021} only.
  \item [$^{(*)}$] Metallicities of the clusters provided for the first time.
  \item [$^{(**)}$] Ages of the clusters provided for the first time.
  \item [$^{(P)}$] Radius of the cluster from \cite{Pietrzynski1999b}.
  \item [$^{(1)}$] There may be two stellar populations with similar age but very
  distinct metallicity. One is metal-rich with $\mathrm{[Fe/H]} \sim 0.16$~dex.
  Further studies are required to confirm this finding.
  \item [$^{(2)}$] There may be two young stellar populations with ages of
  $\mathrm{log(age)} \sim 6.65$ and $\sim 7.27$. The average age is given.
  \item [$^{(3)}$] There may be two stellar populations:
  a~younger one, having age around $\mathrm{log(age)} \sim 7.25$ and an older one
  with $\sim 7.75$. The average age is given.
  \item [$^{(4)}$] There is another stellar population visible with
  $\mathrm{[Fe/H]}=0.08 \pm 0.02$ ($0.13$)~dex and $\mathrm{log(age)} = 7.86 \pm 0.12$,
  but most probably these are field objects.
  \item [$^{(5)}$] Possible two stellar populations. The given metallicity
  represents an average value.
  \end{itemize}}
}
%
%-------------------------------------------------------------

%-------------------------------------------------------------
%                                             Two column Table
%-------------------------------------------------------------
%
\longtab[2]{
\begin{longtable}{lcclccc}
\caption{\label{tab:reddF} Fields surrounding star clusters in the LMC.} \\
\hline\hline
Nb & RA & DEC & $E(B-V)_F$ & $\mathrm{[Fe/H]}_F$ & N$_F$ & IQR \\
 & (h:mm:ss.ss) & (dd:mm:ss.ss) & (mag) & (dex) & & (dex) \\
\hline
\endfirsthead
\caption{continued.}\\
\hline\hline
Nb & RA & DEC & $E(B-V)_F$ & $\mathrm{[Fe/H]}_F$ & N$_F$ & IQR \\
 & (h:mm:ss.ss) & (dd:mm:ss.ss) & (mag) & (dex) & & (dex) \\
\hline
\endhead
\hline
%\endfoot
%%
1   & 4:37:31.00 & -70:35:02.01 & 0.148 & -0.39$\pm$0.04 (0.12)  & 65 & 0.48 \\ % field_NGC1651
2   & 4:37:49.59 & -69:01:45.60 & 0.110 & -0.31$\pm$0.03 (0.12) & 79 & 0.41 \\ % field_KMHK21, SL8
3   & 4:45:22.75 & -83:59:48.00 & 0.160$^{(a)}$ & -1.43$\pm$0.05 (0.11) & 10 & 0.17 \\ % field_NGC1841
4   & 4:50:36.70 & -69:59:06.00 & 0.124 & -0.42$\pm$0.03 (0.11) & 88 & 0.48 \\ % field_NGC1711
5   & 4:54:16.85 & -70:26:30.01 & 0.110 & -0.65$\pm$0.03 (0.12) & 157 & 0.54 \\ % field_NGC1754
6   & 4:57:22.68 & -62:32:05.00 & 0.046$^{(S)}$ & -0.10$\pm$0.08 (0.12) & 11 & 0.34 \\ % field_ESO85-21
7   & 4:59:07.55 & -67:44:43.01 & 0.087 & -0.36$\pm$0.03 (0.12) & 203 & 0.63 \\ % field_NGC1786
8   & 4:59:45.62 & -69:48:06.00 & 0.103 & -0.31$\pm$0.02 (0.12) & 268 & 0.47 \\ % field_NGC1795
9   & 5:00:26.45 & -68:46:22.00 & 0.086 & -0.20$\pm$0.02 (0.11) & 405 & 0.44 \\ % field_LMC5
10  & 5:01:04.01 & -69:05:03.30 & 0.106 & -0.31$\pm$0.02 (0.11) & 496 & 0.50 \\ % field_LMC8
11  & 5:03:05.78 & -69:02:14.91 & 0.090 & -0.41$\pm$0.02 (0.11) & 402 & 0.69 \\ % field_LMC35
12  & 5:04:38.89 & -69:20:26.01 & 0.078 & -0.29$\pm$0.02 (0.11) & 422 & 0.56 \\ % field_LMC61
13  & 5:04:56.79 & -70:01:08.41 & 0.118 & -0.36$\pm$0.02 (0.11) & 323 & 0.52 \\ % field_LMC65
14  & 5:05:06.71 & -69:24:14.30 & 0.088 & -0.42$\pm$0.02 (0.11) & 408 & 0.49 \\ % field_NGC1835
15  & 5:05:09.53 & -68:57:23.81 & 0.082 & -0.57$\pm$0.02 (0.11) & 766 & 0.58 \\ % field_LMC71
16  & 5:05:19.03 & -68:44:14.71 & 0.076 & -0.33$\pm$0.02 (0.11) & 446 & 0.59 \\ % field_LMC78
17  & 5:05:24.63 & -68:30:02.01 & 0.116 & -0.31$\pm$0.02 (0.11) & 360 & 0.51 \\ % field_LMC80
18  & 5:05:35.52 & -68:37:42.01 & 0.086 & -0.36$\pm$0.03 (0.11) & 262 & 0.55 \\ % field_NGC1836
19  & 5:05:39.65 & -68:38:12.01 & 0.083 & -0.38$\pm$0.03 (0.11) & 265 & 0.56 \\ % field_LMC83
20  & 5:05:55.33 & -68:37:43.01 & 0.076 & -0.31$\pm$0.02 (0.11) & 310 & 0.58 \\ % field_LMC90
21  & 5:06:08.56 & -68:26:44.99 & 0.102 & -0.18$\pm$0.03 (0.11) & 309 & 0.58 \\ % field_LMC97
22  & 5:06:47.11 & -68:36:59.39 & 0.097 & -0.11$\pm$0.02 (0.13) & 319 & 0.56 \\ % field_LMC111
23  & 5:06:54.37 & -68:43:07.99 & 0.079 & -0.12$\pm$0.02 (0.13) & 410 & 0.69 \\ % field_LMC113
24  & 5:07:08.18 & -68:58:22.99 & 0.111 & -0.18$\pm$0.02 (0.13) & 503 & 0.63 \\ % field_LMC118
25  & 5:07:30.11 & -67:19:26.30 & 0.091 & -0.15$\pm$0.03 (0.13) & 170 & 0.51 \\ % field_LMC126
26  & 5:07:34.90 & -67:27:38.91 & 0.074 & -0.65$\pm$0.03 (0.12) & 125 & 0.43 \\ % field_NGC1846
27  & 5:07:36.71 & -68:32:30.00 & 0.109 &  0.02$\pm$0.03 (0.13) & 330 & 0.71 \\ % field_SL244
28  & 5:07:44.71 & -71:11:00.00 & 0.102 & -0.18$\pm$0.03 (0.13) & 143 & 0.46 \\ % field_NGC1848
29  & 5:08:06.45 & -69:16:04.00 & 0.117 &  0.12$\pm$0.02 (0.13) & 473 & 0.66 \\ % field_H88-152
30  & 5:08:45.52 & -68:45:39.00 & 0.110 & -0.08$\pm$0.03 (0.13) & 323 & 0.67 \\ % field_NGC1850
31  & 5:08:54.32 & -68:45:13.89 & 0.120 & -0.25$\pm$0.03 (0.13) & 407 & 0.71 \\ % field_LMC145
32  & 5:09:24.89 & -68:51:47.49 & 0.109 & -0.20$\pm$0.03 (0.12) & 379 & 0.79 \\ % field_LMC155
33  & 5:09:55.75 & -68:54:06.20 & 0.086 & -0.35$\pm$0.03 (0.13) & 383 & 0.68 \\ % field_NGC1858_f1
34  & 5:10:38.90 & -69:02:30.99 & 0.111 & -0.12$\pm$0.02 (0.12) & 564 & 0.74 \\ % field_LMC185
35  & 5:11:39.72 & -68:43:36.00 & 0.088 & -0.15$\pm$0.03 (0.13) & 544 & 0.76 \\ % field_NGC1863
36  & 5:13:38.56 & -65:27:52.00 & 0.046$^{(S)}$ & -0.60$\pm$0.03 (0.12) & 14 & 0.18 \\ % field_NGC1866
37  & 5:15:36.83 & -69:28:24.50 & 0.068 & -0.16$\pm$0.02 (0.12) & 795 & 0.69 \\ % field_LMC270
38  & 5:15:40.09 & -69:16:51.00 & 0.088 & -0.16$\pm$0.02 (0.12) & 654 & 0.78 \\ % field_LMC273
39  & 5:16:41.08 & -69:39:24.40 & 0.066 & -0.33$\pm$0.01 (0.12) & 1112 & 0.59 \\ % field_NGC1898
40  & 5:17:07.85 & -69:21:35.50 & 0.120 & -0.14$\pm$0.02 (0.12) & 559 & 0.77 \\ % field_15.1
41  & 5:18:02.20 & -69:43:35.90 & 0.083 & -0.13$\pm$0.02 (0.12) & 780 & 0.66 \\ % field_28.1
42  & 5:18:17.73 & -69:36:57.21 & 0.094 & -0.27$\pm$0.02 (0.12) & 633 & 0.79 \\ % field_40.1
43  & 5:20:05.27 & -63:28:50.01 & 0.040$^{(S)}$ & -0.53$\pm$0.06 (0.11) & 31 & 0.40 \\ % field_SL388
44  & 5:20:23.35 & -69:35:03.00 & 0.076 &  0.07$\pm$0.02 (0.13) & 771 & 0.76 \\ % field_LMC369
45  & 5:20:30.38 & -69:32:09.00 & 0.071 & -0.05$\pm$0.02 (0.13) & 790 & 0.78 \\ % field_LMC375
46  & 5:21:57.61 & -67:57:18.00 & 0.101 & -0.41$\pm$0.09 (0.15) & 26  & 0.55 \\ % field_LH47
47  & 5:22:03.24 & -69:15:18.00 & 0.075 &  0.09$\pm$0.03 (0.13) & 467 & 0.78 \\ % field_LMC404
48  & 5:22:14.21 & -69:30:41.01 & 0.056 & -0.20$\pm$0.02 (0.12) & 775 & 0.67 \\ % field_LMC407
49  & 5:22:28.74 & -67:53:42.01 & 0.133 & -0.27$\pm$0.05 (0.13) & 126 & 0.88 \\ % field_LH48
50  & 5:24:19.71 & -66:24:12.01 & 0.139$^{(S)}$ & -0.17$\pm$0.08 (0.12) & 3 & 0.13 \\ % field_LH52
51  & 5:24:32.75 & -69:54:04.31 & 0.084 & -0.10$\pm$0.02 (0.13) & 724 & 0.76 \\ % field_LMC436
52  & 5:24:33.34 & -69:44:43.11 & 0.070 & -0.08$\pm$0.02 (0.13) & 713 & 0.79 \\ % field_LMC438
53  & 5:25:00.89 & -69:26:03.11 & 0.085 &  0.08$\pm$0.03 (0.13) & 468 & 0.86 \\ % field_LMC446
54  & 5:26:05.99 & -66:14:00.00 & 0.103$^{(S)}$ & -0.74$\pm$0.06 (0.14) & 77 & 0.75 \\ % field_LH53
55  & 5:26:11.57 & -67:29:54.00 & 0.082$^{(S)}$ & -0.30$\pm$0.04 (0.13) & 117 & 0.57 \\ % field_NGC1955
56  & 5:26:48.44 & -69:50:17.01 & 0.077 & -0.05$\pm$0.02 (0.13) & 991 & 0.69 \\ % field_LMC481
57  & 5:27:03.83 & -69:51:51.01 & 0.066 & -0.15$\pm$0.02 (0.12) & 880 & 0.63 \\ % field_LMC487
58  & 5:27:37.54 & -69:58:14.01 & 0.090 & -0.18$\pm$0.02 (0.13) & 873 & 0.57 \\ % field_LMC496
59  & 5:28:44.72 & -66:14:14.01 & 0.056$^{(S)}$ & -0.45$\pm$0.03 (0.12) & 134 & 0.41 \\ % field_NGC1978
60  & 5:28:49.69 & -71:38:00.01 & 0.104 & -0.30$\pm$0.03 (0.12) & 264 & 0.51 \\ % field_SL505
61  & 5:28:50.50 & -69:51:44.02 & 0.084 & -0.24$\pm$0.02 (0.12) & 843 & 0.77 \\ % fiel/d_48.1
62  & 5:29:05.86 & -69:48:30.01 & 0.083 & -0.05$\pm$0.02 (0.12) & 729 & 0.65 \\ % field_LMC516
63  & 5:29:48.82 & -63:38:58.66 & 0.040$^{(S)}$ & -0.47$\pm$0.05 (0.12) & 57 & 0.58 \\ % field_SL509
64  & 5:30:10.13 & -69:45:09.60 & 0.086 & -0.43$\pm$0.02 (0.11) & 578 & 0.55 \\ % field_NGC2005
65  & 5:31:34.82 & -69:56:43.41 & 0.105 & -0.14$\pm$0.02 (0.12) & 733 & 0.68 \\ % field_LMC559
66  & 5:31:41.67 & -72:08:48.01 & 0.088$^{(S)}$ & -0.42$\pm$0.02 (0.13) & 208 & 0.46 \\ % field_SL555
67  & 5:31:56.41 & -70:09:32.50 & 0.081 & -0.43$\pm$0.02 (0.12) & 836 & 0.62 \\ % field_NGC2019
68  & 5:32:02.67 & -64:14:30.01 & 0.056$^{(S)}$ & -0.63$\pm$0.04 (0.12) & 59 & 0.43 \\ % field_SL549
69  & 5:32:11.74 & -66:27:00.01 & 0.052$^{(S)}$ & -0.71$\pm$0.06 (0.13) & 60 & 0.68 \\ % field_LH72
70$^{(c)}$  & -- & -- & -- & -- & -- & -- \\
71  & 5:33:21.62 & -69:57:21.01 & 0.108 & -0.18$\pm$0.02 (0.12) & 576 & 0.72 \\ % field_LMC585
72  & 5:37:37.50 & -70:13:56.01 & 0.142 & -0.25$\pm$0.03 (0.12) & 392 & 0.78 \\ % field_LMC648
73  & 5:44:32.66 & -70:59:35.31 & 0.151 & -0.20$\pm$0.04 (0.13) & 260 & 0.90 \\ % field_LMC715
74  & 5:53:16.88 & -69:32:00.01 & 0.103 & -0.35$\pm$0.05 (0.13) & 107 & 0.69 \\ % field_NGC2136
75  & 5:58:32.05 & -65:28:38.00 & 0.049$^{(S)}$ & -0.44$\pm$0.05 (0.12) & 54 & 0.50 \\ % field_NGC2155
76  & 6:02:01.52 & -60:31:20.01 & 0.030$^{(a)}$ & -0.52$\pm$0.28 (0.12) & 17 & 1.14 \\ % field_ESO121-3
77  & 6:08:15.62 & -62:59:15.01 & 0.050$^{(a)}$ & -0.53$\pm$0.10 (0.12) & 18 & 0.68 \\ % field_SL842
78  & 6:11:31.35 & -69:07:17.00 & 0.069 & -0.41$\pm$0.04 (0.12) & 39 & 0.33 \\ % field_NGC2210
79  & 6:13:26.81 & -70:41:45.16 & 0.123 & -0.31$\pm$0.04 (0.13) & 44 & 0.39 \\ % field_SL862
80  & 6:25:48.64 & -68:55:12.00 & 0.074$^{(S)}$ & -0.20$\pm$0.05 (0.12) & 41 & 0.43 \\ % field_NGC2249
81  & 6:30:11.53 & -64:19:25.99 & 0.040$^{(a)}$ & -0.09$\pm$0.32 (0.12) & 10 & 1.64 \\ % field_NGC2257
\hline
\end{longtable}
\tablefoot{
  Nb: ID number, which corresponds to the field number in column~5 from
  Table~\ref{tab:phot};
  RA, DEC: equatorial coordinates of the center of the field for epoch J2000;
  $E(B-V)_F$: reddening adopted for the field stars;
  $\mathrm{[Fe/H]}_F$: mean metallicity of the field stars (systematic errors are
  given in the parentheses);
  N$_F$: number of stars used for the mean metallicity calculation;
  IQR: interquartile range of the field distribution.
  \begin{itemize}\scriptsize
  \itemsep0em
  \item [$^{(S)}$] Calculated based only on \citet{Skowron2021}.
  \item [$^{(a)}$] Adopted from the literature.
  \item [$^{(c)}$] No field stars (as defined in Sec.~\ref{ssec:sel}) in the field
   of cluster LH77.
\end{itemize}}
}
%
%-------------------------------------------------------------

\section{Observing log}

Table~\ref{tab:lmc} presents the observing log for the fields analyzed in this
work.

%-------------------------------------------------------------
%                      A rotated Two column Table in landscape
%-------------------------------------------------------------
% \setcounter{table}{0} \renewcommand{\thetable}{A.\arabic{table}}
\longtab[1]{
\begin{landscape}
\centering
\begin{longtable}{llllrlll}
\caption{\label{tab:lmc} Observing log.} \\
\hline\hline
Cluster & RA & DEC & Date & T$_{exp}$ ($y$,$b$,$v$) & Airmass ($y$,$b$,$v$)
& Seeing ($y$,$b$,$v$) & Other name  \\
& (hh:mm:ss.s) & (dd:mm:ss.s) & & (s) &  & (arcsec) &\\
\hline
\endfirsthead
\caption{continued.}\\
\hline\hline
Cluster & RA & DEC & Date & T$_{exp}$ ($y$,$b$,$v$) & Airmass ($y$,$b$,$v$)
& Seeing ($y$,$b$,$v$) & Other name  \\
& (hh:mm:ss.s) & (dd:mm:ss.s) & & (s) &  & (arcsec) &\\
\hline
\endhead
\hline

NGC1651    & 04:37:32 & -70:35:07 & 2008~Dec~18 & 120;200;500 & 1.53;1.53;1.54 & 1.03;0.91;0.98
& SL7, ESO55-30, KMHK20 \\
KMHK21      & 04:37:51 & -69:01:45 & 2009~Jan~16 & 100;180;350 & 1.34;1.34;1.33 & 0.83;0.88;0.80
& SL8 \\
NGC1841    & 04:45:23 & -83:59:49 & 2009~Jan~16 & 100;180;350 & 1.73;1.73;1.73 & 1.06;0.87;0.97
& ESO4-15 \\
NGC1711    & 04:50:37 & -69:59:06 & 2009~Jan~16 & 100;180;350 & 1.38;1.38;1.37 & 0.83;0.96;0.83
& SL554, KMHK145 \\
KMHK156     & 04:51:00 & -70:01:24 & 2009~Jan~16 & 100;180;350 & 1.38;1.38;1.37 & 0.83;0.96;0.83
& \\
NGC1754    & 04:54:17 & -70:26:29 & 2008~Dec~19 &  90;140;350 & 1.53;1.53;1.53 & 0.70;0.70;0.75
& SL91, ESO56-25, KMHK247 \\
ESO85-21   & 04:57:22 & -62:32:05 & 2009~Jan~16 & 100;180;400 & 1.27;1.28;1.28 & 0.68;0.61;0.71
& SL126, KMHK322 \\
NGC1786    & 04:59:06 & -67:44:42 & 2008~Dec~19 &  90;140;350 & 1.51;1.52;1.52 & 0.68;0.78;0.80
& SL149, ESO56-39, KMHK385 \\
NGC1795    & 04:59:46 & -69:48:04 & 2008~Dec~19 &  90;140;350 & 1.37;1.38;1.38 & 0.79;0.91;0.78
& SL165, ESO56-44, KMHK411 \\
KMHK421    & 05:00:26 & -68:46:23 & 2009~Jan~16 & 100;180;400 & 1.39;1.39;1.38 & 0.73,0.76,0.86
& OGLE-CL~LMC~5 \\
H88~87      & 05:00:43 & -69:07:14 & 2009~Jan~16 & 100;180;350 & 1.41;1.41;1.47 & 0.71;0.82;0.95
& \\
NGC1804    & 05:01:04 & -69:04:57 & 2009~Jan~16 & 100;180;350 & 1.41;1.41;1.47 & 0.71;0.82;0.95
& OGLE-CL~LMC~8, SL172 , ESO56-46 \\
SL191       & 05:03:05 & -69:02:12 & 2009~Jan~16 & 100;180;400 & 1.52;1.52;1.53 & 0.86;1.03;0.99
& OGLE-CL~LMC~35 \\
H88~104     & 05:04:20 & -69:21:27 & 2009~Jan~16 & 100;140;400 & 1.58;1.57;1.55 & 0.86;0.93;0.97
& OGLE-CL~LMC~538, KMK88~4 \\
H88~107     & 05:04:26 & -69:21:06 & 2009~Jan~16 & 100;140;400 & 1.58;1.57;1.55 & 0.86;0.93;0.97
& OGLE-CL~LMC~57 \\
BRHT3b      & 05:04:31 & -69:21:19 & 2009~Jan~16 & 100;140;400 & 1.58;1.57;1.55 & 0.86;0.93;0.97
& OGLE-CL~LMC~59, H88~108, KMK88~7  \\
NGC1830    & 05:04:39 & -69:20:37 & 2009~Jan~16 & 100;140;400 & 1.58;1.57;1.55 & 0.86;0.93;0.97
& OGLE-CL~LMC~61, SL207, ESO56-56 \\
SL211       & 05:04:49 & -68:55:23 & 2009~Jan~16 &  90;140;350 & 1.67;1.66;1.64 & 0.93;0.87;0.84
& \\
BSDL555     & 05:04:51 & -68:59:14 & 2009~Jan~16 &  90;140;350 & 1.67;1.66;1.64 & 0.93;0.87;0.84
& OGLE-CL~LMC~64 \\
KMHK521     & 05:04:56 & -70:01:09 & 2009~Jan~16 &  90;160;380 & 1.60;1.61;1.62 & 0.84;0.88;0.85
& OGLE-CL~LMC~65, SL65 \\
BSDL565     & 05:05:01 & -68:45:01 & 2009~Jan~16 &  90;140;350 & 1.68;1.69;1.70 & 0.83;0.87;0.99
& OGLE-CL~LMC~66 \\
NGC1835    & 05:05:05 & -69:24:14 & 2008~Dec~19 &  90;140;350 & 1.46;1.47;1.47 & 0.72;0.72;0.72
& OGLE-CL~LMC~69, SL215, ESO56-58 \\
H88~119     & 05:05:07 & -68:57:35 & 2009~Jan~16 &  90;140;350 & 1.67;1.66;1.64 & 0.93;0.87;0.84
& OGLE-CL~LMC~71 \\
H88~120     & 05:05:11 & -69:22:18 & 2008~Dec~19 &  90;140;350 & 1.46;1.47;1.47 & 0.72;0.72;0.72
& OGLE-CL~LMC~74, KMK88~10 \\
BSDL577     & 05:05:13 & -68:44:26 & 2009~Jan~16 &  90;140;350 & 1.68;1.69;1.70 & 0.83;0.87;0.99
& OGLE-CL~LMC~75 \\
BSDL581     & 05:05:17 & -68:43:12 & 2009~Jan~16 &  90;140;350 & 1.68;1.69;1.70 & 0.83;0.87;0.99
& OGLE-CL~LMC~77 \\
BSDL582    & 05:05:19 & -68:29:23 & 2009~Jan~16 &  90;140;350 & 1.77;1.76;1.73 & 0.76;0.85;0.89
&  \\
HS107       & 05:05:19 & -68:44:06 & 2009~Jan~16 &  90;140;350 & 1.68;1.69;1.70 & 0.83;0.87;0.99
& OGLE-CL~LMC~78 \\
SOI343      & 05:05:23 & -68:30:00 & 2009~Jan~16 &  90;140;350 & 1.77;1.76;1.73 & 0.76;0.85;0.89
& OGLE-CL~LMC~80, SL218 \\
BSDL591    & 05:05:32 & -68:39:09 & 2009~Jan~16 &  90;140;350 & 1.79;1.79;1.81 & 0.81;0.86;0.96
& \\
NGC1836    & 05:05:35 & -68:37:42 & 2009~Jan~16 &  90;140;350 & 1.79;1.79;1.81 & 0.81;0.86;0.96
& OGLE-CL~LMC~81, SL223 \\
HS109       & 05:05:35 & -68:42:52 & 2009~Jan~16 &  90;140;350 & 1.68;1.69;1.70 & 0.83;0.87;0.99
& OGLE-CL~LMC~82 \\
BRHT4b      & 05:05:40 & -68:38:22 & 2009~Jan~16 &  90;140;350 & 1.88;1.87;1.84 & 0.85;0.95;1.00
& OGLE-CL~LMC~83 \\
HS111       & 05:05:44 & -68:30:20 & 2009~Jan~16 &  90;140;350 & 1.77;1.76;1.73 & 0.76;0.85;0.89
& OGLE-CL~LMC~85 \\
BSDL599    & 05:05:46 & -68:35:35 & 2009~Jan~16 &  90;140;350 & 1.79;1.79;1.81 & 0.81;0.86;0.96
& \\
BSDL603     & 05:05:54 & -68:37:46 & 2009~Jan~16 &  90;140;350 & 1.90;1.91;1.93 & 0.82;0.89;0.86
& OGLE-CL~LMC~90 \\
NGC1839    & 05:06:02 & -68:37:36 & 2009~Jan~16 &  90;140;350 & 2.02;2.01;1.97 & 0.98;0.93;0.92
& OGLE-CL~LMC~93, SL226, ESO53-63 \\
NGC1838    & 05:06:07 & -68:26:42 & 2009~Jan~17 &  90;140;350 & 1.28;1.28;1.28 & 0.95;1.16;1.05
& OGLE-CL~LMC~97, SL225, ESO56-64 \\
BSDL623     & 05:06:22 & -68:35:34 & 2009~Jan~16 &  90;140;350 & 1.90;1.91;1.93 & 0.82;0.89;0.86
& \\
OGLE-CL~LMC~111 & 05:06:47 & -68:37:05 & 2009~Jan~17 &  90;140;380 & 1.28;1.28;1.28 & 0.98;1.09;1.16
& HS118 \\
BSDL646    & 05:06:53 & -68:34:52 & 2009~Jan~17 &  90;140;380 & 1.28;1.28;1.28 & 0.98;1.09;1.16
& \\
OGLE-CL~LMC~113 & 05:06:54 & -68:43:07 & 2009~Jan~17 &  90;200;400 & 1.29;1.29;1.29 & 0.96;1.13;1,13
& SL234 \\
NGC1847    & 05:07:08 & -68:58:17 & 2009~Jan~17 & 100;160;350 & 1.30;1.30;1.29 & 0.92;0.97;1.02
& OGLE-CL~LMC~118, SL240, ESO56-66 \\
NGC1848    & 05:07:27 & -71:11:43 & 2009~Jan~17 & 100;160;400 & 1.36;1.36;1.36 & 0.97;1.05;1.15
& SL247, ESO56-68, KMHK580 \\
BSDL664     & 05:07:28 & -68:58:32 & 2009~Jan~17 & 100;160;350 & 1.30;1.30;1.29 & 0.92;0.97;1.02
& OGLE-CL~LMC~124 \\
NGC1844    & 05:07:30 & -67:19:24 & 2009~Jan~17 & 100;140;400 & 1.27;1.27;1.27 & 0.89;0.96;1.03
& OGLE-CL~LMC~126, SL242, ESO85-48, KMHK556 \\
NGC1846    & 05:07:34 & -67:27:36 & 2008~Dec~18 & 120;200;500 & 1.59;1.60;1.61 & 0.94;1.00;1.01
& OGLE-CL~LMC~128, SL243, ESO56-67, KMHK557 \\
KMHK565     & 05:07:35 & -71:10:03 & 2009~Jan~17 & 100;160;400 & 1.36;1.36;1.36 & 0.97;1.05;1.15
&  \\
SL244       & 05:07:37 & -68:32:31 & 2009~Jan~17 & 100;160;400 & 1.30;1.30;1.30 & 1.04;1.18;1.17
&  \\
H88~152     & 05:08:06 & -69:15:52 & 2009~Jan~17 & 100;160;400 & 1.34;1.34;1.33 & 1.19;1.26;1.25
& OGLE-CL~LMC~136, KMHK24 \\
SL256      & 05:08:10 & -71:10:23 & 2009~Jan~17 & 100;160;400 & 1.36;1.36;1.36 & 0.97;1.05;1.15
&  \\
NGC1850A    & 05:08:39 & -68:45:32 & 2009~Jan~17 & 100;160;400 & 1.34;1.34;1.34 & 1.28;1.22;1.34
& \\
            &  &  & 2009~Jan~18 & 100;100;400 & 1.28;1.28;1.28 & 0.84;0.89;0.89 &  \\
NGC1850    & 05:08:44 & -68:45:33 & 2009~Jan~17 & 100;160;400 & 1.34;1.34;1.34 & 1.28;1.22;1.34
& OGLE-CL~LMC~142, SL261, ESO56-70 \\
            &  &  & 2009~Jan~18 & 100;100;400 & 1.28;1.28;1.28 & 0.84;0.89;0.89 &  \\
BRHT5b      & 05:08:53 & -68:45:08 & 2009~Jan~17 & 100;160;400 & 1.34;1.34;1.34 & 1.28;1.22;1.34
& OGLE-CL~LMC~145, H88~156 \\
            &  &  & 2009~Jan~18 & 100;100;400 & 1.28;1.28;1.28 & 0.84;0.89;0.89 &  \\
H88~165     & 05:09:16 & -68:44:01 & 2009~Jan~18 & 100;100;400 & 1.28;1.28;1.28 & 0.84;0.89;0.89
& OGLE-CL~LMC~152 \\
NGC1854    & 05:09:19 & -68:50:50 & 2009~Jan~18 & 120;200;450 & 1.29;1.29;1.29 & 0.99;0.91;0.97
& OGLE-CL~LMC~154, SL265, ESO56-72 \\
BSDL745     & 05:09:24 & -68:51:46 & 2009~Jan~18 & 120;200;450 & 1.29;1.29;1.29 & 0.99;0.91;0.97
& OGLE-CL~LMC~155 \\
BSDL748     & 05:09:27 & -68:51:02 & 2009~Jan~18 & 120;200;450 & 1.29;1.29;1.29 & 0.99;0.91;0.97
& OGLE-CL~LMC~156 \\
NGC1858    & 05:09:56 & -68:53:59 & 2009~Jan~18 & 120;200;450 & 1.29;1.30;1.30 & 0.73;0.84;1.11
& OGLE-CL~LMC~164, SL274 \\
           &  &  & 2009~Jan~18 & 120;200;350 & 1.30;1.30;1.30 & 0.80;0.93;1.12 & \\
 H88~177   & 05:10:17 & -68:55:41 & 2009~Jan~18 & 120;200;450 & 1.29;1.30;1.30
& 0.73;0.84;1.11 & OGLE-CL~LMC~177 \\
           &  &  & 2009~Jan~18 & 120;200;350 & 1.30;1.30;1.30 & 0.80;0.93;1.12 & \\
BRHT48b     & 05:10:20 & -68:52:45 & 2009~Jan~18 & 120;200;450 & 1.29;1.30;1.30 & 0.73;0.84;1.11
& OGLE-CL~LMC~176, KMK88~32, H88~178 \\
            &  &  & 2009~Jan~18 & 120;200;350 & 1.30;1.30;1.30 & 0.80;0.93;1.12 & \\
H88~180     & 05:10:29 & -68:56:03& 2009~Jan~18 & 120;200;450 & 1.29;1.30;1.30 & 0.73;0.84;1.11
& OGLE-CL~LMC~180 \\
            &  &  & 2009~Jan~18 & 120;200;350 & 1.30;1.30;1.30 & 0.80;0.93;1.12 & \\
BRHT48a    & 05:10:30 & -68:52:21 & 2009~Jan~18 & 120;200;450 & 1.29;1.30;1.30 & 0.73;0.84;1.11
& OGLE-CL~LMC~179, HS153 \\
           &  &  & 2009~Jan~18 & 120;200;350 & 1.30;1.30;1.30 & 0.80;0.93;1.12 & \\
OGLE-CL~LMC~185 & 05:10:39 & -69:02:31 & 2009~Jan~18 & 100;180;400 & 1.32;1.32;1.31 & 0.95;0.99;0.93
& SL288 \\
NGC1863    & 05:11:39 & -68:43:48 & 2009~Jan~18 & 100;180;400 & 1.32;1.32;1.32 & 1.15;0.89;1.00
& OGLE-CL~LMC~206, SL299, ESO56-77 \\
NGC1866    & 05:13:39 & -65:27:54 & 2008~Dec~18 & 120;200;500 & 1.32;1.33;1.33 & 1.30;1.07;1.46
& SL319, ESO85-52, KMHK664 \\
BRHT8b     & 05:15:37 & -69:28:23 & 2009~Jan~18 & 100;180;400 & 1.35;1.34;1.34 & 0.85;0.88;1.13
& OGLE-CL~LMC~270, SL341 \\
OGLE-CL~LMC~273* & 05:15:40.26 & -69:16:50.7 & 2009~Jan~18 & 100;180;400 & 1.35;1.35;1.35
& 0.95;1.12;1.05 & \\
NGC1894     & 05:15:51 & -69:28:09 & 2009~Jan~18 & 100;180;400 & 1.35;1.34;1.34 & 0.85;0.88;1.13
& OGLE-CL~LMC~278, SL344, ESO56-89, BRHT8a \\
H88~236     & 05:15:56 & -69:27:16 & 2009~Jan~18 & 100;180;400 & 1.35;1.34;1.34 & 0.85;0.88;1.13
& OGLE-CL~LMC~280 \\
NGC1898    & 05:16:42 & -69:39:22 & 2008~Dec~19 &  90;140;350 & 1.37;1.37;1.37 & 0.71;0.77;0.81
& OGLE-CL~LMC~292, SL350, ESO56-90 \\
NGC1903     & 05:17:22 & -69:20:17 & 2008~Dec~17 & 60;100;300 & 1.36;1.36;1.36 & 0.73;0.77;0.88
& OGLE-CL~LMC~309, SL356, ESO56-93, BRHT9a \\
BRHT9b      & 05:17:24 & -69:22:35 & 2008~Dec~17 & 60;100;300 & 1.36;1.36;1.36 & 0.73;0.77;0.88
& OGLE-CL~LMC~311, SL357 \\
H88~255     & 05:17:27 & -69:21:28 & 2008~Dec~17 & 60;100;300 & 1.36;1.36;1.36 & 0.73;0.77;0.88
& OGLE-CL~LMC~312 \\
OGLE-CL~LMC~318 & 05:17:48 & -69:38:43 & 2008~Dec~17 & 60;100;300 & 1.44;1.45;1.50 & 0.80;0.83;0.97
& SL363 \\
OGLE-CL~LMC~321 & 05:17:55 & -69:34:53 & 2008~Dec~17 & 60;100;300 & 1.44;1.45;1.50 & 0.80;0.83;0.97
& HS213 \\
ESO85-72    & 05:20:05 & -63:28:49 & 2009~Jan~16 &  90;140;350 & 2.01;2.02;2.04 & 0.99;0.78;1.02
& SL388, KMHK773 \\
BSDL1291    & 05:20:14 & -69:34:56 & 2009~Jan~17 & 100;160;400 & 1.39;1.39;1.38 & 1.08;1.19;1.15
& \\
OGLE-CL~LMC~369 & 05:20:22 & -69:35:10 & 2009~Jan~17 & 100;160;400 & 1.39;1.39;1.38 & 1.08;1.19;1.15
& SL402 \\
H88~283     & 05:20:25 & -69:34:15 & 2009~Jan~17 & 100;160;400 & 1.39;1.39;1.38 & 1.08;1.19;1.15
& OGLE-CL~LMC~371 \\
NGC1926     & 05:20:35 & -69:31:31 & 2009~Jan~17 & 100;160;400 & 1.40;1.40;1.41 & 1.08;1.11;1.14
& OGLE-CL~LMC~379, SL403, ESO56-105 \\
NGC1935    & 05:21:58 & -67:57:20 & 2009~Jan~17 & 100;160;400 & 1.41;1.41;1.39 & 1.08;1.16;1.17
& \\
OGLE-CL~LMC~404 & 05:22:00 & -69:15:16 & 2009~Jan~17 & 100;160;400 & 1.44;1.44;1.45 & 1.02;1.09;1.28
& SL419 \\
LH47        & 05:22:07 & -67:56:35 & 2009~Jan~17 & 100;160;400 & 1.41;1.41;1.39 & 1.08;1.16;1.17
& NGC1935 \\
OGLE-CL~LMC~407 & 05:22:11 & -69:30:48 & 2009~Jan~17 & 100;160;400 & 1.49;1.48;1.47 & 1.13;1.20;1.29
& SL423 \\
BSDL1411   & 05:22:17 & -69:28:17 & 2009~Jan~17 & 100;160;400 & 1.49;1.48;1.47 & 1.13;1.20;1.29
& \\
NGC1937    & 05:22:29 & -67:53:40 & 2009~Jan~17 & 100;180;450 & 1.48;1.48;1.49 & 1.12;1.11;1.16
& LH48 \\
OGLE-CL~LMC~431 & 05:24:21 & -69:46:25 & 2009~Jan~17 & 110;180;450 & 1.77;1.75;1.72 & 1.15;1.23;1.30
& HS275 \\
OGLE-CL~LMC~438 & 05:24:36 & -69:44:44 & 2009~Jan~17 & 110;180;450 & 1.77;1.75;1.72 & 1.15;1.23;1.30
& SL449 \\
BSDL1576    & 05:24:42 & -69:53:15 & 2009~Jan~17 & 110;180;450 & 1.67;1.68;1.69 & 1.13;1.14;1.19
& OGLE-CL~LMC~440 \\
BSDL1592    & 05:24:57 & -69:51:43 & 2009~Jan~17 & 110;180;450 & 1.67;1.68;1.69 & 1.13;1.14;1.19
& OGLE-CL~LMC~444 \\
BSDL1588    & 05:24:58 & -69:25:26 & 2009~Jan~17 & 110;180;450 & 1.78;1.79;1.80 & 1.21;1.22;1.41
& OGLE-CL~LMC~445 \\
OGLE-CL~LMC~446 & 05:25:01 & -69:25:58 & 2009~Jan~17 & 110;180;450 & 1.78;1.79;1.80 & 1.21;1.22;1.41
& SL453 \\
BSDL1597    & 05:25:05 & -69:52:25 & 2009~Jan~17 & 110;180;450 & 1.67;1.68;1.69 & 1.13;1.14;1.19
& OGLE-CL~LMC~449 \\
NGC1950    & 05:24:33 & -69:54:08 & 2009~Jan~17 & 110;180;450 & 1.67;1.68;1.69 & 1.13;1.14;1.19
& OGLE-CL~LMC~436, SL450, ESO56-116 \\
BSDL1601    & 05:25:08 & -69:43:06 & 2009~Jan~17 & 110;180;450 & 1.77;1.75;1.72 & 1.15;1.23;1.30
& OGLE-CL~LMC~450 \\
SL457       & 05:25:25 & -69:26:37 & 2009~Jan~17 & 110;180;450 & 1.78;1.79;1.80 & 1.21;1.22;1.41
& \\
NGC1948    & 05:25:51 & -66:15:51 & 2009~Jan~17 & 110;180;450 & 1.63;1.62;1.59 & 1.14;1.21;1.33
& LH52, SL458 \\
NGC1955   & 05:26:12 & -67:29:56 & 2009~Jan~18 & 100;180;400 & 1.36;1.36;1.37 & 0.83;0.86;0.95
& SL467, KMHK888, LH54 \\
BSDL1674  & 05:26:15 & -67:29:56 & 2009~Jan~18 & 100;180;400 & 1.36;1.36;1.37 & 0.83;0.86;0.95
& \\
LH53       & 05:26:16 & -66:07:51 & 2009~Jan~18 & 100;180;400 & 1.34;1.33;1.32 & 0.72;0.87;0.97
& LH53s \\
KMK88~56    & 05:26:32 & -69:48:03 & 2009~Jan~18 & 100;180;400 & 1.44;1.43;1.42 & 0.80;0.93;0.97
& OGLE-CL~LMC~476 \\
             & & & 2009~Jan~18 & 100;180;400 & 1.45;1.46;1.46 & 0.78;0.77;0.81 & \\
NGC1969    & 05:26;32 & -69:50:29 & 2009~Jan~18 & 100;180;400 & 1.44;1.43;1.42 & 0.80;0.93;0.97
& OGLE-CL~LMC~477, SL479, ESO56-124 \\
           & & & 2009~Jan~18 & 100;180;400 & 1.45;1.46;1.46 & 0.78;0.77;0.81 & \\
OGLE-CL~LMC~478* & 05:26:35.30 & -69:49:23.1 & 2009~Jan~18 & 100;180;400 & 1.44;1.43;1.42
& 0.80;0.93;0.97 & \\
               & & & 2009~Jan~18 & 100;180;400 & 1.45;1.46;1.46 & 0.78;0.77;0.81 & \\
BSDL1759   & 05:26:45 & -69:48:11 & 2009~Jan~18 & 100;180;400 & 1.44;1.43;1.42 & 0.80;0.93;0.97
& \\
NGC1971    & 05:26:45 & -69:51:07 & 2009~Jan~18 & 100;180;400 & 1.44;1.43;1.42 & 0.80;0.93;0.97
& OGLE-CL~LMC~480, SL481, ESO56-128, BRHT12a \\
             &  &  & 2009~Jan~18 & 100;180;400 & 1.45;1.46;1.46 & 0.78;0.77;0.81 & \\
             &  &  & 2009~Jan~18 & 100;180;400 & 1.50;1.50;1.48 & 0.67;0.72;0.81 & \\
NGC1972    & 05:26:48 & -69:50:18 & 2009~Jan~18 & 100;180;400 & 1.44;1.43;1.42 & 0.80;0.93;0.97
& OGLE-CL~LMC~481, SL480, ESO56-129, BRHT12b \\
            & & & 2009~Jan~18 & 100;180;400 & 1.45;1.46;1.46 & 0.78;0.77;0.81 & \\
            & & & 2009~Jan~18 & 100;180;400 & 1.50;1.50;1.48 & 0.67;0.72;0.81 & \\
KMK88~57    & 05:26:52 & -69:48:57 & 2009~Jan~18 & 100;180;400 & 1.44;1.43;1.42 & 0.80;0.93;0.97
& OGLE-CL~LMC~483 \\
BSDL1783   & 05:27:02 & -69:50:26 & 2009~Jan~18 & 100;180;400 & 1.44;1.43;1.42 & 0.80;0.93;0.97
& \\
             & & & 2009~Jan~18 & 100;180;400 & 1.45;1.46;1.46 & 0.78;0.77;0.81 & \\
BSDL1785    & 05:27:04 & -69:52:01 & 2009~Jan~18 & 100;180;400 & 1.45;1.46;1.46 & 0.78;0.77;0.81
& OGLE-CL~LMC~487 \\
            & & & 2009~Jan~18 & 100;180;400 & 1.50;1.50;1.48 & 0.67;0.72;0.81 & \\
            & & & 2009~Jan~18 & 100;180;400 & 1.44;1.43;1.42 & 0.80;0.93;0.97 & \\
BSDL1807    & 05:27:27 & -69:52:14 & 2009~Jan~18 & 100;180;400 & 1.45;1.46;1.46 & 0.78;0.77;0.81
& OGLE-CL~LMC~491 \\
BSDL1821    & 05:27:35 & -69:53:47 & 2009~Jan~18 & 100;180;400 & 1.45;1.46;1.46 & 0.78;0.77;0.81
& OGLE-CL~LMC~495 \\
NGC1986    & 05:27:38 & -69:57:49 & 2009~Jan~18 & 100;180;350 & 1.51;1.52;1.53 & 0.68;0.70;0.68
& OGLE-CL~LMC~496, SL489, ESO56-134 \\
BSDL1858    & 05:28:02 & -69:55:49 & 2009~Jan~18 & 100;180;350 & 1.51;1.52;1.53 & 0.68;0.70;0.68
& OGLE-CL~LMC~498 \\
OGLE-CL~LMC~500 & 05:28:02 & -69:59:10 & 2009~Jan~18 & 100;180;350 & 1.51;1.52;1.53 & 0.68;0.70;0.68
& HS310 \\
OGLE-CL~LMC~512* & 05:28:44 & -69:50:03 & 2008~Dec~17 & 60;100;300 & 1.49;1.50;1.50 & 0.86;0.83;0.95
& HS321, SL504 \\
            & & & 2009~Jan~18 &  90;140;350 & 1.64;1.63;1.61 & 0.95;0.93;0.85 & \\
NGC1978    & 05:28:45 & -66:14:10 & 2009~Jan~18 &  90;140;350 & 1.52;1.52;1.50 & 0.69;0.72;0.77
& SL501, ESO85-90, KMHK944 \\
KMHK960    & 05:28:50 & -71:37:58 & 2009~Jan~18 &  90;140;350 & 1.60;1.60;1.61 & 0.69;0.68;0.77
& SL505 \\
BSDL1928    & 05:29:03 & -69:48:38 & 2009~Jan~18 &  90;140;350 & 1.64;1.63;1.61 & 0.95;0.93;0.85
& OGLE-CL~LMC~516 \\
OGLE-CL~LMC~525  & 05:29:35 & -69:46:37 & 2008~Dec~19 &  90;140;350 & 1.37;1.37;1.37 & 0.84;0.83;0.78
& SL514 \\
              &  &  & 2009~Jan~18 &  90;140;350 & 1.64;1.63;1.61 & 0.95;0.93;0.85 & \\
ESO85-91   & 05:29:48 & -63:38:58 & 2008~Dec~19 &  90;140;350 & 1.45;1.46;1.46 & 0.76;0.80;0.79
& SL509, KMHK957 \\
NGC2005    & 05:30:09 & -69:45:08 & 2008~Dec~19 &  90;140;350 & 1.37;1.37;1.37 & 0.84;0.83;0.78
& OGLE-CL~LMC~538, SL518, ESO56-137 \\
OGLE-CL~LMC~540  & 05:30:12 & -69:47:31 & 2008~Dec~19 &  90;140;350 & 1.37;1.37;1.37 & 0.84;0.83;0.78
& HS332 \\
NGC~2016    & 05:31:38 & -69:56:45 & 2009~Jan~18 &  90;140;350 & 1.73;1.73;1.74 & 0.83;0.88;0.89
& OGLE-CL~LMC~559, SL547, ESO56-142 \\
KMHK1046    & 05:31:42 & -72:08:46 & 2009~Jan~18 &  90;140;350 & 1.82;1.81;1.79 & 0.73;0.73;0.88
& SL555 \\
BSDL2205   & 05:31:50 & -69:55:14 & 2009~Jan~18 &  90;140;350 & 1.73;1.73;1.74 & 0.83;0.88;0.89
& \\
BSDL2212   & 05:31:52 & -69:58:50 & 2009~Jan~18 &  90;140;350 & 1.73;1.73;1.74 & 0.83;0.88;0.89
& \\
NGC2019    & 05:31:56 & -70:09:34 & 2008~Dec~19 &  90;140;350 & 1.39;1.39;1.39 & 0.73;0.79;0.81
& OGLE-CL~LMC~565, SL554, ESO56-145 \\
KMHK1013    & 05:32:03 & -64:14:32 & 2009~Jan~18 &  90;140;350 & 1.82;1.83;1.84 & 0.75;0.84;0.85
& SL549 \\
LH72       & 05:32:19 & -66:26:19 & 2009~Jan~18 &  90;140;350 & 1.93;1.91;1.88 & 0.74;0.83;0.84
& SL553 \\
OGLE-CL~LMC~585  & 05:33:21 & -69:57:13 & 2009~Jan~18 &  90;140;350 & 2.04;2.03;1.99 & 1.02;1.02;1.01
& SL574 \\
LH77       & 05:33:29 & -66:59:34 & 2009~Jan~18 &  90;140;350 & 1.94;1.95;1.97 & 0.96;0.90;0.97
& \\
OGLE-CL~LMC~591  & 05:33:45 & -69:54:57 & 2009~Jan~18 &  90;140;350 & 2.04;2.03;1.99 & 1.02;1.02;1.01
& HS353 \\
NGC2028    & 05:33:48 & -69:57:06 & 2009~Jan~18 &  90;140;350 & 2.04;2.03;1.99 & 1.02;1.02;1.01
& OGLE-CL~LMC~594, SL575, ESO56-152, LH80 \\
BSDL2624    & 05:37:26 & -70:13:21 & 2009~Jan~18 &  90;180;400 & 2.04;2.05;2.07 & 0.85;0.94;1.01
& \\
NGC2065    & 05:37:37 & -70:14:09 & 2009~Jan~18 &  90;180;400 & 2.04;2.05;2.07 & 0.85;0.94;1.01
& OGLE-CL~LMC~648, SL626, ESO57-2, KMHK1160 \\
NGC2111    & 05:44:32 & -70:59:36 & 2009~Jan~17 & 140;200;600 & 1.84;1.85;1.87 & 1.62;1.35;1.42
& OGLE-CL~LMC~715, SL699, ESO57-35, BRHT21a \\
KMHK1489    & 05:52:57 & -69:31:51 & 2009~Jan~17 & 110;180;450 & 1.96;1.97;1.99 & 1.23;1.26;1.10
& \\
NGC2136    & 05:52:59 & -69:29:33 & 2009~Jan~17 & 110;180;450 & 1.96;1.97;1.99 & 1.23;1.26;1.10
& SL762, ESO57-48, KMHK1490 \\
NGC2155    & 05:58:33 & -65:28:37 & 2008~Dec~18 & 120;200;500 & 1.36;1.36;1.37 & 0.92;0.97;1.02
& SL804, ESO86-45, KMHK1563 \\
ESO121-3   & 06:02:02 & -60:31:24 & 2008~Dec~18 & 120;200;500 & 1.33;1.33;1.34 & 1.02;1.07;0.84
& KMHK1591 \\
ESO86-61    & 06:08:15 & -62:59;15 & 2009~Jan~18 & 100;180;400 & 2.09;2.06;2.01 & 0.88;0.91;0.96
& SL842, KMHK1652 \\
KMHK1679    & 06:10:53 & -69:08:24 & 2008~Dec~19 &  90;140;350 & 1.41;1.41;1.41 & 0.76;0.84;0.77
& LW421 \\
NGC2210    & 06:11:31 & -69:07:18 & 2008~Dec~19 &  90;140;350 & 1.41;1.41;1.41 & 0.76;0.84;0.77
& SL858, ESO57-71, KMHK1782 \\
ESO57-75    & 06:13:27 & -70:41:45 & 2009~Jan~18 & 100;180;400 & 2.02;2.02;2.04 & 0.89;0.91;0.98
& SL862, KMHK1692 \\
NGC2249    & 06:25:49 & -68:55:12 & 2008~Dec~18 & 120;200;500 & 1.32;1.32;1.32 & 0.97;1.05;0.86
& SL893, ESO57-82, KMHK1755 \\
NGC2257    & 06:30:12 & -64:19:34 & 2008~Dec~18 & 180;300;500 & 1.21;1.21;1.22 & 1.02;1.04;0.92
& SL895, ESO87-24, KMHK1756 \\
\hline
\end{longtable}
\tablefoot{
  Cluster: name of the cluster;
  RA, DEC: equatorial coordinates of the cluster for epoch J2000 from \citet{Bica1999};
  Date: date of observation;
  T$_{\mathrm{exp}}$:  exposure time of filter $y$, $b$, and $v$;
  Airmass: airmass of observations;
  Seeing: average seeing;
  Other name: other name of the clusters in use.}
\end{landscape}
}
%-------------------------------------------------------------

\section{Literature values}

Table~\ref{tab:lit} summarizes the literature values for studied star clusters.

%-------------------------------------------------------------
%                              Table longer than a single page
%                                            and in landscape,
%                    in the preamble, use: \usepackage{lscape}
%-------------------------------------------------------------

% All long tables will be placed automatically at the end of the document
%
% \setcounter{table}{1} \renewcommand{\thetable}{A.\arabic{table}}
\longtab[1]{
% \begin{landscape}
\begin{longtable}{llll}
\caption{\label{tab:lit} Reddenings, metallicities and ages of star clusters from the literature.}\\
\hline\hline
Cluster & $E(B-V)$  & [Fe/H] & log(Age) \\
& (mag) & (dex) & (yr) \\
\hline
\endfirsthead
\caption{continued.}\\
\hline\hline
Name & $E(B-V)$ & [Fe/H] & log(Age) \\
& (mag) & (dex) & (yr) \\
\hline
\endhead
\hline
%\endfoot
%%
NGC1651         & 0.11$^{33}$, 0.10$^{15}$, 0.14$^{20}$ & -0.37$\pm$0.02$^{1}$, -0.63$\pm$0.04$^{22}$,  & 9.3$^{1,14}$, 9.20$\pm$0.10$^{15}$ \\
               & 0.04$^{22}$, 0.07$^{56}$, 0.157$^{55}$ & -0.28$\pm$0.02$^{22}$, -0.58$\pm$0.02$^{22}$, -0.4$^{15}$ & 9.24$\pm$0.06$^{22}$, 9.30$\pm$0.10$^{22}$ \\
               & 0.098$^{58,*}$ & -0.07$\pm$0.10$^{27}$, -0.82$\pm$0.44$^{29}$ & 9.26$^{+0.07}_{-0.08}\, ^{27}$, 9.34$\pm$0.08$^{29}$ \\
               &  & -0.70$\pm$0.10$^{33}$, -0.30$\pm$0.03$^{34}$ & 9.30$\pm$0.03$^{33}$, 9.30$\pm$0.04$^{56}$ \\
               &  & -1.05$\pm$0.15$^{56}$ & \\
\hline \\
KMHK21         & 0.04$^{17,44}$, 0.07$^{56}$, 0.133$^{55}$ & -0.50$\pm$0.20$^{17}$, -0.40$\pm$0.20$^{30}$ & 9.26$\pm$0.08$^{17}$, 9.26$^{+0.07}_{-0.10}\,^{30}$ \\
               & 0.115$^{58,*}$ & -0.50$\pm$0.30$^{44}$, -0.85$\pm$0.10$^{56}$ & 9.21$^{+0.09}_{-0.12}\,^{44}$, 9.26$^{+0.06}_{-0.08}\,^{56}$ \\
               &  & -0.35$^{17}$ (field), -0.30$^{30}$ (field) & \\
\hline \\
NGC1841         & 0.20$^{12}$, 0.11$^{12}$ & -2.20$\pm$0.20$^{12}$, -1.96$\pm$0.12$^{59}$ & 10.10$\pm$0.02$^{14}$, 10.14$^{+0.05}_{-0.04}\,^{59}$ \\
\hline \\
NGC1711        & 0.14$^{8}$, 0.09$^{22}$, 0.07$^{52}$ & -0.40$^{8}$, -0.57$\pm$0.06$^{22}$ & 7.79$^{8}$, 7.70$\pm$0.05$^{22}$ \\
               & 0.06$^{52}$, 0.141$^{55}$, 0.144$^{58,*}$ & -0.78$\pm$0.17$^{38}$, -0.57$\pm$0.11$^{40}$ (FeI) & 8.26$^{+0.22}_{-0.48}\,^{38}$, 7.70$^{52}$ \\
               &  & -0.88$\pm$0.19$^{40}$ (FeII), -0.06$\pm$0.05$^{52}$ & \\
               &  & -0.53$\pm$0.42$^{22}$ (field) & \\
\hline \\
KMHK156        & 0.123$^{55}$, 0.126$^{58,*}$ & & \\
\hline \\
NGC1754        & 0.09$^{21}$, 0.138$^{55}$, 0.113$^{58,*}$ & -1.42$\pm$0.15$^{21}$, -1.37$^{+0.21}_{-0.12}\,^{26}$ & 10.19$\pm$0.07$^{21}$, 9.85$^{+0.08}_{-0.15}\,^{26}$ \\
               &  & -1.38$^{+0.13}_{-0.14}\,^{26}$, -1.44$\pm$0.02$^{29}$ & 10.15$^{+0.03}_{-0.02}\,^{26}$, 9.84$\pm$0.04$^{29}$ \\
               &  & -1.48$\pm$0.09$^{37}$ & 10.15$^{37}$ \\
\hline \\
ESO85-21       & 0.01$^{17}$, 0.0$^{44}$, 0.053$^{58,*}$ & -0.45$\pm$0.20$^{17,30}$, -0.45$\pm$0.30$^{44}$ & 9.34$^{+0.04}_{-0.06}\,^{17}$, 9.34$^{+0.08}_{-0.09}\,^{30}$ \\
               &  & -0.58$\pm$0.01$^{45}$, -0.45$^{17}$ (field) & 9.37$^{+0.08}_{-0.09}\,^{44}$, 9.42$\pm$0.01$^{45}$ \\
               &  & -0.50$^{30}$ (field) & \\
\hline \\
NGC1786        & 0.09$^{12}$, 0.108$^{55}$, 0.092$^{58,*}$ & -1.87$\pm$0.20$^{1}$, -2.10$\pm$0.30$^{12}$ & 10.18$^{+0.04}_{-0.02}\,^{26}$, 10.09$^{37}$ \\
               &  & -1.58$^{+0.13}_{-0.12}\,^{26}$, -1.63$^{+0.11}_{-0.12}\,^{26}$ & \\
               &  & -1.75$\pm$0.02$^{36}$, -1.77$\pm$0.08$^{37}$ & \\
\hline \\
NGC1795        & 0.096$^{44}$, 0.10$^{48}$, 0.07$^{56}$ & -0.23$\pm$0.20$^{1}$, -0.69$\pm$0.42$^{29}$ & 8.90-9.04$^{1}$, 9.30$\pm$0.11$^{29}$ \\
               & 0.116$^{55}$, 0.110$^{58,*}$ & -0.47$\pm$0.10$^{37}$, -0.10$\pm$0.11$^{48}$ & 9.11$^{37}$, 9.20$^{+0.07}_{-0.09}\,^{44}$ \\
               &  & -0.90$\pm$0.15$^{56}$ & 8.95$\pm$0.07$^{48}$, 9.18$^{+0.05}_{-0.07}\,^{56}$ \\
\hline \\
KMHK421         & 0.15$^{35}$, 0.092$^{55}$, 0.095$^{58,*}$ & & 8.10$^{35}$ \\
\hline \\
H88~87          & 0.20$^{35}$, 0.102$^{55}$, 0.091$^{58,*}$ & & 8.20$^{35}$ \\
\hline \\
NGC1804         & 0.20$^{35}$, 0.126$^{55}$, 0.114$^{58,*}$ & & 7.80$^{35}$ \\
\hline \\
SL191           & 0.107$^{55}$, 0.094$^{58,*}$ & & \\
\hline \\
H88~104         & 0.090$^{55}$, 0.080$^{58,*}$ & & $>$9.20$^{24}$ \\
\hline \\
H88~107         & 0.090$^{55}$, 0.080$^{58,*}$ & & \\
\hline \\
BRHT3b          & 0.086$^{55}$, 0.080$^{58,*}$ & & 8.80$\pm$0.10$^{24}$ \\
\hline \\
NGC1830         & 0.089$^{55}$, 0.081$^{58,*}$ & -1.02$^{+0.19}_{-0.40}\,^{26}$, -1.30$^{+0.51}_{-0.17}\,^{26}$ & 8.40$\pm$0.10$^{24}$, 9.09$^{+0.16}_{-0.24}\,^{26}$ \\
                &  &  & 9.18$^{+0.15}_{-0.10}\,^{26}$ \\
\hline \\
SL211           & 0.086$^{55}$, 0.074$^{58,*}$ & & \\
\hline \\
BSDL555         & 0.108$^{55}$, 0.090$^{58,*}$ & & \\
\hline \\
KMHK521         & 0.15$^{35}$, 0.150$^{55}$, 0.115$^{58,*}$ & & 7.70$^{35}$ \\
\hline \\
BSDL565         & 0.07$^{35}$, 0.098$^{55}$, 0.078$^{58,*}$ & & 8.00$\pm$0.05$^{24}$, 8.30$^{17}$ \\
\hline \\
NGC1835         & 0.08$^{21}$, 0.113$^{55}$, 0.083$^{58,*}$ & -1.80$\pm$0.20$^{1}$, -1.62$\pm$0.15$^{21}$ & $>$10.20$^{1}$, 10.21$\pm$0.08$^{21}$ \\
                &  & -1.40$^{+0.18}_{-0.13}\,^{26}$, -1.74$^{+0.22}_{-0.14}\,^{26}$ & 9.92$^{+0.08}_{-0.09}\,^{26}$, 10.10$^{+0.05}_{-0.09}\,^{26}$ \\
\hline \\
H88~119         & 0.098$^{55}$, 0.084$^{58,*}$ & & \\
\hline \\
H88~120         & 0.103$^{55}$, 0.088$^{58,*}$ & & \\
\hline \\
BSDL577         & 0.095$^{55}$, 0.078$^{58,*}$ & & 7.90$\pm$0.05$^{24}$ \\
\hline \\
BSDL581         & 0.097$^{55}$, 0.074$^{58,*}$ & & 7.90$\pm$0.10$^{24}$ \\
\hline \\
BSDL582         & 0.122$^{55}$, 0.132$^{58,*}$ & & \\
\hline \\
HS107           & 0.095$^{55}$, 0.075$^{58,*}$ & & 8.00$\pm$0.10$^{24}$ \\
\hline \\
SOI343          & 0.10$^{35}$, 0.120$^{55}$, 0.131$^{58,*}$ & -0.40$\pm$0.20$^{30}$, -0.25$^{30}$ (field) & 7.60$^{17}$, 7.70$^{+0.08}_{-0.10}\,^{30}$ \\
\hline \\
BSDL591         & 0.094$^{55}$, 0.092$^{58,*}$ & & \\
\hline \\
NGC1836         & 0.06$^{30,44}$, 0.04$^{48}$, 0.095$^{55}$ & 0.00$\pm$0.20$^{30}$, 0.00$\pm$0.30$^{44}$ & 8.50$\pm$0.05$^{24}$, 8.60$^{+0.10}_{-0.12}\,^{30,44}$ \\
                & 0.098$^{58,*}$ & -0.40$\pm$0.29$^{48}$ & 8.80$\pm$0.10$^{48}$ \\
\hline \\
HS109           & 0.03$^{35}$, 0.089$^{55}$, 0.081$^{58,*}$ & & 8.00$^{17}$ \\
\hline \\
BRHT4b          & 0.03$^{35}$, 0.06$^{44}$, 0.01$^{48}$ & -0.40$\pm$0.20$^{30}$, -0.40$\pm$0.30$^{44}$ & 7.80$\pm$0.05$^{24}$, 8.00$^{+0.08}_{-0.10}\,^{30,44}$ \\
                & 0.095$^{55}$, 0.093$^{58,*}$ & -0.10$\pm$0.11$^{48}$ & 7.95$^{17}$, 8.45$\pm$0.08$^{48}$ \\
                &  & -0.25$^{30}$ (field) & \\
\hline \\
HS111           & 0.138$^{55}$, 0.140$^{58,*}$ & & \\
\hline \\
BSDL599         & 0.107$^{55}$, 0.099$^{58,*}$ & & \\
\hline \\
BSDL603         & 0.13$^{35}$, 0.092$^{55}$, 0.080$^{58,*}$ & & 7.80$\pm$0.10$^{24}$, 7.80$^{35}$ \\
\hline \\
NGC1839         & 0.06$^{44}$, 0.04$^{48}$, 0.091$^{55}$ & -0.40$\pm$0.20$^{30}$, -0.40$^{44}$ & 8.00$\pm$0.05$^{24}$, 8.10$^{+0.08}_{-0.10}\,^{30}$ \\
                & 0.080$^{58,*}$ & -0.01$\pm$0.12$^{48}$, -0.25$^{30}$ (field) & 7.90$^{35}$, 8.11$^{0.08}_{-0.07}\,^{44}$ \\
                &  &  & 8.30$\pm$0.50${44}$ \\
\hline \\
NGC1838         & 0.06$^{44,48}$, 0.120$^{55}$, 0.117$^{58,*}$ & -0.40$\pm$0.20$^{30}$, -0.40$^{44}$ & 8.00$^{+0.08}_{-0.10}\,^{30}$, 7.18$\pm$0.12$^{44}$ \\
                &  & -0.01$\pm$0.09$^{48}$, -0.25$^{30}$ (field) & 8.20$\pm$0.10$^{48}$ \\
\hline \\
BSDL623         & 0.08$^{35}$, 0.101$^{55}$, 0.107$^{58,*}$ & & 8.00$^{35}$ \\
\hline \\
OGLE-CL~LMC~111 & 0.11$^{35}$, 0.121$^{55}$, 0.100$^{58,*}$ & & 8.00$^{35}$ \\
\hline \\
BSDL646         & 0.125$^{55}$, 0.105$^{58,*}$ & & \\
\hline \\
OGLE-CL~LMC~113 & 0.03$^{35}$, 0.100$^{55}$, 0.078$^{58,*}$ & & 7.95$^{35}$ \\
\hline \\
NGC1847         & 0.20$^{35}$, 0.16$^{43}$, 0.04$^{52}$ & -0.4$^{43}$, -0.91$\pm$0.09$^{52}$ & 7.30$^{35}$, 7.76$^{43}$, 7.70$^{52}$ \\
                & 0.06$^{52}$, 0.131$^{55}$, 0.124$^{58,*}$ & & \\
\hline \\
NGC1848         & 0.123$^{55}$, 0.110$^{58,*}$ & & \\
\hline \\
BSDL664         & 0.134$^{55}$, 0.138$^{58,*}$ & & \\
\hline \\
NGC1844         & 0.15$^{35}$, 0.15$^{42}$, 0.04$^{52}$ & -0.2$^{42}$, -0.6$^{42}$, -0.50$\pm$0.11$^{52}$ & 7.90$^{35}$, 8.18$^{42}$ \\
                & 0.06$^{52}$, 0.102$^{55}$, 0.096$^{58,*}$ & & \\
\hline \\
NGC1846         & 0.02$^{48,59}$, 0.081$^{55}$, 0.07$^{56}$, & -0.70$\pm$0.20$^{1}$, -0.80$\pm$0.14$^{26}$ & 9.49$^{+0.01}_{-0.07}\,^{26}$, 9.23$^{+0.14}_{-0.10}\,^{26}$ \\
                & 0.078$^{58,*}$ & -0.75$\pm$0.20$^{26}$, -1.40$\pm$0.05$^{29}$ & 9.50$\pm$0.05$^{29}$, 9.15$^{+0.08}_{-0.10}\,^{44}$ \\
                &  & -0.01$\pm$0.09$^{48}$, -0.70$\pm$0.08$^{54}$ & 9.00$\pm$0.05$^{48}$, 9.15$^{+0.05}_{-0.07}\,^{56}$ \\
                &  & -0.90$\pm$0.15$^{56}$, -0.49$\pm$0.08$^{59}$ & 9.23$\pm$0.01$^{59}$ \\
\hline \\
KMHK565         & 0.10$^{35}$, 0.107$^{55}$, 0.094$^{58,*}$ & & 7.30$^{35}$ \\
\hline \\
SL244           & 0.06$^{28,44}$, 0.03$^{48}$, 0.118$^{55}$ & -0.70$\pm$0.20$^{28}$, -0.75$\pm$0.30$^{28}$ & 9.11$^{+0.09}_{-0.11}\,^{28}$, 9.18$^{+0.12}_{-0.18}\,^{28}$ \\
                & 0.125$^{58,*}$ & -0.40$\pm$0.20$^{30}$, -0.30$\pm$0.30$^{44}$ & 9.08$^{+0.08}_{-0.10}\,^{30}$, 9.14$^{+0.04}_{-0.05}\,^{44}$ \\
                &  & -0.28$\pm$0.16$^{48}$, -0.25$^{30}$ (field) & 9.15$\pm$0.09$^{48}$ \\
\hline \\
H88~152         & 0.144$^{55}$, 0.130$^{58,*}$ & & \\
\hline \\
SL256           & 0.15$^{35}$, 0.122$^{55}$, 0.101$^{58,*}$ & & 7.80$^{35}$ \\
\hline \\
NGC1850A        & 0.140$^{55}$, 0.106$^{58,*}$ & -0.4$^{7,8}$ & 6.95$\pm$0.05$^{7}$, 6.82$^{8}$ \\
\hline \\
NGC1850         & 0.17$^{3}$, 0.18$^{7,8}$, 0.20$^{35}$ & -0.26$^{3}$, -0.12$\pm$0.03$^{4}$, & 7.70$\pm$0.10$^{7}$, 7.78$^{+0.07}_{-0.08}\,^{7}$ \\
                & 0.06$^{52}$, 0.16$^{60}$, 0.12$^{59}$ & -0.4$^{7,8}$, -0.53$\pm$0.04$^{52}$ & 7.90$^{8,52}$, 7.90$\pm$0.05$^{24}$ \\
                & 0.141$^{55}$, 0.106$^{58,*}$ & -0.31$\pm$0.20$^{59}$ & 7.10$^{35}$, 7.93$^{+0.18}_{-0.33}\,^{60}$ \\
                &  &  & 7.95$^{+0.03}_{-0.02}\,^{59}$ \\
\hline \\
BRHT5b          & 0.152$^{55}$, 0.127$^{58,*}$ & & 8.00$\pm$0.10$^{24}$ \\
\hline \\
H88~165         & 0.20$^{35}$, 0.151$^{55}$, 0.137$^{58,*}$ & & 8.00$\pm$0.10$^{24}$, 8.20$^{35}$ \\
\hline \\
NGC1854         & 0.122$^{55}$, 0.110$^{58,*}$, 0.21$^{57,*}$ & Z=0.007$^{57}$ & 7.85$\pm$0.05$^{24}$, 7.78$^{57}$ \\
\hline \\
BSDL745         & 0.118$^{55}$, 0.118$^{58,*}$ & & 7.60$\pm$0.05$^{24}$ \\
\hline \\
BSDL748         & 0.115$^{55}$, 0.095$^{58,*}$ & & 7.80$\pm$0.05$^{24}$ \\
\hline \\
NGC1858         & 0.15$^{7,8}$, 0.105$^{55}$, 0.086$^{58,*}$ & -0.4$^{7,8}$, Z=0.007$^{57}$ & 6.90$^{7}$, 6.88$^{8}$, 6.70$^{57}$ \\
                & 0.19$^{57}$ & & 7.30$^{57}$ \\
\hline \\
H88~177         & 0.098$^{55}$, 0.084$^{58,*}$ & & 8.20$\pm$0.05$^{24}$ \\
\hline \\
BRHT48b         & 0.098$^{55}$, 0.078$^{58,*}$ & & 8.30$\pm$0.10$^{24}$ \\
\hline \\
H88~180         & 0.109$^{55}$, 0.116$^{58,*}$ & & \\
\hline \\
BRHT48a         & 0.100$^{55}$, 0.077$^{58,*}$ & & 8.35$\pm$0.10$^{24}$ \\
\hline \\
OGLE-CL~LMC~185 & 0.10$^{35}$, 0.137$^{55}$, 0.123$^{58,*}$ & & 7.70$^{35}$ \\
\hline \\
NGC1863         & 0.15$^{35}$, 0.06$^{44,52}$, 0.08$^{48}$ & -0.40$\pm$0.20$^{30}$, -0.40$^{44}$ & 7.70$^{+0.08}_{-0.10}\,^{30}$, 7.80$^{35}$ \\
                & 0.05$^{52}$, 0.119$^{55}$, 0.086$^{58,*}$ & -0.01$\pm$0.09$^{48}$, -0.53$\pm$0.09$^{52}$ & 7.60$^{+0.10}_{-0.12}\,^{44}$, 8.00$\pm$0.09$^{48}$ \\
                &  & -0.40$^{30}$ (field) & 7.60$^{52}$ \\
\hline \\
NGC1866         & 0.07$^{2,8}$, 0.06$^{10,38}$, 0.1$^{47}$, & -0.43$\pm$0.18$^{10}$, -0.38$^{10}$, -0.4$^{8,47}$ & 8.14$^{8}$, 8.0$^{10,23}$, 8.48$\pm$0.77$^{29}$ \\
                & 0.15$^{60}$, 0.060$^{58,*}$ & -0.50$\pm$0.10$^{23}$, -0.43$\pm$0.04$^{39}$ & 8.00-8.48$^{38}$, 8.26$\pm$0.20$^{46}$ \\
                &  & 0.00$\pm$0.04$^{38}$, -0.27$\pm$0.16$^{40}$ & \\
                &  & -0.33$\pm$0.07$^{46}$ & \\
\hline \\
BRHT8b          & 0.092$^{55}$, 0.054$^{58,*}$ & & 6.70$\pm$0.10$^{24}$ \\
\hline \\
OGLE-CL~LMC~273 & 0.108$^{55}$, 0.087$^{58,*}$ & & \\
\hline \\
NGC1894         & 0.089$^{55}$, 0.054$^{58,*}$ & & 7.85$\pm$0.05$^{24}$ \\
\hline \\
H88~236         & 0.096$^{55}$, 0.067$^{58,*}$ & & \\
\hline \\
NGC1898         & 0.07$^{21}$, 0.093$^{55}$, 0.057$^{58,*}$ & -1.37$\pm$0.15$^{1}$, -1.18$\pm$0.16$^{21}$ & 10.13$\pm$0.08$^{21}$, 10.04$^{+0.07}_{-0.04}\,^{26}$ \\
                &  & -1.27$^{+0.20}_{0.15}\,^{26}$, -1.33$^{+0.33}_{0.15}\,^{26}$ & 10.14$\pm$0.04$^{26}$ \\
                &  & -1.23$\pm$0.05$^{32}$ (FeI), -0.81$\pm$0.13$^{32}$ (FeII) & \\
\hline \\
NGC1903         & 0.143$^{55}$, 0.145$^{58,*}$ & & 7.88$\pm$0.03$^{24}$ \\
\hline \\
BRHT9b          & 0.139$^{55}$, 0.137$^{58,*}$ & & 9.00$\pm$0.08$^{24}$ \\
\hline \\
H88~255         & 0.142$^{55}$, 0.137$^{58,*}$ & & 8.00$\pm$0.05$^{24}$ \\
\hline \\
OGLE-CL~LMC~318 & 0.095$^{55}$, 0.056$^{58,*}$ & -0.90$\pm$0.15$^{56}$ & 9.35$\pm$0.02$^{56}$ \\
\hline \\
OGLE-CL~LMC~321 & 0.093$^{55}$, 0.051$^{58,*}$ & & 8.10$\pm$0.07$^{24}$ \\
\hline \\
ESO85-72        & 0.03$^{17,44}$, 0.04$^{56}$, 0.053$^{58,*}$ & -0.65$\pm$0.20$^{17,30}$, -0.65$\pm$0.30$^{44}$ & 9.34$\pm$0.03$^{17}$, 9.34$\pm$0.03$^{17}$ \\
                &  & -0.58$\pm$0.06$^{45}$, -0.95$\pm$0.10$^{56}$ & 9.38$^{+0.08}_{-0.10}\,^{30}$, 9.38$^{+0.08}_{-0.12}\,^{44}$ \\
                &  & -0.60$^{17}$ (field), -0.55$^{30}$ (field) & 9.42$\pm$0.10$^{45}$, 9.34$\pm$0.06$^{56}$ \\
\hline \\
BSDL1291        & 0.091$^{55}$, 0.077$^{58,*}$ & & \\
\hline \\
OGLE-CL~LMC~369 & 0.088$^{55}$, 0.077$^{58,*}$ & & 8.30$\pm$0.07$^{24}$ \\
\hline \\
H88~283         & 0.089$^{55}$, 0.073$^{58,*}$ & & \\
\hline \\
NGC1926         & 0.092$^{55}$, 0.059$^{58,*}$ & & 8.00$\pm$0.10$^{24}$ \\
\hline \\
NGC1935         & 0.092$^{55}$, 0.115$^{58,*}$ & & \\
\hline \\
OGLE-CL~LMC~404 & 0.102$^{55}$, 0.063$^{58,*}$ & & 8.35$\pm$0.05$^{24}$ \\
\hline \\
LH47            & 0.11$^{11}$, 0.123$^{55}$, 0.114$^{58,*}$ & -0.4$^{11}$ & 6.30$^{11}$ \\
\hline \\
OGLE-CL~LMC~407 & 0.085$^{55}$, 0.038$^{58,*}$ & & 8.20$\pm$0.10$^{24}$ \\
\hline \\
BSDL1411        & 0.086$^{55}$, 0.043$^{58,*}$ & & \\
\hline \\
NGC1937         & 0.11$^{11}$, 0.125$^{55}$, 0.180$^{58,*}$ & -0.4$^{11}$ & 6.30$^{11}$ \\
\hline \\
OGLE-CL~LMC~431 & 0.086$^{55}$, 0.064$^{58,*}$ & & 8.00$\pm$0.05$^{24}$ \\
\hline \\
OGLE-CL~LMC~438 & 0.082$^{55}$, 0.072$^{58,*}$ & & \\
\hline \\
BSDL1576        & 0.096$^{55}$, 0.101$^{58,*}$ & & \\
\hline \\
BSDL1592        & 0.096$^{55}$, 0.125$^{58,*}$ & & 8.00$\pm$0.08$^{24}$ \\
\hline \\
BSDL1588        & 0.111$^{55}$, 0.074$^{58,*}$ & & \\
\hline \\
OGLE-CL~LMC~446 & 0.110$^{55}$, 0.075$^{58,*}$ & & 8.30$\pm$0.05$^{24}$ \\
\hline \\
BSDL1597        & 0.107$^{55}$, 0.162$^{58,*}$ & & \\
\hline \\
NGC1950         & 0.094$^{55}$, 0.100$^{58,*}$ & & 8.70$\pm$0.08$^{24}$ \\
\hline \\
BSDL1601        & 0.104$^{55}$, 0.096$^{58,*}$ & & 8.20$\pm$0.10$^{24}$ \\
\hline \\
SL457           & 0.107$^{55}$, 0.095$^{58,*}$ & & \\
\hline \\
NGC1948         & 0.20$^{13}$, 0.136$^{58,*}$ & -0.30$^{9}$, -0.4$^{13}$ & 7.00$^{9}$, 6.70-7.00$^{13}$ \\
\hline \\
NGC1955         & 0.09$^{18}$, 0.108$^{58,*}$ & -0.40$^{18}$ & 7.19$\pm$0.05$^{18}$ \\
\hline \\
BSDL1674        & 0.108$^{58,*}$ & & 7.00$^{35}$ \\
\hline \\
LH53            & 0.130$^{58,*}$ & -0.30$^{9}$ & 7.00$^{9}$ \\
\hline \\
KMK88~56        & 0.110$^{55}$, 0.078$^{58,*}$ & & 8.40$\pm$0.10$^{24}$ \\
\hline \\
NGC1969         & 0.093$^{55}$, 0.076$^{58,*}$ & & 7.80$\pm$0.05$^{24}$ \\
\hline \\
OGLE-CL~LMC~478 & 0.102$^{55}$, 0.077$^{58,*}$ & & 8.00$\pm$0.10$^{24}$ \\
\hline \\
BSDL1759        & 0.104$^{55}$, 0.076$^{58,*}$ & & \\
\hline \\
NGC1971         & 0.088$^{55}$, 0.076$^{58,*}$ & & 8.00$\pm$0.05$^{24}$ \\
\hline \\
NGC1972         & 0.062$^{44}$, 0.085$^{55}$, 0.076$^{58,*}$ & -0.44$\pm$0.30$^{44}$ & 7.80$\pm$0.10$^{24}$, 8.20$^{+0.20}_{-0.09}\,^{44}$ \\
\hline \\
KMK88~57        & 0.096$^{55}$, 0.065$^{58,*}$ & & \\
\hline \\
BSDL1783        & 0.087$^{55}$, 0.057$^{58,*}$ & & \\
\hline \\
BSDL1785        & 0.093$^{55}$, 0.057$^{58,*}$ & & \\
\hline \\
BSDL1807        & 0.084$^{55}$, 0.050$^{58,*}$ & & \\
\hline \\
BSDL1821        & 0.086$^{55}$, 0.049$^{58,*}$ & & 8.60$\pm$0.20$^{24}$ \\
\hline \\
NGC1986         & 0.05$^{52}$, 0.06$^{52}$, 0.107$^{55}$ & -0.46$\pm$0.06$^{52}$ & 8.00$^{52}$ \\
                & 0.090$^{58,*}$ & & \\
\hline \\
BSDL1858        & 0.119$^{55}$, 0.086$^{58,*}$ & & \\
\hline \\
OGLE-CL~LMC~500 & 0.106$^{55}$, 0.081$^{58,*}$ & & 8.20$\pm$0.10$^{24}$ \\
\hline \\
OGLE-CL~LMC~512 & 0.101$^{55}$, 0.067$^{58,*}$ & & \\
\hline \\
NGC1978         & 0.09$^{38}$, 0.07$^{51,53}$, 0.05$^{59}$ & -0.42$\pm$0.04$^{1}$, -0.96$\pm$0.15$^{23}$ & 9.30$^{1,50,59}$, 9.30$\pm$0.02$^{23}$ \\
                & 0.074$^{58,*}$ & -0.21$^{+0.21}_{-0.26}\,^{26}$, -0.58$^{+0.16}_{-0.15}\,^{26}$ & 9.18$^{+0.08}_{-0.12}\,^{26}$, 9.13$^{+0.43}_{-0.53}\,^{26}$ \\
                &  & -0.72$\pm$0.01$^{29}$, -0.38$\pm$0.02$^{31,34}$ (FeI) & 9.41$\pm$0.06$^{29}$, 9.30$^{+0.10}_{-0.12}\,^{38}$ \\
                &  & -0.26$\pm$0.0.2$^{31}$ (FeII), -0.71$\pm$0.08$^{38}$ & 9.75$^{49}$ \\
                &  & -0.54$\pm$0.19$^{40}$ (FeI), 0.05$\pm$0.14$^{40}$ (FeII) & \\
                &  & -0.35$^{50}$, -0.50$^{49}$, $\sim$-1.0$^{53}$ & \\
                &  & -0.43$\pm$0.06$^{53}$, -0.49$\pm$0.10$^{59}$ & \\
\hline \\
KMHK960         & 0.07$^{28,44}$, 0.15$^{35}$, 0.03$^{48}$ & -0.50$\pm$0.20$^{8}$, -0.70$\pm$0.30$^{8}$ & 8.95$^{+0.09}_{-0.10}\,^{28}$, 9.18$^{+0.08}_{-0.10}\,^{28,44}$ \\
                & 0.121$^{55}$, 0.112$^{58,*}$ & -0.40$\pm$0.20$^{30}$, -0.5$^{44}$ & 8.95$^{+0.08}_{-0.09}\,^{30}$, 8.80$^{35}$ \\
                &  & -0.70$\pm$0.30$^{44}$, -0.28$\pm$0.16$^{48}$ & 9.19$^{+0.09}_{-0.11}\,^{44}$, 8.90$\pm$0.10$^{48}$ \\
                &  & -0.15$^{30}$ (field) & \\
\hline \\
BSDL1928        & 0.106$^{55}$, 0.076$^{58,*}$ & & \\
\hline \\
OGLE-CL~LMC~525 & 0.120$^{55}$, 0.092$^{58,*}$ & & \\
\hline \\
ESO85-91        & 0.03$^{17,44}$, 0.05$^{56}$, 0.062$^{58,*}$ & -0.85$\pm$0.20$^{17}$, -0.65$\pm$0.20$^{30}$ & 9.08$^{+0.10}_{-0.12}\,^{17}$, 9.08$^{+0.08}_{-0.10}\,^{30}$ \\
                &  & -0.54$\pm$0.09$^{37}$, -0.85$\pm$0.30$^{44}$ & 9.08$^{37}$, 9.14$\pm$0.01$^{45}$ \\
                &  & -1.18$\pm$0.08$^{45}$, -1.10$\pm$0.15$^{56}$ & 9.11$^{+0.12}_{-0.15}\,^{44}$, 9.08$^{+0.10}_{-0.13}\,^{56}$ \\
                &  & -0.50$^{17}$ (field), -0.35$^{30}$ (field) & \\
\hline \\
NGC2005         & 0.10$^{21,38}$, 0.112$^{55}$, 0.069$^{58,*}$ & -1.92$\pm$0.20$^{1}$, -1.35$\pm$0.16$^{21}$ & 10.19$\pm$0.20$^{21}$, $>$9.20$^{24}$ \\
                & 0.10$^{21,38}$, 0.112$^{55}$, 0.069$^{58,*}$ & -1.51$^{+0.12}_{-0.31}\,^{26}$, -1.34$^{+0.26}_{-0.32}\,^{26}$ & 9.80$^{+0.07}_{-0.17}\,^{26}$, 10.20$^{+0.06}_{-0.04}\,^{26}$ \\
                &  & -1.80$\pm$.10$^{32}$ (FeI), -1.33$\pm$.09$^{32}$ (FeII), & $>$10$^{38}$, $>$5$^{38}$ \\
                &  & -1.52$\pm$0.06$^{38}$, -1.54$\pm$0.04$^{40}$ (FeI) & \\
                &  & -1.27$\pm$0.03$^{40}$ (FeII) & \\
\hline \\
OGLE-CL~LMC~540 & 0.138$^{55}$, 0.089$^{58,*}$ & & \\
\hline \\
NGC2016         & 0.138$^{55}$, 0.104$^{58,*}$ & & \\
\hline \\
KMHK1046        & 0.07$^{28,44,56}$, 0.02$^{35,48}$ & -0.70$\pm$0.20$^{28,30}$, -0.75$\pm$0.30$^{28}$ & 9.20$^{+0.12}_{-0.16}\,^{28}$, 9.20$^{+0.08}_{-0.07}\,^{28}$ \\
                & 0.116$^{58,*}$ & -0.70$^{44}$, -0.75$\pm$0.30$^{44}$ & 9.28$^{+0.08}_{-0.10}\,^{30}$, 8.80$^{35}$ \\
                &  & -0.70$\pm$0.58$^{48}$, -1.10$\pm$0.10$^{56}$ & 9.20$^{+0.08}_{-0.09}\,^{44}$, 9.23$^{+0.09}_{-0.12}\,^{44}$ \\
                &  & -0.40$^{30}$ (field) & 9.30$\pm$0.20$^{48}$, 9.23$\pm$0.05$^{56}$ \\
\hline \\
BSDL2205        & 0.122$^{55}$, 0.107$^{58,*}$ & & \\
\hline \\
BSDL2212        & 0.132$^{55}$, 0.104$^{58,*}$ & & \\
\hline \\
NGC2019         & 0.06$^{21,38}$, 0.096$^{55}$, 0.083$^{58,*}$ & -1.80$\pm$0.20$^{1}$, -1.23$\pm$0.15$^{21}$ & 10.21$\pm$0.09$^{21}$, $>$9.20$^{24}$ \\
                &  & -1.41$^{+0.40}_{-0.20}\,^{26}$, -1.44$^{+0.16}_{-0.37}\,^{26}$ & 10.20$^{+0.06}_{-0.09}\,^{26}$, 10.12$^{+0.11}_{-0.03}\,^{26}$ \\
                &  & -1.37$\pm$0.07$^{32}$, -1.10$\pm$0.16$^{32}$ & $>$10$^{38}$, $>$7$^{38}$ \\
                &  & -1.64$\pm$0.05$^{38}$, -1.67$\pm$0.03$^{40}$ & \\
                &  & -1.65$\pm$0.04$^{40}$ & \\
\hline \\
KMHK1013        & 0.04$^{28,44,48}$, 0.05$^{56}$ & -0.90$\pm$0.20$^{28}$, -0.70$\pm$0.20$^{30}$ & 9.30$^{+0.10}_{-0.12}\,^{28,44}$, 9.30$^{+0.06}_{-0.07}\,^{28}$ \\
                & 0.074$^{58,*}$ & -0.90$^{44}$, -0.18$\pm$0.22$^{48}$ & 9.23$^{+0.08}_{-0.07}\,^{30}$, 9.22$^{+0.14}_{-0.22}\,^{44}$ \\
                &  & -1.10$\pm$0.10$^{56}$ & 9.20$\pm$0.20$^{48}$, 9.23$^{+0.07}_{-0.06}\,^{56}$ \\
                &  & -0.70$^{30}$ (field) & \\
\hline \\
LH72            & 0-0.17$^{16}$, 0.069$^{58,*}$ & -0.60$^{16}$ & 6.70-7.18$^{16}$ \\
\hline \\
OGLE-CL~LMC~585 & 0.138$^{55}$, 0.119$^{58,*}$ & & \\
\hline \\
LH77            & 0.06$^{18}$, 0.056$^{58,*}$ & -0.40$^{18}$ & 7.20$\pm$0.14$^{18}$ \\
\hline \\
OGLE-CL~LMC~591 & 0.133$^{55}$, 0.123$^{58,*}$ & & \\
\hline \\
NGC2028         & 0.121$^{55}$, 0.106$^{58,*}$ & & \\
\hline \\
BSDL2624        & 0.157$^{55}$, 0.125$^{58,*}$ & & \\
\hline \\
NGC2065         & 0.10$^{52}$, 0.06$^{52}$, 0.156$^{55}$ & -0.40$\pm$0.06$^{52}$ & 8.00$^{52}$ \\
                & 0.151$^{58,*}$ & & \\
\hline \\
NGC2111         & 0.159$^{55}$, 0.190$^{58,*}$ & & 8.20$\pm$0.05$^{24}$ \\
\hline \\
KMHK1489        & 0.129$^{55}$, 0.126$^{58,*}$ & & 8.20$^{35}$ \\
\hline \\
NGC2136         & 0.10$^{5,22}$, 0.09$^{22}$, 0.15$^{35}$ & -0.5$^{5}$, -0.56$\pm$0.03$^{22}$, & 6.90$^{5}$, 8.16$\pm$0.05$^{22}$ \\
                & 0.07$^{52}$, 0.06$^{52}$, 0.148$^{55}$ &  -0.55$\pm$0.06$^{22}$, -0.40$\pm$0.01$^{41}$ & 8.00$\pm$0.10$^{22}$, 7.91$\pm$0.02$^{29}$ \\
                & 0.123$^{58,*}$ & -0.48$^{43}$, -0.51$\pm$0.08$^{52}$ & 7.30$^{35}$, 8.00$^{41}$, 8.09$^{43,52}$ \\
\hline \\
NGC2155         & 0.07$^{2}$, 0.02$^{33}$, 0.05$^{44}$, & -0.55$\pm$0.20$^{1}$, -1.08$\pm$0.12$^{19}$ & 9.40$^{1}$, 9.60$\pm$0.03$^{19}$ \\
                & 0.03$^{51}$, 0.04$^{56}$, 0.02$^{59}$ & -0.68$^{25}$, -0.68$\pm$0.20$^{25}$ & 9.50$\pm$0.05$^{25}$, 9.43$\pm$0.26$^{29}$ \\
                & 0.065$^{58,*}$ & -0.44$\pm$0.86$^{29}$, -0.80$\pm$0.20$^{30}$ & 9.56$^{+0.08}_{-0.10}\,^{30}$, 9.48$\pm$0.03$^{33}$ \\
                &  & -0.70$\pm$0.10$^{33,56}$, -0.90$^{44}$, -0.66$^{51}$ & 9.51$^{+0.07}_{-0.10}\,^{44}$, 9.40$^{+0.08}_{-0.10}\,^{51}$ \\
                &  & -1.0$\pm$0.10$^{56}$, -0.59$\pm$0.12$^{59}$ & 9.48$^{+0.04}_{-0.05}\,^{56}$, 9.48$^{+0.03}_{-0.04}\,^{59}$ \\
                &  & -0.75$^{30}$ (field) & \\
\hline \\
ESO121-3        & 0.03$^{17,44}$ & -0.93$\pm$0.20$^{1}$, -1.05$\pm$0.20$^{17}$ & 10.0$^{1}$, 9.96$^{14}$, 9.93$\pm$0.01$^{17}$ \\
                &  & -1.01$\pm$0.15$^{19}$, -0.91$\pm$0.16$^{23}$ & 9.98$\pm$0.10$^{19}$, 9.20$^{23}$ \\
                &  & -1.05$\pm$0.30$^{44}$, -1.40$\pm$0.05$^{45}$ & 9.93$^{+0.01}_{-0.02}\,^{44}$, 9.99$\pm$0.01$^{45}$ \\
\hline \\
ESO86-61        & 0.03$^{35,44}$, 0.05$^{56}$ & -0.36$\pm$0.20$^{1}$, -0.65$\pm$0.20$^{17}$ & 9.50$^{1}$, 9.30$\pm$0.09$^{14}$ \\
                &  & -0.60$\pm$0.20$^{30}$, -0.60$\pm$0.30$^{44}$ & 9.34$\pm$0.03$^{35}$, 9.34$^{+0.08}_{-0.09}\,^{30}$ \\
                &  & -1.00$\pm$0.10$^{56}$, -0.55$^{17}$ (field) & 9.31$^{+0.09}_{-0.11}\,^{44}$, 9.34$\pm$0.06$^{56}$ \\
                &  & -0.60$^{30}$ (field) & \\
\hline \\
KMHK1679        & 0.076$^{55}$, 0.082$^{58,*}$ &  & \\
\hline \\
NGC2210         & 0.09$^{12}$, 0.078$^{55}$, 0.074$^{58,*}$ & -2.2$\pm$0.20$^{12}$, -1.97$\pm$0.20$^{16}$ & 10.20$^{23}$, 9.58$\pm$0.05$^{29}$ \\
                &  & -1.75$\pm$0.10$^{23}$, -1.16$\pm$0.20$^{29}$ & \\
                &  & -1.65$\pm$0.04$^{36}$ & \\
\hline \\
ESO57-75        & 0.09$^{35,44}$, 0.07$^{56}$, 0.118$^{55}$ & -0.85$\pm$0.20$^{17}$, -0.75$\pm$0.20$^{30}$ & 9.26$\pm$0.03$^{35}$, 9.26$^{+0.07}_{-0.10}\,^{30}$ \\
                & 0.159$^{58,*}$ & -0.47$\pm$0.07$^{37}$, -0.85$\pm$0.30$^{44}$ & 9.23$^{37}$, 9.26$^{+0.06}_{-0.08}\,^{44,56}$ \\
                &  & -0.90$\pm$0.15$^{56}$, -0.60$^{17}$ (field) & \\
                &  & -0.55$^{30}$ (field) & \\
\hline
NGC2249         & 0.25$^{6,8}$, 0.01$^{33}$, 0.098$^{58,*}$ & -0.4$^{8}$, -0.40$\pm$0.02$^{29}$ & 8.74$^{6}$, 8.54$\pm$0.10$^{8}$ \\
                &  & -0.45$\pm$0.10$^{33}$ & 8.44$\pm$0.30$^{29}$, 9.00$\pm$0.03$^{33}$ \\
\hline \\
NGC2257         & 0.06$^{20}$, 0.04$^{22}$, 0.04$^{59}$ & -1.63$\pm$0.21$^{22}$, -0.85$\pm$0.10$^{22}$ & $10.0-10.3^{22}$, 10.11$^{+0.05}_{-0.09}\,^{59}$ \\
                &  & -1.95$\pm$0.04$^{36}$, -1.64$\pm$0.11$^{59}$ & \\
\hline \\
\end{longtable}
\tablefoot{
  Cluster: name of the cluster;
  E(B-V): reddening;
  [Fe/H]: metallicity;
  log(Age): logarithm of age.

 \begin{itemize}\scriptsize
 \itemsep0em
   \item [$^{(*)}$] E(V-I).
   \item[] References:
   (1) \citet{Olszewski1991} (low-resolution spectroscopy, CaT);
   (2) \citet{Bertelli1992} (optical photometry);
   (3) \citet{Fischer1993} (optical photometry);
   (4) \citet{JT1994} (medium-resolution spectroscopy);
   (5) \citet{SHR1994} (Str\"omgren photometry);
   (6) \citet{Vallenari1994a} (optical photometry);
   (7) \citet{Vallenari1994b} (optical photometry);
   (8) \citet{Girardi1995} (optical photometry);
   (9) \citet{Hill1995} (UV and optical photometry);
   (10) \citet{Hilker1995} (Str\"omgren photometry);
   (11) \citet{OM1995} (optical photometry);
   (12) \citet{Brocato1996} (RGB/HB method);
   (13) \citet{Will1996} (high-resolution spectroscopy);
   (14) \citet{Geisler1997} (Washington photometry);
   (15) \citet{Mould1997} (HST photometry);
   (16) \citet{Olsen1997} (low-resolution spectroscopy);
   (17) \citet{Bica1998} (Washington photometry);
   (18) \citet{DH1998} (optical photometry);
   (19) \citet{Sarajedini1998} (HST photometry, RGB slope);
   (20) \citet{Schlegel1998} (optical photometry);
   (21) \citet{Olsen1998} (HST photometry);
   (22) \citet{Dirsch2000} (Str\"omgren photometry);
   (23) \citet{Hill2000} (high-resolution spectroscopy);
   (24) \citet{PU2000} (optical photometry);
   (25) \citet{Rich2001} (HST photometry);
   (26) \citet{Beasley2002} (low-resolution spectroscopy, CaT);
   (27) \citet{Sarajedini2002} (NIR photometry, RGB slope);
   (28) \citet{Geisler2003} (Washington photometry);
   (29) \citet{LR2003} (integrated spectroscopy);
   (30) \citet{Piatti2003b} (Washington photometry);
   (31) \citet{Ferraro2006} (high-resolution spectroscopy);
   (32) \citet{Johnson2006} (high-resolution spectroscopy);
   (33) \citet{Kerber2007} (HST photometry);
   (34) \citet{Mucciarelli2008} (high-resolution spectroscopy);
   (35) \citet{Glatt2010} (optical photometry);
   (36) \citet{Mucciarelli2010} (high-resolution spectroscopy);
   (37) \citet{Sharma2010} (low-resolution spectroscopy);
   (38) \citet{Colucci2011} (integrated spectroscopy);
   (39) \citet{Mucciarelli2011} (high-resolution spectroscopy);
   (40) \citet{Colucci2012} (integrated spectroscopy);
   (41) \citet{Mucciarelli2012} (high-resolution spectroscopy);
   (42) \citet{Milone2013} (HST photometry);
   (43) \citet{Niederhofer2015} (HST photometry);
   (44) \citet{Palma2015} (Washington photometry);
   (45) \citet{Pieres2016} (DES photometry);
   (46) \citet{Lemasle2017} (high-resolution spectroscopy);
   (47) \citet{Milone2017} (HST photometry);
   (48) \citet{Perren2017} (Washington photometry, ASteCA);
   (49) \citet{Martocchia2018a} (HST photometry);
   (50) \citet{Martocchia2018b} (HST photometry);
   (51) \citet{Martocchia2019} (HST photometry);
   (52) \citet{Piatti2019} (Str\"omgren photometry);
   (53) \citet{PB2019} (Str\"omgren photometry);
   (54) \citet{Song2019} (high-resolution spectroscopy);
   (55) \citet{Gorski2020} (optical photometry);
   (56) \citet{Piatti2020} (Str\"omgren photometry);
   (57) \citet{DeMarchi2021} (HST photometry);
   (58) \citet{Skowron2021} (optical photometry).
   (59) \citet{Song2021} (high-resolution spectroscopy).
   (60) \citet{Yang2021} (HST photometry).
 \end{itemize}
}
% \end{landscape}
}
%
%-------------------------------------------------------------

\section{Examples of metallicity determinations}
\label{sec:appexample}

Figures \ref{fig:ngc2249} to \ref{fig:ngc1971} show the examples of star
clusters where only one or two stars were taken for metallicity determination.

%-------------------------------------------------------------
%                                             Two column Figure
%-------------------------------------------------------------
   \begin{figure*}
   \resizebox{\hsize}{!}
            {\includegraphics[]{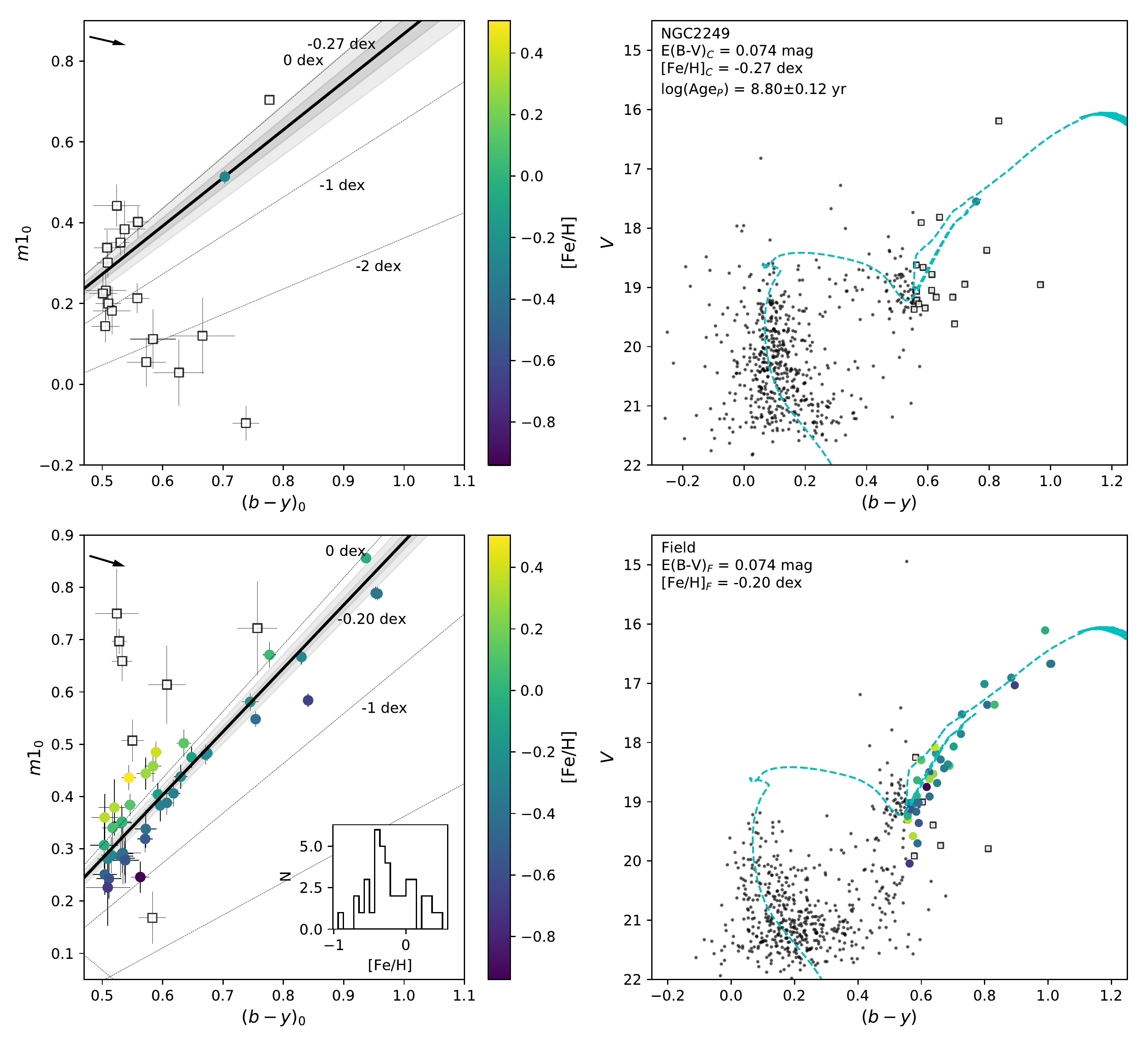}}
      \caption{Reddening-corrected two-color diagrams (left panels) and reddened
               CMDs (right panels) for NGC2249 (upper panels) and the surrounding
               field stars (lower panels).
               The symbols are the same as in Fig.~\ref{fig:ngc1651}.
              }
      \label{fig:ngc2249}
   \end{figure*}
%
%-------------------------------------------------------------

%-------------------------------------------------------------
%                                             Two column Figure
%-------------------------------------------------------------
   \begin{figure*}
   \resizebox{\hsize}{!}
            {\includegraphics[]{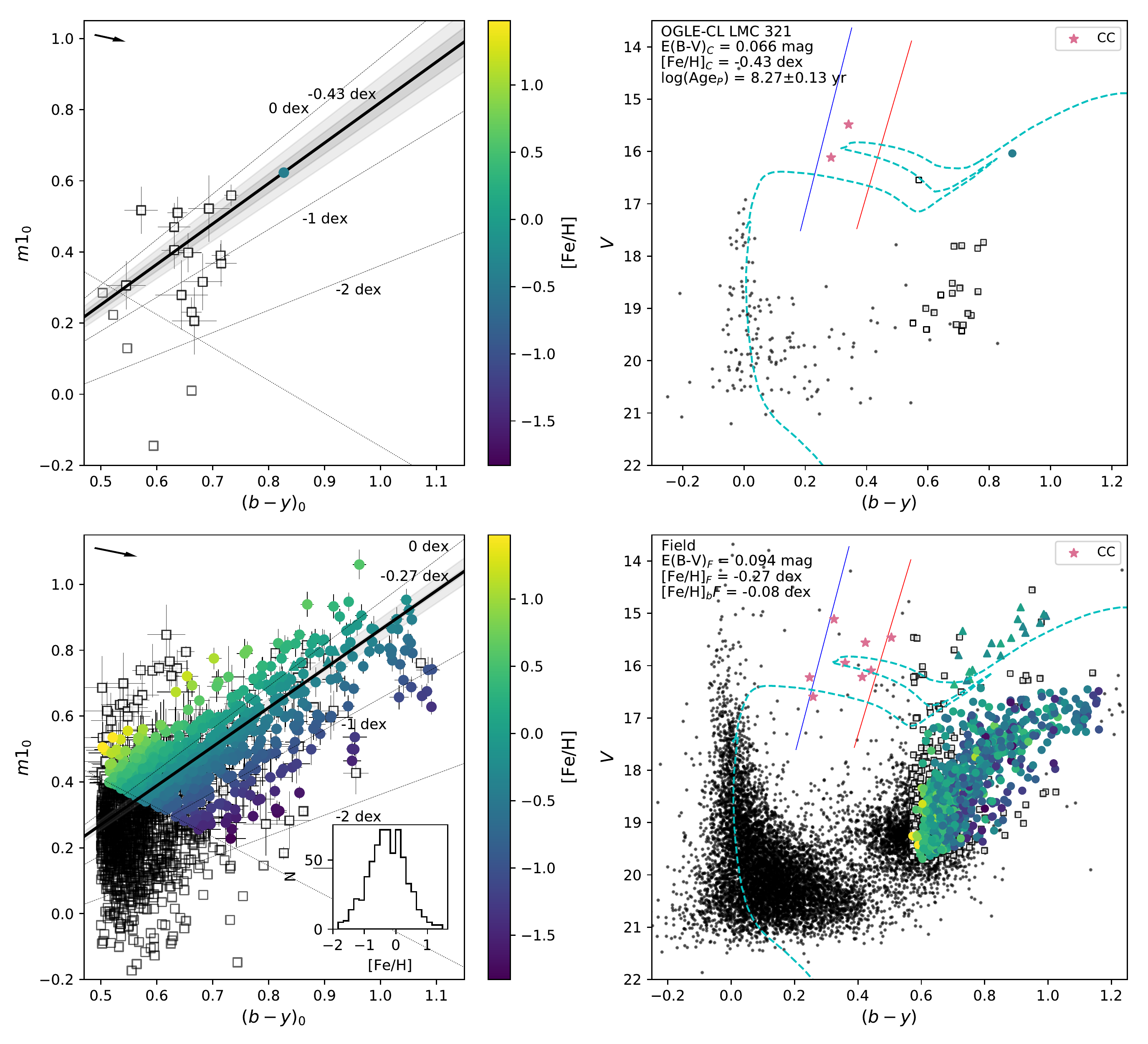}}
      \caption{Reddening-corrected two-color diagrams (left panels) and reddened
               CMDs (right panels) for OGLE-CL~LMC~321 (upper panels) and the surrounding
               field stars (lower panels).
               The symbols are the same as in Fig.~\ref{fig:ngc1903}.
              }
      \label{fig:lmc321}
   \end{figure*}
%
%-------------------------------------------------------------

%-------------------------------------------------------------
%                                             Two column Figure
%-------------------------------------------------------------
   \begin{figure*}
   \resizebox{\hsize}{!}
            {\includegraphics[]{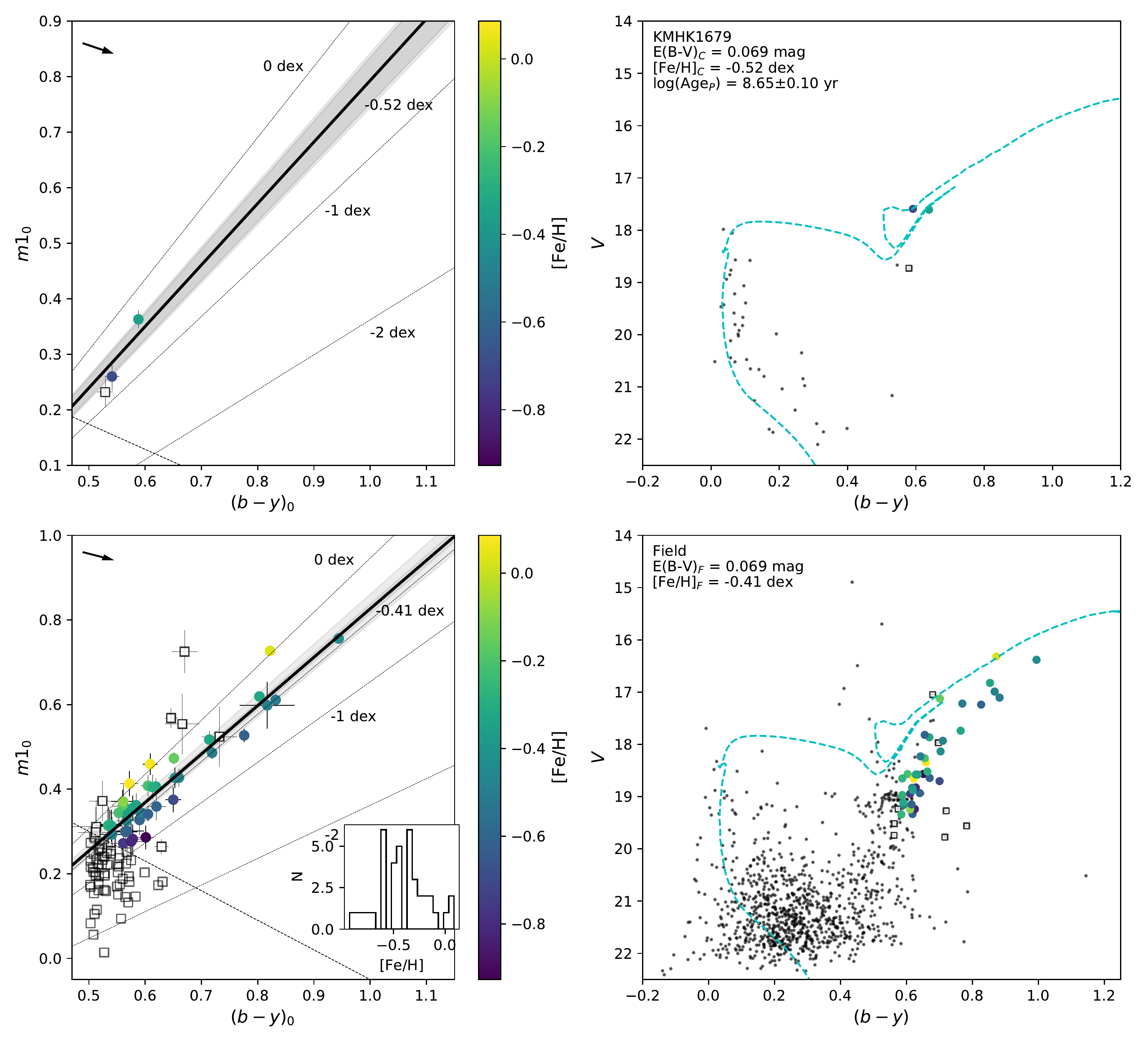}}
      \caption{Reddening-corrected two-color diagrams (left panels) and reddened
               CMDs (right panels) for KMHK1679 (upper panels) and the surrounding
               field stars (lower panels).
               The rest of the symbols are the same as in Fig.~\ref{fig:ngc1651}.
              }
      \label{fig:kmhk1679}
   \end{figure*}
%
%-------------------------------------------------------------

%-------------------------------------------------------------
%                                             Two column Figure
%-------------------------------------------------------------
   \begin{figure*}
   \resizebox{\hsize}{!}
            {\includegraphics[]{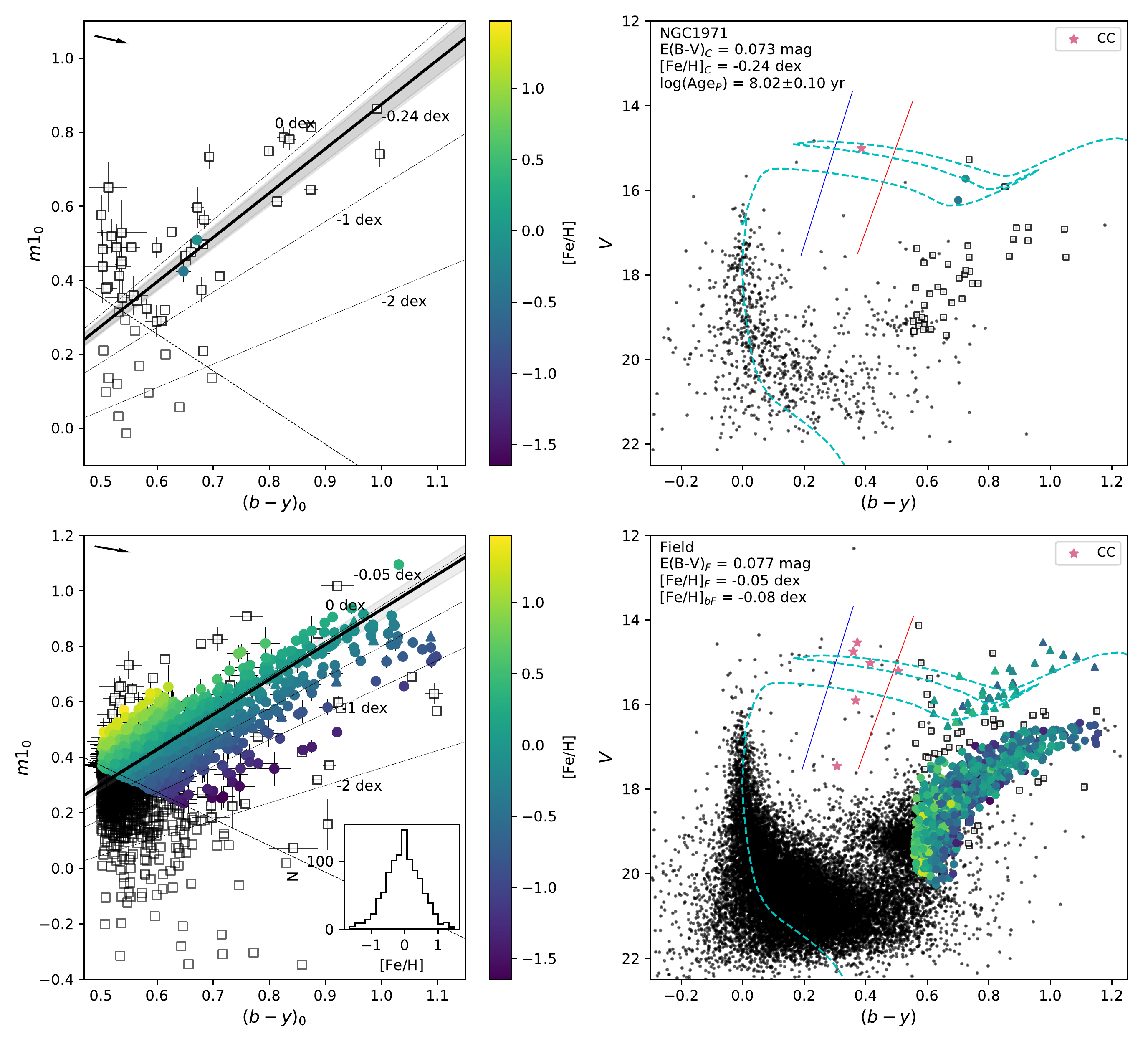}}
      \caption{Reddening-corrected two-color diagrams (left panels) and reddened
               CMDs (right panels) for NGC1971 (upper panels) and the surrounding
               field stars (lower panels).
               The symbols are the same as in Fig.~\ref{fig:ngc1903}.
              }
      \label{fig:ngc1971}
   \end{figure*}
%
%-------------------------------------------------------------

\end{appendix}

\end{document}